\documentclass[preprint,3p,times,onecolumn]{elsarticle}

\journal{Combustion and Flame}

\usepackage{latexsym,amssymb,graphicx,deluxetable,setspace}

\def\eps@scaling{1.0}% 
\newcommand\epsscale[1]{\gdef\eps@scaling{#1}}% 
\newcommand\plotone[1]{% 
 \centering 
 \leavevmode 
 \includegraphics[width={\eps@scaling\textwidth}]{#1}% 
}% 
\newcommand\plottwo[2]{% 
 \centering 
 \leavevmode 
 \textwidth=.45\textwidth 
 \includegraphics[width={\eps@scaling\textwidth}]{#1}% 
 \hfil
 \includegraphics[width={\eps@scaling\textwidth}]{#2}% 
}% 
\newcommand\plotfiddle[7]{% 
 \centering 
 \leavevmode 
 \vbox\@to#2{\rule{\z@}{#2}}% 
 \includegraphics[% 
  scale=#4, 
  angle=#3, 
  origin=c 
 ]{#1}% 
}% 

\newif\ifAMStwofonts
\ifAMStwofonts \else % If no AMS fonts

\fi

\newcommand{\pd}[2]{ { \partial {#1} \over \partial {#2} } }

\newcommand{\vect}[1]{\mathbf{#1}}   % Vector as bold
\newcommand{\sfrac}[2]{\,{}^{#1}\!/_{#2}}

\newcommand{\beq}{\begin{equation}}
\newcommand{\eeq}{\end{equation}}
\newcommand{\bdm}{\begin{displaymath}}
\newcommand{\edm}{\end{displaymath}}
\newcommand{\ie}{i.e., }
\newcommand{\ea}{et al.}

\begin{document}

\begin{frontmatter}



\title{The Interaction of High-Speed Turbulence with Flames: Turbulent
Flame Speed}

\author{A.Y.~Poludnenko\corref{ayp}}
\ead{apol@lcp.nrl.navy.mil}
\author{E.S.~Oran}
\cortext[ayp]{Corresponding author. Fax: +1 202 767 4798.}
\address{Laboratory for Computational Physics and Fluid Dynamics,
Naval Research Laboratory, Washington, DC 20375, USA}

\begin{abstract}

Direct numerical simulations of the interaction of a premixed flame
with driven, subsonic, homogeneous, isotropic, Kolmogorov-type
turbulence in an unconfined system are used to study the mechanisms
determining the turbulent flame speed, $S_T$, in the thin reaction
zone regime. High intensity turbulence is considered with the
r.m.s. velocity 35 times the laminar flame speed, $S_L$, resulting in
the Damk\"ohler number $Da = 0.05$. The simulations were performed
with Athena-RFX, a massively parallel, fully compressible, high-order,
dimensionally unsplit, reactive-flow code. A simplified
reaction-diffusion model, based on the one-step Arrhenius kinetics,
represents a stoichiometric H$_2$-air mixture under the assumption of
the Lewis number $Le = 1$. Global properties and the internal
structure of the flame were analyzed in an earlier paper, which showed
that this system represents turbulent combustion in the thin reaction
zone regime. This paper demonstrates that: (1) The flame brush has a
complex internal structure, in which the isosurfaces of higher fuel
mass fractions are folded on progressively smaller scales. (2) Global
properties of the turbulent flame are best represented by the
structure of the region of peak reaction rate, which defines the flame
surface. (3) In the thin reaction zone regime, $S_T$ is predominantly
determined by the increase of the flame surface area, $A_T$, caused by
turbulence. (4) The observed increase of $S_T$ relative to $S_L$
exceeds the corresponding increase of $A_T$ relative to the surface
area of the planar laminar flame, on average, by $\approx \! 14\%$,
varying from only a few percent to as high as $\approx \! 30\%$. (5)
This exaggerated response is the result of tight flame packing by
turbulence, which causes frequent flame collisions and formation of
regions of high flame curvature $\gtrsim 1/\delta_L$, or ``cusps,''
where $\delta_L$ is the thermal width of the laminar flame. (6) The
local flame speed in the cusps substantially exceeds its laminar
value, which results in a disproportionately large contribution of
cusps to $S_T$ compared with the flame surface area in them. (7) A
criterion is established for transition to the regime significantly
influenced by cusp formation.  In particular, at Karlovitz numbers $Ka
\gtrsim 20$, flame collisions provide an important mechanism
controlling $S_T$, in addition to the increase of $A_T$ by large-scale
motions and the potential enhancement of diffusive transport by
small-scale turbulence.

\end{abstract}

\end{frontmatter}



\section{Introduction}
%-------------------------------------------------------------------------------
\label{Intro}

One of the fundamental questions of turbulent combustion research
concerns our ability to understand and predict the rate of energy
release, or equivalently the burning speed, of a turbulent flame. This
can be achieved if two key aspects of the combustion process can be
determined: (1) its local properties, namely the local speed of flame
propagation, and (2) its global characteristics, i.e., the overall
structure of the turbulent flame that connects the local burning
velocity with the turbulent flame speed. In general, such local and
global characteristics are not universal for the combustion
process. They can vary substantially both in space and time due to the
unsteadiness, inhomogeneity, and anisotropy of the turbulent flow
associated with the system geometry, presence of walls and boundaries,
change in the flow conditions, etc.

Our present understanding of turbulent combustion is largely based on
the concept originally proposed almost 70 years ago by Damk\"ohler
\cite{Damkohler} (also see reviews by
\cite{Bilger,Driscoll,Lipatnikov1,Peters}). This concept is
formulated for a turbulent flame that is, on average, a planar,
quasi-one-dimensional structure propagating in a steady, homogeneous,
and isotropic background turbulent flow. According to Damk\"ohler's
suggestion, two qualitatively distinct regimes of turbulence-flame
interaction can be identified, and they are typically referred to as
``large-scale'' and ``small-scale'' turbulence
\cite{Damkohler,Peters}.

In the first regime, the overall structure of the turbulent flame is
determined by turbulent motions on large scales greater than the flame
width, $\delta_L$. These motions stretch and fold the flame, thereby
increasing its surface area, $A_T$. At the same time, they do not
affect the local burning velocity, $S_n$, with which the flame
propagates normal to its surface at each point and which remains equal
to the speed, $S_L$, of the unperturbed laminar flame. Consequently,
the speed, $S_T$, of the turbulent flame grows proportionally to the
increase of the flame surface area, i.e., $S_T = S_L(A_T/A_L)$, where
$A_L$ is the surface area of the planar laminar flame. The limitations
of this assumption were known practically at the time when this model
was proposed \cite{Zeldovich0}. In particular, when the Lewis number
$Le \ne 1$, $S_n$ can vary with the curvature of the flame. As a
result, the stretching and folding action of scales $\lambda >
\delta_L$ ends up affecting $S_n$. The relation between $A_T$ and
$S_T$, however, can be generalized by incorporating the effects of
flame strain and curvature in terms of the stretch factor $I$
(\cite{Driscoll}, also see \cite{Law} for the review of the theory of
flame stretch),
\beq
\frac{S_T}{S_L} = I \frac{A_T}{A_L}.
\label{e:Damk0}
\eeq

In the second regime, large-scale wrinkling of the flame by turbulence
is absent. Instead, small-scale motions, which are energetic enough,
penetrate and disrupt the internal flame structure. Thus molecular
diffusivity and thermal conduction become enhanced, or completely
dominated, by the turbulent transport associated with such small-scale
turbulence. As a result, in this regime, the turbulent flame speed is
suggested to be \cite{Damkohler}
\beq
\frac{S_T}{S_L} \sim \bigg(\frac{D_T}{D}\bigg)^{\sfrac{1}{2}},
\label{e:Damk1}
\eeq
where $D$ and $D_T$ are, respectively, the molecular and the effective
turbulent diffusivities.

These two regimes, in effect, represent two limits of low and high
turbulent intensity. In the first case, large-scale turbulence must be
slow enough so that the Kolmogorov scale $\eta > \delta_L$, thus
causing the flow to remain laminar on scales $\lambda \lesssim
\delta_L$. This combustion regime corresponds to the ``wrinkled'' and
``corrugated flamelets.'' In the opposite limit, the small-scale
turbulence regime is realized when turbulent motions on all scales in
the system are sufficiently fast to form an effectively
one-dimensional (1D) structure analogous to the planar laminar flame,
but which has a much larger width and speed. This second regime is
often associated with either a ``thin reaction zone'' flame
\cite{Peters} or a ``broken reaction zone'' (``distributed'') flame
\cite{Lipatnikov1,Schetinkov,Aspden,Aspden2}. As a consequence of
considering these two limits, description of the turbulent flame
becomes significantly simplified. The large-scale regime assumes very
simple local flame properties, i.e., those of the laminar flame, but a
complex global structure of a highly convolved flame surface. The
small-scale regime, on the other hand, trades the simplicity of the
global structure of a planar flame for the complexity of its internal
structure, which is pronouncedly distinct from that of the laminar
flame. Therefore, each of these regimes has only one dynamically
important quantity that completely determines $S_T$, namely $A_T$ for
the large-scale turbulence and $D_T$ for the small-scale turbulence.

This picture then naturally raises the following question. How does
the transition occur between these two limiting regimes of combustion?
At some intermediate turbulent intensities, the flow will cease to be
laminar on scales $\lambda \lesssim \delta_L$. As a result, turbulence
will begin to penetrate the flame, disrupting it and causing its
internal structure to differ from that of the laminar flame. At the
same time, the flame can remain sufficiently thin to form, under the
action of high-intensity large-scale motions, a highly convolved and
tightly packed configuration. The picture outlined above would then
imply a gradual shift with increasing turbulent intensity from the
large-scale turbulence being the dominant process controlling $S_T$ to
the small-scale one. Over a substantial range of turbulent
intensities, however, it must be their combined action that determines
the magnitude of $S_T$.

Of particular interest, in this respect, is the ``thin reaction zone''
regime, in which only the preheat zone is disrupted by turbulence. The
presence of turbulent motions on scales $\lambda \lesssim \delta_L$
suggests that the molecular and thermal transport between the reaction
and preheat zones are affected by turbulence. This should cause $S_n$
to differ from $S_L$, which is characteristic of the small-scale
turbulence regime. On the other hand, if the structure of the reaction
zone is the same as in the laminar case, as occurs at the large-scale
boundary of this regime, then $S_n$ must equal $S_L$. Indeed, $S_n =
(1/\rho_0)\int \rho \dot{Y}d\zeta$, where $\rho_0$ is the fuel
density, $\dot{Y}$ is the reaction rate, and $\zeta$ is the coordinate
normal to the flame. Since $\dot{Y} \ne 0$ only in the reaction zone
where it does not differ from the laminar flame, it follows that $S_n
= S_L$. What does then determine $S_T$ in the thin reaction zone
regime? Is it folding and stretching of the flame by turbulence,
implying the validity of eq.~(\ref{e:Damk0}), or the microscopic
turbulent transport represented by eq.~(\ref{e:Damk1}), or both? How
does the balance between the two processes change with turbulent
intensity?  Finally, is the effect of high-speed turbulence on the
flame limited to the two processes discussed above or do other
mechanisms also play a role?

Before these questions can be addressed, several conceptual
difficulties must first be resolved. Most importantly, the definition
of the flame surface and its area must be revisited. This question,
which has a seemingly simple answer at low turbulent intensities,
becomes quite nontrivial when the flow is no longer laminar on scales
$\lambda \lesssim \delta_L$. In this case, different regions of the
flame are affected differently by turbulence and, thus, the
isosurfaces of temperature or fuel mass fraction are not necessarily
parallel. Furthermore, without an accurate definition of the flame
surface, it is not clear what the local burning speed of the flame is,
since this quantity typically describes flame propagation normal to
its surface and determining its local value requires integration of
the distribution of $\dot{Y}$ normal to the flame surface, as was
discussed above.

In recent years, the advent of direct numerical simulations (DNS) has
provided a tool permitting the mechanisms that determine $S_T$ to be
directly investigated. DNS have been used by a number of groups to
study the dynamics and properties of turbulent flames
\cite{Aspden}-\cite{Poludnenko} (also see a review by
\cite{Lipatnikov2}). With the exception of the work by Aspden et
al. \cite{Aspden,Aspden2} that analyzed the fully distributed
thermonuclear flame, these studies generally assume relatively low
turbulent intensities. Several authors also did explicitly consider
the relation between $S_T$ and $A_T$.  Bell et al. \cite{Bell}
analyzed the three-dimensional (3D) DNS of a statistically planar
turbulent methane flame and determined that $I$ deviates from unity by
$\lesssim \! 10\%$. A similar estimate was obtained by Khokhlov
\cite{Khokhlov} in 3D simulations of Rayleigh-Taylor driven
thermonuclear flames in degenerate matter. Both of these studies,
however, considered reactive mixtures in which $Le \ne 1$ and,
therefore, $I$ would not be expected to be exactly equal to one. In
this context, the result of \cite{Khokhlov} is particularly
interesting since degenerate matter is characterized by $Le \gg 1$ and
typically in excess of $10^3$. Furthermore, both Bell et al.
\cite{Bell} and Khokhlov \cite{Khokhlov} considered turbulent
intensities which were too low to allow small-scale motions to
penetrate efficiently the interior of the flame. For instance, in
\cite{Bell}, the characteristic turbulent velocity was $\approx \!
4.3S_L$, and in \cite{Khokhlov}, it reached values $\lesssim \!
12S_L$. Consequently, no effects of the microscopic turbulent
transport would be expected in these cases.

In this respect, the work of Hawkes \& Chen \cite{Hawkes} is of
particular interest as it considered the validity of
eq.~(\ref{e:Damk0}) in the thin reaction zone regime. They analyzed
simulations of lean, statistically flat turbulent methane flames at
turbulent intensities $\lesssim \! 28.5S_L$ and found values of $I$
within a few percent of unity, with the exception of the
methane-hydrogen flame for which $I = 1.18$ was determined.
Interpretation of the implications of these results for the thin
reaction zone regime, however, is complicated by the two-dimensional
(2D) nature of their simulations. Moreover, the turbulent integral
scale was $< \! \delta_L$, which substantially suppressed flame
wrinkling and prevented the development of a highly convolved flame
which would typically be expected to form in a realistic setting.

The studies performed to-date, thus, generally show that, at low
turbulent intensities, the increase in $S_T$ is indeed almost
completely determined by the growth of $A_T$. At the same time, they
leave open the questions discussed above regarding the mechanisms
controlling $S_T$ in the thin reaction zone regime, let alone at
higher turbulent intensities.

The objective of this paper is to begin addressing these questions
systematically by first considering the interaction of a statistically
planar, turbulent flame with the steady, homogeneous, isotropic
turbulence. We assume sufficiently high turbulent intensity to
represent the regime that is borderline between thin and broken
reaction zones, according to the traditional combustion regime
diagrams \cite{Peters,Williams,Williams2} (cf. Fig.~2 in
\cite{Poludnenko}). In other words, we consider the highest-intensity
turbulence, which has been hypothesized to allow for the existence of
a flame with the internal structure of the reaction zone that is
essentially not affected by turbulent transport. The turbulent
r.m.s. velocity of the flow in cold fuel is $U_{rms} \approx 35S_L$,
leading to the Damk\"ohler number $Da = 0.05$ and Gibson scale $L_G
\approx 3\times10^{-4}\delta_L$. Here we focus on the $Le = 1$
situation to exclude thermodiffusive effects as a potential source of
variations of the local flame speed. Developing rigorous understanding
of this simpler case will then serve as a starting point for the study
of more realistic situations represented by more complex chemistry.

This paper continues the analysis of the simulations first presented
in \cite{Poludnenko}. The primary focus of \cite{Poludnenko} was a
detailed study of flame properties and evolution in the presence of
such high-speed turbulence. In particular, a method was presented
which allowed the direct determination of the internal structure of
the flame based on its actual 3D configuration inside the flame
brush. The analysis showed that the preheat zone is broadened by
turbulence while the reaction-zone structure remains virtually
identical to that of the planar laminar flame. Therefore, this study
demonstrated that, even under the action of such intense turbulence,
the system can be classified as being in the thin reaction zone
regime.

Our starting point in this work is the question of the definition of
the flame surface area in the high-speed turbulent flow. Answering
this will then allow us to determine whether a relation between $S_T$
and $A_T$, similar to eq.~(\ref{e:Damk0}), can indeed be established
and, thus, which processes control $S_T$ in the thin reaction zone
regime.

\section{Numerical method and simulations performed}
%-------------------------------------------------------------------------------
\label{Methods}

\subsection{Physical and numerical models}
\label{ModelMethod}

Here we summarize the physical model, the numerical method used, and
key aspects of the simulation setup. A more detailed discussion can be
found in \cite{Poludnenko}.

We solve the system of unsteady, compressible, reactive flow
equations,
\begin{eqnarray}
\pd{\rho}{t} + \nabla \cdot (\rho \vect{u} )          & = & 0,
\label{Euler1} \\
\pd{\rho \vect{u}}{t} + \nabla \cdot ( \rho \vect{u} \otimes \vect{u}) 
				+ \nabla P            & = & 0,
\label{Euler2} \\
\pd{E}{t} + \nabla \cdot \big((E+P) \vect{u}\big) 
        - \nabla \cdot \big(\mathcal{K} \nabla T\big) & = & -\rho q \dot{Y},
\label{Euler3} \\
\pd{\rho Y}{t} + \nabla \cdot \big(\rho Y \vect{u}\big)
	- \nabla \cdot \big(\rho D \nabla Y\big)      & = & \rho \dot{Y}.
\label{Euler4}
\end{eqnarray}
Here $\rho$ is the mass density, $\vect{u}$ is the velocity, $E$ is
the energy density, $P$ is the pressure, $Y$ is the mass fraction of
the reactants, $q$ is the chemical energy release, and $\dot{Y}$ is
the reaction source term. The coefficients of thermal conduction,
$\mathcal{K}$, and molecular diffusion, $D$, are
\beq
D = D_0 \frac{T^n}{\rho}, \quad
\frac{\mathcal{K}}{\rho C_p} = \kappa_0 \frac{T^n}{\rho},
\label{e:CondDiff}
\eeq
where $D_0$, $\kappa_0$, and $n$ are constants, and $C_p = \gamma
R/M(\gamma-1)$ is the specific heat at constant pressure. The equation
of state is that of an ideal gas. Chemical reactions are modeled using
the first-order Arrhenius kinetics
\begin{equation}
\frac{dY}{dt} \equiv \dot{Y} = -\rho Y B\exp \Big(-\frac{Q}{RT}\Big),
\label{e:ReacModel}
\end{equation}
where $B$ is the pre-exponential factor, $Q$ is the activation energy,
and $R$ is the universal gas constant.

Table~\ref{t:Params} summarizes the parameters of the physical model
used as well as the resulting properties of the planar laminar flame
\cite{Poludnenko}. These parameters are based on a simplified
reaction-diffusion model designed to represent the stoichiometric
H$_2$-air mixture \cite{Gamezo}. We refer to \cite{Poludnenko,Gamezo}
for the detailed discussion of the limitations of applicability of
this model for the description of the actual H$_2$-air mixture.

Eqs.~(\ref{Euler1})-(\ref{e:ReacModel}) are solved using the code
Athena-RFX \cite{Poludnenko} -- the reactive flow extension of the
magnetohydrodynamic code Athena \cite{Athena, Gardiner}. Athena-RFX is
a fixed-grid, massively parallel, finite-volume, fully conservative
code. It implements a variant \cite{Gardiner} of the fully unsplit
corner-transport upwind (CTU) algorithm \cite{Colella} and its 3D
extension presented in \cite{Saltzman}, in conjunction with the PPM
spatial reconstruction \cite{PPM} and the approximate nonlinear HLLC
Riemann solver. Overall, the code achieves 3$^{rd}$-order accuracy in
space and 2$^{nd}$-order accuracy in time.  A detailed description and
an extensive suite of tests of the hydrodynamic integration algorithm
can be found in \cite{Athena, Gardiner}. Implementation of the
reactive-diffusive extensions in Athena-RFX, along with the results of
tests including convergence studies, is discussed in detail in
\cite{Poludnenko,Poludnenko2}.

The simulations presented here model flame interaction with steady,
homogeneous, isotropic turbulence, described by the classical
Kolmogorov theory \cite{K41}. Turbulence driving is implemented using
a spectral method \cite{Poludnenko,Poludnenko2} similar to the one
used in \cite{Stone,Lemaster}. In this method, velocity perturbations
$\delta \hat{\vect{u}}(\vect{k})$ are initialized in Fourier space,
and each component $\delta \hat{u}'_i(\vect{k})$ is an independent
realization of a Gaussian random field superimposed with the desired
energy injection spectrum. Subsequently, nonsolenoidal components of
$\delta \hat{\vect{u}}(\vect{k})$ are removed to ensure that $\nabla
\cdot \delta \vect{u}(\vect{x}) = 0$. An inverse Fourier transform of
$\delta \hat{\vect{u}}(\vect{k})$ gives $\delta \vect{u}(\vect{x})$,
the velocity perturbation field in the physical space. The $\delta
\vect{u}(\vect{x})$ is normalized to provide a constant rate
$\varepsilon$ of kinetic-energy injection, and the total momentum in
the perturbation field is subtracted from $\delta \vect{u}(\vect{x})$
to ensure that no net momentum is added to the domain, \ie $\int \rho
\delta \vect{u} = 0$. The resulting velocity perturbations are added to
the flow field $\vect{u}(\vect{x})$ at every time step, and the
overall perturbation pattern is regenerated at every time interval
$\Delta t_{vp} \sim 10\Delta x/c_s$, where $c_s$ is the maximum
sound speed in the domain and $\Delta x$ is the computational cell
size.

\begin{figure}[t]
\centering
\includegraphics[clip, width=0.5\textwidth]{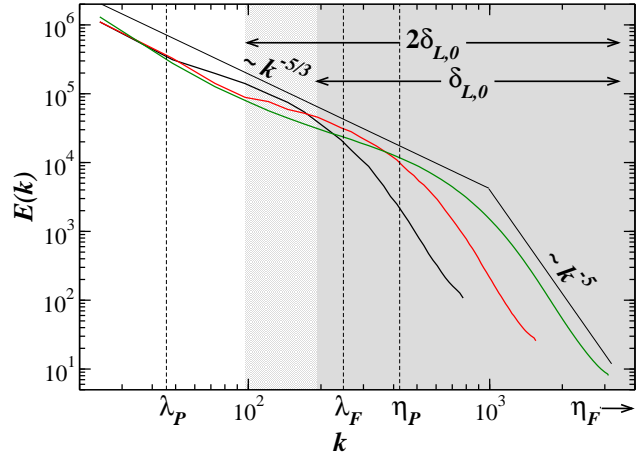}
\caption{Instantaneous kinetic energy spectral density for simulations
S1 (black), S2 (red), and S3 (green) at a time immediately prior to
ignition. Shaded regions illustrate scales associated with thermal
width $\delta_{L,0}$ and full width $2\delta_{L,0}$ of the laminar
flame.  Vertical dashed lines show wavenumbers corresponding to the
Taylor microscales in the product, $\lambda_P$, and fuel, $\lambda_F$,
as well as the Kolmogorov scale in the product, $\eta_P$, based on the
value of the viscosity coefficient corresponding to the $Sc = Pr = 1$
condition. The wavenumber corresponding to the Kolmogorov scale in the
fuel, $\eta_F = 1.18\times 10^{-3}$ cm, is located outside the range
of the graph. (Reproduced from \cite{Poludnenko}.)}
\label{f:FlameSpectra}
\end{figure}

Energy is injected only at the scale of the domain width, $L$, to
obtain the Kolmogorov-type spectrum. The resulting turbulence is
statistically steady, isotropic, and homogeneous with the inertial
range of the energy cascade extending all the way to the energy
injection scale (Fig.~\ref{f:FlameSpectra}). Moreover, since the
velocity perturbation field is divergence-free, no compressions or
rarefactions are artificially induced as a result of driving.

Eqs.~(\ref{Euler1})-(\ref{Euler4}) do not explicitly include physical
viscosity, but instead rely on numerical viscosity to provide the
kinetic-energy dissipation. By systematically varying the numerical
resolution in this approach, the energy spectrum can be extended to
progressively smaller scales $\lambda \ll \delta_{L,0}$ while
maintaining the spectrum constant on larger scales.  This is
illustrated in Fig.~\ref{f:FlameSpectra}, which shows instantaneous
kinetic energy spectra in the cold nonreactive flow immediately before
the flame is initiated in the computational domain in three
simulations with progressively increasing resolution. Such approach
allows us to vary only the intensity of the small-scale turbulent
motions, and thus to investigate their effects by differentiating them
from the effects of large scales. Furthermore, it shows the range of
scales that must be resolved in a numerical simulation in order to
capture accurately the evolution of the turbulent flame. This is
particularly important in the context of high-speed turbulent reactive
flows, in which it is typically impossible to resolve the Kolmogorov
scale. The analysis presented in \cite{Poludnenko} demonstrated that
scales $\lambda \ll \delta_{L,0}$ have virtually no effect on the
properties of the turbulent flame with the exception of the degree of
flame surface wrinkling on the fuel side. This showed that the
evolution of the turbulent flame can be accurately reproduced without
the need to resolve scales $\lambda \ll \delta_{L,0}$. We refer to
\cite{Poludnenko} for a detailed discussion of this issue and, in
particular, for a discussion of the applicability of the results to
the actual stoichiometric H$_2$-air mixture.

\subsection{Summary of simulations}
%-------------------------------------------------------------------------------
\label{Simulations}

Key parameters of the simulations discussed here are summarized in
Table~\ref{t:Runs}. The main difference between the three calculations
is the resolution, which progressively increases from $\Delta x =
\delta_{L,0}/8$ in S1 to $\Delta x = \delta_{L,0}/32$ in S3. The domain
in all cases was initialized with uniform density $\rho_0$ and
temperature $T_0$ (see Table~\ref{t:Params}). In S1 and S2, initial
fluid velocities were set to 0. In contrast, the velocity field in S3
was initialized with the ideal energy spectrum $\mathcal{E}(k) \propto
k^{-5/3}$ extending from the energy injection scale $L$ to the
numerical Kolmogorov scale $\eta = 2\Delta x$. This initial spectrum
was normalized to ensure that at $t = 0$ the total kinetic energy in
the domain was equal to its predicted steady-state value
\cite{Poludnenko2}. In S1 and S2, the flow field was allowed to evolve
for the time $t_{ign} = 3\tau_{ed}$, and in S3 for the time $t_{ign} =
2\tau_{ed}$, to develop the steady-state turbulent flow field. The
corresponding instantaneous spectral density of the specific kinetic
energy immediately prior to $t_{ign}$ in S1-S3 is shown in
Fig.~\ref{f:FlameSpectra}. At the time $t_{ign}$, a planar laminar
flame with its front parallel to the $x\!-\!y$ plane\footnote{The
longest dimension of the domain is along the $z$-axis.} was
initialized in the domain with the values of $\rho$, $T$, and $Y$
based on the exact laminar flame solution. The velocity field was not
modified, thereby preserving its structure that developed during the
equilibration stage.

Prior to $t_{ign}$, all domain boundaries were periodic. At $t_{ign}$,
boundary conditions (BCs) along the left and right $z$-boundaries were
switched to zero-order extrapolation in order to prevent pressure
build-up inside the domain and to accommodate the resulting global
fluid flow. Reflections of the acoustic perturbations caused by such
BCs can be neglected in comparison with the acoustic noise generated
by the turbulent flow field itself, and overall such choice of the BCs
was found not to introduce any unphysical artifacts and thus not to
affect the solution accuracy. Furthermore, any potential effect of the
outflow BCs was minimized by the large length-to-width ratio of the
computational domain, which allowed the flame brush to remain
sufficiently far from the boundaries at all times \cite{Poludnenko}.
Hereafter, for simplicity we refer to the moment of ignition as $t =
0$.

Energy is injected into the domain with the constant rate
$\varepsilon$ per unit volume for the total duration of the
simulations to provide steady driving of the turbulent flow. The value
of $\varepsilon$ was chosen to produce a high-intensity turbulence
field that, however, was weak enough to minimize the probability of
creating weak transonic shocklets arising from the intermittency in
turbulent flow. The characteristic turbulent velocities and the
turbulent integral scale of the steady-state flow before the moment of
ignition are listed in Table~\ref{t:Runs}. Corresponding values of the
Damk\"ohler number and the Gibson scale, given in the table, show that
$Da \ll 1$ and $L_G \ll \delta_{L,0}$. Prior analysis of the resulting
structure of the turbulent flame demonstrated that, despite the high
turbulent intensity, small-scale turbulence fails to penetrate the
internal structure of the reaction zone, and the system evolves in the
thin reaction zone regime \cite{Poludnenko}.

The following two key global characteristics of the turbulent flame
are used extensively in this work. The width of the turbulent flame
brush is defined as
\beq
\delta_T = z_{1,max} - z_{0,min},
\label{e:deltaT}
\eeq
where $z_{0,min}$ and $z_{1,max}$ are defined as
\beq
\begin{array}{l}
z_{0,min} = \max(z): Y(x,y,z) < 0.05 \ \forall \ (x, y, z < z_{0,min}), \\
z_{1,max} = \min(z): Y(x,y,z) > 0.95 \ \forall \ (x, y, z > z_{1,max}).
\end{array}
\label{e:z0z1}
\eeq
In other words, $z_{0,min}$ marks the rightmost $x$-$y$-plane, to the
left of which is pure product, while $z_{1,max}$ marks the leftmost
$x$-$y$-plane, to the right of which is pure fuel. This is illustrated
in Fig.~\ref{f:isoS}. The turbulent flame speed is defined as
\beq
S_T = \frac{\dot{m}_R}{\rho_0 L^2},
\label{e:ST}
\eeq
where $\dot{m}_R$ is the total rate of fuel consumption inside the
flame brush, \ie the total mass of reactants which is transformed to
product per unit time. The detailed discussion of this choice of the
definition of $S_T$ can be found in \cite{Poludnenko}.

Kinetic energy dissipation in the turbulent flow in the domain causes
gradual heating of the fuel. The resulting relative increase in the
internal energy, $E_t$, and temperature of the nonreactive flow in one
eddy turnover time, $\tau_{ed}$, can be accurately estimated as
\cite{Poludnenko,Poludnenko2}
\beq
\frac{E_t - E_{t,0}}{E_{t,0}} = \frac{T - T_0}{T_0} =
D_K\gamma (\gamma - 1)Ma_F^2.
\label{e:dEdT}
\eeq
Here the constant $D_K = 0.5$ and $E_{t,0}$ is the thermal energy in
the domain at $t = 0$. Based on values of $\gamma$ from
Table~\ref{t:Params} and $Ma_F$ from Table~\ref{t:Runs}, $T$ increases
by $\approx \! 1.8$ K over $\tau_{ed}$. Consequently, by the end of
the simulation, the fuel temperature in the domain rises from $T_0 =
293$ K to $\approx \! 320\!-\!330$ K.\footnote{Temperature increase
occurs, on average, at constant density.} As a result of such change
in the thermodynamic state of fuel, the corresponding laminar flame
properties also vary with time. In particular, this concerns the
instantaneous thermal width, $\delta_L(t)$, and speed, $S_L(t)$, of
the laminar flame, which will deviate from their initial values
$\delta_{L,0}$ and $S_{L,0}$ corresponding to the fuel temperature
$T_0$ and pressure $P_0$ (Table~\ref{t:Params}).

In order to determine $\delta_L$ and $S_L$ in simulations S1-S3, we
found at each time $T_{aver}(t)$ and $P_{aver}(t)$ averaged over all
cells located between the planes $z = z_{1,max} + L$ and $z =
z_{1,max} + 2L$. This provided the average current thermodynamic state
of fuel entering the flame brush. Based on $T_{aver}$ and $P_{aver}$,
an exact laminar flame solution was constructed, which gave the
instantaneous values of $\delta_L(t)$ and $S_L(t)$. This analysis
demonstrated that in the simulations, $\delta_L$ decreases by
$\lesssim 15\%$. The $S_L$ changes more substantially increasing by
$\lesssim 25\%$. Thus, in order to consider accurately the nature of
the turbulent flame speed in the high-speed turbulent reactive flows,
it is important to account for the increase in the local burning speed
of the flame due to the turbulent fuel heating. Consequently, in our
further analysis, we relate $S_T$ to $S_L$ rather than $S_{L,0}$.

Detailed discussion of the turbulent flame evolution in the
simulations is given in \cite{Poludnenko}. Here, for completeness, we
reproduce in Fig.~\ref{f:LV} for all three calculations the time
histories of $\delta_T$, normalized by $\delta_L$, as well as $S_T$,
normalized by $S_L$ (cf. Fig.~4 in \cite{Poludnenko}). 

Finally, Table~\ref{t:Props} lists the time-averaged values of both
$\delta_T/\delta_L$ and $S_T/S_L$ \cite{Poludnenko} as well as of
other key characteristics of the turbulent flame discussed below.
Table~\ref{t:Props} also shows the corresponding order of
self-convergence for each quantity listed. Since the computational
cell size decreases progressively by a factor of 2 for each
simulation, the order of self-convergence $O(\phi)$ of a variable
$\phi$ is defined as
\beq
O(\phi) = \log_2\Bigg(\frac{|\phi_{S1} - \phi_{S3}|}
{|\phi_{S2}-\phi_{S3}|}\Bigg).
\eeq

\section{Surface area and density of the fuel mass fraction isosurfaces}
%-------------------------------------------------------------------------------
\label{AreaDensity}

\subsection{Isosurface area}
\label{A:Isosurf}

Fig.~\ref{f:ASigma}a,b shows the evolution of the surface areas
$A_{0.01}$ and $A_{0.99}$ of the isosurfaces of the fuel mass fraction
$Y = 0.01$ and $Y = 0.99$, respectively. All isosurfaces are
constructed using the ``marching cubes'' algorithm. These isosurfaces
represent the outer boundaries that separate the flame brush from pure
product and pure fuel. All values are normalized by the domain
cross-section $L^2$ corresponding to the surface area of the planar
laminar flame unperturbed by turbulence. In principle, the
normalization should be performed over some average surface area of
the flame brush, thus reflecting its overall large-scale shape. With
sufficient accuracy, however, the turbulent flame in the system
considered here can be viewed, on average, as a planar propagating
front, and thus we normalize over its area $L^2$.

Fig.~\ref{f:ASigma}a shows that $A_{0.01}$ exhibits substantial
variability, similar to $\delta_T$ and $S_T$ (cf. Fig.~\ref{f:LV}).
There is a clear correspondence between the peaks and troughs of
$A_{0.01}$ and $S_T$ and, to a lesser extent, $\delta_T$.  Moreover,
in a manner similar to $\delta_T$ and $S_T$, $A_{0.01}$ varies less
with increasing resolution, and it appears on average to have
converged.

Such correlation between $A_{0.01}$ and $S_T$ reflects the
evolutionary cycle of the flame brush, discussed in \cite{Poludnenko}.
In the turbulent flame, which on average is in a steady state, the
influx of fresh fuel and the rate of its consumption never perfectly
balance each other. Instead, periodically either one or the other
process dominates. In particular, an increase in the flame surface
area inside the brush, and the associated increase in $\delta_T$,
leads to a higher $S_T$. This causes rapid consumption of fuel inside
the brush, decreasing both the flame surface area and the overall
width of the flame brush. The result is a slower, less convolved
flame, which is thus more susceptible to the action of turbulence.
This leads to increased folding and stretching of the flame, and the
cycle repeats.

The behavior of $A_{0.99}$ is pronouncedly different. There appears to
be no correlation between variations in $A_{0.99}$ and either
$\delta_T$ or $S_T$. Moreover, with increasing resolution, the
magnitude of the variations increases, and there is no evidence of
convergence of the growing average values of $A_{0.99}$. We will
consider the correlation between $A_Y$ and both $S_T$ and $\delta_T$
in a more quantitative form in \S~\ref{FlameArea}.

These conclusions regarding the change of $A_{0.01}$ and $A_{0.99}$
with resolution are supported by Fig.~\ref{f:ASigma}e, which shows the
full normalized time-averaged distributions $\overline{A}_Y/L^2$ for
all values of $Y$. In simulation S1, $\overline{A}_{0.01}$ and
$\overline{A}_{0.99}$ are nearly equal. With increasing resolution,
however, they diverge to the point that, in S3, there is almost a
factor of 2 difference between them. Values of $\overline{A}_{0.01}$
steadily decrease with resolution and they indeed demonstrate
convergence. At the same time, $\overline{A}_{0.99}$ increases with
resolution, and the difference between S2 and S3 is only marginally
smaller than between S1 and S2.

This behavior is part of a broader qualitative change in the overall
shape of the distribution of $\overline{A}_Y$ that occurs around $Y
\approx 0.5$, \ie at the boundary of the reaction and preheat zones.
The lower resolution of S1 suppresses small-scale turbulent motions,
which, in turn, causes less wrinkling of the isosurfaces in the
preheat zone. As a result, the flame-brush surface appears similar
both on the product and fuel sides (cf. Fig.~3 in \cite{Poludnenko}).
With higher resolution, however, isosurface wrinkling becomes more
pronounced with increasing distance from the reaction zone, and the
distribution of $\overline{A}_Y$ develops a distinctive inverted-S
shape, as seen in Fig.~\ref{f:ASigma}e. The consequence of this was
observed in \cite{Poludnenko}, which reported a much more highly
convolved flame-brush surface on the fuel side in calculations S2 and
S3. We will discuss the role of small-scale turbulence in further
detail in \S~\ref{D:SmallScale}.

Profiles of $\overline{A}_Y$ in S2 and S3 are closer to each other in
the reaction zone than in the preheat zone (Fig.~\ref{f:ASigma}e).
Table~\ref{t:Props} shows that in the region of peak reaction rate,
i.e., at $Y \approx 0.15$, $\overline{A}_Y/L^2$ exhibits the
3$^{rd}$-order convergence. This is substantially faster than what
would be expected for an overall 2$^{nd}$-order-accurate
code. Moreover, in all three calculations, $\overline{A}_Y$ varies the
slowest inside the reaction zone, \ie for $Y \approx 0.15 - 0.6$. This
shows that isosurfaces of $Y$, on average, follow each other
relatively closely within the reaction zone, which thus can be viewed
as being uniformly folded and stretched by turbulence as a coherent
structure. This is consistent with the very low variability in the
reaction zone of the instantaneous profiles of $Y$ and $T$, which
represent the internal flame structure
\cite{Poludnenko}.

\subsection{Isosurface density}
%-------------------------------------------------------------------------------
\label{D:Isosurf}

The analysis of the isosurface area distribution does not show how the
flame is organized inside the flame brush and, in particular, how
tightly it is packed. More specifically, the question arises whether
variations in the isosurface area seen in Fig.~\ref{f:ASigma}a,b can
be accounted for only by the change in the turbulent flame thickness,
or whether they are also related to how the flame is folded inside the
flame brush.

Consider the surface density of the fuel mass fraction isosurfaces,
defined as the isosurface area for a given $Y$ normalized by the
effective flame-brush volume,
\beq
\Sigma_Y = \frac{A_Y}{\delta_T L^2}.
\label{e:Sigma}
\eeq
A detailed discussion of this definition of $\Sigma_Y$ and, in
particular, of the choice of the uniform normalization by the full
flame-brush volume is given in \ref{AppA}.

The evolution of $\Sigma_{0.01}$ and $\Sigma_{0.99}$ is shown in
Fig.~\ref{f:ASigma}c,d. The overall trends are similar to those
exhibited by $A_{0.01}$ and $A_{0.99}$. In particular, $\Sigma_Y$ is
lower on the product side of the flame. Similar to $A_Y$, $\Sigma_Y$
rapidly converges with resolution at lower values of $Y$ and shows
virtually no convergence in the preheat zone. Most important, there is
substantial variability in $\Sigma_Y$, especially on the fuel side,
where $\Sigma_Y$ changes by more than a factor of 3 in the highest
resolution case. This tighter packing of the isosurfaces, which occurs
periodically in the course of system evolution, suggests that the
increase in $A_Y$ is associated not only with the increase of
$\delta_T$ but also of $\Sigma_Y$.

Fig.~\ref{f:ASigma}f shows the time-averaged distributions of
$\Sigma_Y$ in simulations S1-S3. Comparison with Fig.~\ref{f:ASigma}e
demonstrates that both $\overline{A}_Y/L^2$ and $\overline{\Sigma}_Y$
show remarkably similar behavior. The time-averaged values of
$\overline{\Sigma}_Y$ are also not constant, but, instead, for
sufficiently high numerical resolution, they progressively increase
through the flame interior from its product side to the fuel side. The
resulting inverted-S shape of the profiles is very similar to that of
$\overline{A}_Y$ with both $\overline{A}_Y$ and $\overline{\Sigma}_Y$
at the two extreme values of $Y$ ($Y = 0.01$ and $0.99$) differing in
S3 by almost a factor of two. Furthermore, similar to
$\overline{A}_Y$, $\overline{\Sigma}_Y$ varies the least in the
interior of the reaction zone, i.e., for $Y \approx 0.15 - 0.6$, while
outside this region it shows a strong dependence on $Y$. Just outside
the reaction zone, individual profiles begin to diverge from each
other, with the variation between them becoming progressively larger
for higher values of $Y$.

In the reaction zone, values of $\overline{\Sigma}_Y$ for all three
calculations are much closer than those of $\overline{A}_Y$. In fact,
Table~\ref{t:Props} shows that $\overline{\Sigma}_Y$ exhibits
4$^{th}$-order convergence compared with 3$^{rd}$-order convergence
for $\overline{A}_Y$. As a result, values of $\overline{\Sigma}_Y$ are
virtually identical for $Y \approx 0.01 - 0.4$ in S2 and S3. On the
other hand, in the preheat zone, the profiles of $\overline{\Sigma}_Y$
diverge more than in the case of $\overline{A}_Y$. This shows that the
effects of small-scale turbulent motions on the reaction and the
preheat zones are, in fact, even more disparate in terms of their
ability to provide tight folding of the isosurfaces, compared with
their ability to increase the isosurface area.

Flame surface density is a quantity used both in experimental and
theoretical combustion research. In particular, $\Sigma_{max}$ is
defined as the surface density in the region of the flame brush with
mean reactedness $\bar{c} = 0.5$, where $c = (T - T_0)/(T_P - T_0)$
and $T_P$ is the post-flame temperature. It can then be shown that
\cite{Driscoll}
\beq
\Sigma_{max} = \frac{A'_{0.5}}{\delta_T A'_L},
\label{e:SigmaMax}
\eeq
where $A'_{0.5}$ is the flame surface area based on $\bar{c} = 0.5$,
and $A'_L$ represents the average area of the flame brush. In a system
described by the one-step Arrhenius kinetics, $\bar{c} = 0.5$
corresponds to $Y = 0.5$. Therefore, $A'_{0.5} \equiv A_{0.5}$. At the
same time, in a planar brush $A'_L = L^2$, as was discussed in
\S~\ref{A:Isosurf}. Thus, $\Sigma_{max}$ is equivalent to
$\Sigma_{0.5}$ as given by eq.~(\ref{e:Sigma}). It is reasonable then
to compare values of $\Sigma_{0.5}$ in simulations S1-S3 with those of
$\Sigma_{max}$ obtained in other experimental and numerical settings.
In particular, in S3, $\overline{\Sigma}_{0.5} = 0.73$ mm$^{-1}$. This
is comparable to the range of values of $\Sigma_{max}
\approx 0.12 - 0.6$ mm$^{-1}$ obtained in a number of experimental
studies using a wide variety of flame configurations (see
\cite{Driscoll} for a summary table). The fact that the value in S3 is
somewhat larger is not surprising, given that the turbulent intensity
in it is substantially higher than in the experiments.

Overall, however, the traditional definition of $\Sigma$
\cite{Driscoll} is not equivalent to the definition of $\Sigma_Y$
used here unless $Y = 0.5$. Therefore, unlike $\Sigma_{max}$,
$\Sigma_{0.5}$ is not necessarily the maximum value of $\Sigma_Y$ for
all values of $Y$. Fig.~\ref{f:ASigma}f shows that while in S1
$\overline{\Sigma}_{0.5}$ is indeed approximately the largest value,
this is not the case in S2 and S3. Nonetheless, in all three
calculations, $\overline{\Sigma}_{0.5}$ is representative of the
values of $\overline{\Sigma}_Y$ throughout the reaction zone with the
difference between $\overline{\Sigma}_{0.5}$ and
$\overline{\Sigma}_{0.15}$ being $\approx \! 6\!-\!9\%$ in the
higher-resolution cases.

\subsection{Relation between $A_Y$ and $\Sigma_Y$}
%-------------------------------------------------------------------------------
\label{D:CorrelASigma}

Fig.~\ref{f:CorrelASigma} shows the relation between the isosurface
area and density in three key regions of the flame: the peak reaction
rate ($Y = 0.15$), the boundary of the preheat and reaction zones ($Y
= 0.5$), and the coldest region of the preheat zone ($Y = 0.95$). A
pronounced linear correlation between $A_Y/L^2$ and $\Sigma_Y$ exists
throughout the flame interior. This demonstrates that at all values of
$Y$, an increase of $A_Y$ is greatly influenced by the associated
tighter packing of the isosurfaces inside the flame brush.

With increasing $Y$ the average $\overline{A}_Y/L^2$ and
$\overline{\Sigma}_Y$ become larger, as was seen in
Fig.~\ref{f:ASigma}e,f, due to the overall shift of the distributions
toward larger values of both $A_Y/L^2$ and $\Sigma_Y$. Note also a
slight increase both in the scatter of the distribution and the slope
of its least squares fit at higher $Y$. Thus, isosurfaces of higher
fuel-mass fractions are indeed consistently more tightly packed than
at lower $Y$. This is in agreement with the observation that the flame
surface is wrinkled on finer scales on the fuel side than on the
product side \cite{Poludnenko}.

\subsection{Distributions of $\overline{A}_Y$ and $\overline{\Sigma}_Y$
and the effects of small-scale turbulence}
%-------------------------------------------------------------------------------
\label{D:SmallScale}

How do these results concerning the distributions of $\overline{A}_Y$
and $\overline{\Sigma}_Y$, shown in Fig.~\ref{f:ASigma}e,f, relate to
previous conclusions \cite{Poludnenko} regarding the inability of the
turbulent cascade to penetrate the flame interior and the role of
small-scale turbulent motions? As was discussed in
\S~\ref{ModelMethod} (also see \cite{Poludnenko} for further details),
the progressive increase in resolution from S1 to S3 extends the
turbulent cascade to smaller scales, and this leads to a substantial
increase in the energy of turbulent motions on scales $\lambda <
\delta_L$ in nonreactive turbulence. This increase in resolution was
found to cause the flame surface on the fuel side to be wrinkled on
progressively finer scales, while remaining virtually unchanged on the
product side. Furthermore, it was determined that the deviation of the
internal turbulent flame structure from that of the planar laminar
flame increases with decreasing temperature. These two results showed
that the effects of small-scale turbulent motions are most pronounced
in the coldest parts of the preheat zone. With increasing temperature,
these effects diminish and they completely disappear once the reaction
rate becomes significant \cite{Poludnenko}.

The results presented above are consistent with this picture. The
steady growth of $\overline{A}_Y/L^2$ and $\overline{\Sigma}_Y$ on the
fuel side of the preheat zone with increasing resolution shows that
more intense small-scale motions are indeed present there, folding
isosurfaces on progressively finer scales. It must be emphasized that
the change in $\overline{A}_Y$ alone does not necessarily mean that it
is caused by finer wrinkling on smaller scales. Only when viewed in
conjunction with the surface density, which shows a very similar
dependence on $Y$, does such increase in $\overline{A}_Y$ serve as a
strong indication of the small-scale wrinkling.

With increasing temperature, not only do $\overline{A}_Y/L^2$ and
$\overline{\Sigma}_Y$ decrease rapidly, but also profiles for S2 and
S3 begin to approach each other. This means that the main difference
between these two simulations, namely the presence of more energetic
small-scale turbulence in S3, is being eliminated and, thereby,
motions on scales $\lambda < \delta_L$ are gradually suppressed. Since
it is these small scales that enhance the diffusive transport, which
in turn broadens the preheat zone, their suppression causes the
internal flame structure to approach that of the planar laminar flame
\cite{Poludnenko}.

As the reaction rate becomes substantial at $Y \lesssim 0.5$, both
$\overline{A}_Y/L^2$ and $\overline{\Sigma}_Y$ become similar in S2
and S3. This suggests that, at this point, only scales $\lambda
\gtrsim \delta_L$ remain energetic. Indeed, these scales are originally the
same in both calculations (cf. Fig.~\ref{f:FlameSpectra}).
Consequently, they generate similar isosurface areas and densities.
Since these scales cannot support small-scale diffusive transport, any
broadening of the internal flame structure effectively disappears at
$Y \lesssim 0.5$ \cite{Poludnenko}. Further decrease in
$\overline{A}_Y/L^2$ and $\overline{\Sigma}_Y$ is the consequence of
the continued suppression of the progressively larger scales $\lambda
> \delta_L$.

There is one important distinction between the distributions of
$\overline{A}_Y/L^2$ and $\overline{\Sigma}_Y$ and the time-averaged
profiles of $Y$ and $T$, which represent the internal flame structure
\cite{Poludnenko}. The $Y$ and $T$ profiles showed very little
variation between simulations S1-S3 not only in the reaction zone, but
also in the preheat zone. This is in contrast with the distributions
$\overline{A}_Y/L^2$ and $\overline{\Sigma}_Y$, which differ
substantially in the preheat zone even between S2 and S3. Such
discrepancy suggests that small scales, $\lambda < \delta_L$,
contribute differently to the turbulent diffusive transport and to the
wrinkling of the isosurfaces. The former is primarily governed by the
largest of these small scales, i.e., scales not much smaller than
$\delta_L$, since these scales are associated with the highest
velocities. The energy contained on these scales is close in all three
simulations and, thus, they produce the same structure of the
broadened preheat zone. At the same time, all scales smaller than
$\delta_L$ contribute to the isosurface wrinkling, and the different
energy content of these scales causes the distributions of
$\overline{A}_Y/L^2$ and $\overline{\Sigma}_Y$ to differ substantially
in the preheat zone.

Qualitatively, both the internal flame structure described in
\cite{Poludnenko} and the distributions of $\overline{A}_Y$ and
$\overline{\Sigma}_Y$ presented here create a consistent picture of
the transformation that turbulence undergoes as it passes through the
flame. Changes in the energy budget between different scales, which
are most certainly accompanied by the development of both anisotropy
and inhomogeneity of the velocity field, are complex. While the
evidence presented here and in \cite{Poludnenko} provide hints about
the nature of this transformation, the details remain unknown. For
instance, are the energy release and the resulting fluid expansion the
only effects responsible for altering the turbulent field and
redistributing energy between different scales?  What are the relative
contributions of various small scales to the increase in $A_Y$ and
$\Sigma_Y$? How does the shift in balance between small and large
scales with increasing temperature change at different turbulent
intensities? All of these questions need to be addressed in future
studies.

\section{What is the flame surface area?}
%-------------------------------------------------------------------------------
\label{FlameArea}

In the wrinkled and corrugated flamelet regimes, in which
eq.~(\ref{e:Damk0}) is typically applicable, the flow is laminar on
scales $\lambda \lesssim \delta_L$. As a result, isosurfaces of
different values of $Y$ are parallel to each other, the flame is
folded by turbulence as a coherent structure, and, therefore, it has a
well-defined surface area $A_T$. Consequently, the stretch factor $I$
has a unique value at each moment in time, and so it directly shows
how much of the increase of $S_T$ can be ascribed to the increase of
$A_T$.

In contrast, in the thin reaction zone regime discussed here, results
presented in \S~\ref{AreaDensity} demonstrate that the flame, folded
inside the flame brush by high-intensity turbulence, has a complex
internal structure. The various parts of the flame have distinctly
different responses of its various parts to the action of turbulence.
Consider, by analogy with eq.~(\ref{e:Damk0}), the factor $I_Y$, which
relates the increase of the turbulent flame speed relative to its
planar laminar value to the increase of $A_Y$ rather than $A_T$,
\beq
\frac{S_T}{S_L} = I_Y \ \frac{A_Y}{L^2}.
\label{e:Damk}
\eeq
In this case, unlike $I$, $I_Y$ is a function of $Y$. In particular,
based on the data shown in Fig.~\ref{f:ASigma}, $I_Y$ can vary by as
much as a factor of 4 depending on the choice of $Y$. As a result,
values of $I_Y$ close to and substantially larger than unity can be
found in different regions of the flame.

This shows that before we can answer the question whether the observed
$S_T$ can be accounted for by the increase in the flame surface area,
we must first determine what this area is in such high-speed turbulent
flows, and whether such a concept is even applicable in this regime.
In particular, what value of $Y$ fully and accurately represents the
overall behavior of the turbulent flame? We address these questions by
considering the correlation of $A_Y$ with two global properties of the
turbulent flame, its speed and width.

\subsection{Relation between $A_Y$ and $S_T$}
%-------------------------------------------------------------------------------
\label{FA:Speed}

Fig.~\ref{f:CorrelAS} shows correlations between $S_T/S_L$ and
$A_Y/L^2$ calculated for the same three representative values of $Y$
as in Fig.~\ref{f:CorrelASigma}, namely $Y = 0.15$, corresponding to
the region of peak reaction rate, $Y = 0.5$, corresponding to the
boundary between the reaction and the preheat zones, and $Y = 0.95$,
corresponding to the coldest part of the preheat zone. Left panels of
Fig.~\ref{f:CorrelAS} show the time-evolution of all quantities, and
the right panels show the corresponding correlation scatter plots.

Since both the local flame speed and the induction time of the
unburned fuel have finite values, there must exist a delay in the
response of $S_T$ to the changes in the flame configuration,
represented by the areas of its isosurfaces. Therefore, the degree of
correlation between $S_T$ and $A_Y$ is a function of a time lag,
$\Delta t$, between these two quantities. For each value of $Y$ in
Fig.~\ref{f:CorrelAS}, we determined the time lag, $\Delta t_c$, that
produced the best correlation. This was done by directly calculating
the cross-correlation between $S_T(t)/S_L$ and $A_Y(t)/L^2$ and
finding its maximum. The value obtained for $\Delta t_c$ was verified
by determining the least squares fit for the distribution of
$S_T(t)/S_L$ vs. $A_Y(t-\Delta t')/L^2$ and by finding $\Delta t_c'$
that maximized the slope of the fit and minimized its residuals. The
values of $A_Y/L^2$ in Fig.~\ref{f:CorrelAS} were then shifted in time
with respect to $S_T/S_L$ by the corresponding time lag $\Delta t_c$
given in the lower right corner of the panels (a), (c), and (e).

The first key conclusion emerging from Fig.~\ref{f:CorrelAS} is that
$S_T$ and $A_Y$ become progressively less correlated with increasing
$Y$. While the correlation is pronounced throughout the reaction zone
and is exceptionally strong near the peak reaction rate, $S_T$ and
$A_Y$ are only very weakly correlated in the colder parts of the
preheat zone. Note that panels (b), (d), and (f) have the same scale,
which shows the relative increase in the scatter of the distribution
and its overall shift to higher values of $A_Y$ in the preheat zone.

The time delays that give the best correlation between the two
quantities are positive, and thus they show that $S_T$ does indeed lag
behind $A_Y$. Moreover, the magnitude of $\Delta t_c$ increases with
increasing $Y$. Qualitatively, this agrees with the fact that $S_T$ at
any given moment is primarily determined by the region of the highest
reaction rate, which corresponds to the values of $Y$ close to 0.15.
Consequently, the time delay between $S_T$ and the area of the
isosurface representing the peak reaction rate is close to zero,
namely $0.04\tau_{ed}$. Since turbulence reorganizes the flame on a
timescale $\sim \! \tau_{ed}$, such a small lag is too short for the
flame structure to change in any significant way. Therefore, the
correlation between $A_{0.15}$ and $S_T$ is very strong.

At the same time, since the reaction rate is fairly low at $Y = 0.5$,
the contribution of this region to the overall turbulent flame speed
is small. Burning still needs to accelerate substantially in this area
in order for it to become the new region of peak reaction rate, which
will then predominantly determine the magnitude of $S_T$. The time
required for this to occur results in a longer time delay, namely
$0.11\tau_{ed}$. During this time, however, turbulence is able to
change the flame structure more strongly than in the case of $Y =
0.15$. As a result, the degree of correlation between $A_{0.5}$ and
$S_T$ decreases.

In the coldest regions of the preheat zone, there is no correlation
between $S_T$ and $A_{0.95}$ at small values of $\Delta t_c$. We found
evidence of weak correlation for a much larger time delay $\Delta t_c
= 0.61\tau_{ed}$. This value, however, is comparable to the time
during which the turbulence completely reorganizes the structure of
the turbulent flame. Therefore, by the time burning reaches the
reactants in the outer regions of the preheat zone, the connection
between their original distribution and the resulting turbulent flame
speed is significantly disrupted.

We also found weak correlation between $A_{0.95}$ and $A_{0.15}$ with
the time lag for $A_{0.95}$ being $\Delta t_c = 0.56\tau_{ed}$ and the
slope of the least squares fit of $0.4907$.\footnote{In contrast, we
were unable to find any statistically significant correlation or
anticorrelation between $\Sigma_{0.15}$ and $\Sigma_{0.95}$ at any
time lag.} This weak correlation between $A_{0.15}$ and $A_{0.95}$ is
consistent with a similarly weak correlation between $S_T$ and
$A_{0.95}$ found at a comparable time lag. In particular, it reflects
the fact that after the time $\approx \!  0.6\tau_{ed}$, burning
reaches the outer part of the preheat zone. As a result, the former $Y
= 0.95$ region becomes the new $Y = 0.15$ region and thus the new site
of the peak reaction rate which now determines $S_T$. Substantial
change in the flame-brush structure during this time, however, again
leads only to weak correlation between $A_{0.15}$ and $A_{0.95}$.

In order to verify this physical meaning of the obtained time delays,
values of $\Delta t_c$ can be compared with the characteristic
induction time, $\tau_{ind}$, in the planar laminar flame. The
$\tau_{ind}$ is the time for a fluid element to reach the peak
reaction rate starting at some initial temperature or, equivalently,
fuel mass fraction. In reactive mixtures with large activation
energies, such as H$_2$-air considered here, $\tau_{ind}$ is also
approximately the time necessary to reach the post-flame temperature,
$T_P$ \cite{Zeldovich}.

The $\tau_{ind}$ can be calculated using the velocity distribution
$U_L(x)$ in the exact planar laminar flame solution as
\beq
\tau_{ind}(Y') = \int_{x(Y')}^{x(Y^*)} \frac{dx}{U_L(x) - U_{\infty}}.
\label{e:tind1}
\eeq
Here $U_{\infty}$ is the fuel velocity at infinity, $Y'$ is the
initial fuel mass fraction of the fluid element, and $Y^*$ is the fuel
mass fraction corresponding to the peak reaction rate. Instead, to
maintain consistency with the numerical simulations, we computed
$\tau_{ind}$ directly by solving the full time-dependent reactive
Navier-Stokes equations (\ref{Euler1})-(\ref{e:ReacModel}) on a
uniform 1D grid with the same numerical integration method used for
S1-S3 (\S~\ref{ModelMethod}). The flame was initialized as a
discontinuity with the fuel and product temperature and density equal
to $T_0$, $\rho_0$ and $T_{P,0}$, $\rho_{P,0}$, respectively, given in
Table~\ref{t:Params}. Constant pressure $P_0$ and zero velocity were
set throughout the domain. We introduced an additional advection
equation describing a scalar $\xi$ that was not subject to diffusion
or reactions and was only advected with the flow. After the
steady-state structure of the flame had been established in the
computational domain, $\xi$ was set to unity in one cell corresponding
to the desired initial $T'$ or, equivalently, $Y'$, and to zero
everywhere else. As the system evolved, the fluid element marked with
the scalar $\xi$ moved through the flame and burned. The time it took
this fluid element to reach peak reaction rate, i.e., $Y = 0.15$, gave
$\tau_{ind}(Y')$.\footnote{The distribution of $\xi$ would spread with
time over several cells due to numerical diffusion. To compensate for
this, we used very high resolution, namely $\Delta x =
\delta_{L,0}/128$, which is much larger than the resolution necessary
to obtain an accurate laminar flame solution, i.e., typically $\Delta
x = \delta_{L,0}/4$ \cite{Poludnenko2}. As a result, the distribution
of $\xi$ was found to spread over the range of values of $Y$ with the
full width at half maximum equal to $\Delta Y \approx 0.05$. We
assumed that the position of the fluid element being tracked was
marked by the maximum of this distribution.  A convergence study with
increased resolution, which caused $\xi$ to be distributed over
progressively narrower range of $Y$, was used to verify that the
obtained values of $\tau_{ind}$ were indeed accurate.}

Using this procedure, we found that for a planar laminar flame,
\beq
\begin{array}{l}
\tau_{ind} \approx 0.27\tau_{ed} \approx 2.46\Delta t_c,
\ \ \textrm{for} \ Y = 0.5, \\
\tau_{ind} \approx 1.29\tau_{ed} \approx 2.12\Delta t_c,
\ \ \textrm{for} \ Y = 0.95.
\end{array}
\label{e:tind2}
\eeq
When turbulent heating of the fuel is taken into account (see
\S~\ref{Simulations}), i.e., when the fuel temperature in this
calculation is increased to $320$ K from $T_0 = 293$ K while
maintaining constant density $\rho_0$, these values of $\tau_{ind}$
decrease by $\approx\!20-25\%$. Further details and, in particular,
the full distributions of $\tau_{ind}$ throughout the flame for these
two fuel temperatures, can be found in \ref{AppB}.

The calculated time lags are, therefore, within a factor of
$\approx\!\!2$ of the corresponding values of $\tau_{ind}$, which,
given the statistical nature of $\Delta t_c$, is reasonable
agreement. This shows that the correlation time delays for the
distributions of $S_T/S_L$ and $A_Y/L^2$ in the turbulent flame can
indeed be associated with the induction times necessary for the
burning to accelerate in regions with higher reactant concentrations.

The values of $\Delta t_c$, however, are substantially lower than the
characteristic propagation time of the laminar flame,
\beq
\tau_{\delta} = \frac{\delta_{L,0}}{S_{L,0}} =
\frac{30}{8}\frac{L}{U} = 3.75\tau_{ed},
\label{e:tsc}
\eeq
where we used the values of $L$ and $U$ given in Table~\ref{t:Runs}.
Even for $Y = 0.95$, $\Delta t_c = 0.61\tau_{ed} \approx
(1/6)\tau_{\delta}$. Furthermore, this $\Delta t_c$ is an order of
magnitude smaller than the time $1.67\delta_{L,0}/S_{L,0} \approx \!
6.3\tau_{ed}$, which is necessary for the flame to propagate over the
characteristic distance $1.67\delta_{L,0}$ separating $Y = 0.95$ and
$Y = 0.15$ in the laminar flame structure (cf. Fig.~7 in
\cite{Poludnenko}).

A good analytical estimate for $\Delta t_c$ throughout the reaction
zone can be obtained using the expression for the adiabatic induction
time \cite{Zeldovich},
\beq
\tau_{ind}^a(Y') = \Big(\frac{C_pT'}{Bq\rho' Y'}\Big)\Big(\frac{RT'}{Q}\Big)
\exp\Big({\frac{Q}{RT'}}\Big),
\label{e:tinda}
\eeq
where $T'$ and $\rho'$ for a given $Y'$ are determined based on the
exact planar laminar flame solution, and the values of $B$, $q$, and
$Q$ are given in Table~\ref{t:Params}. In particular, at the fuel
temperature $T_0$ and density $\rho_0$,
\beq
\begin{array}{ll}
\tau_{ind}^a \approx 0.05\tau_{ed} \approx 1.25\Delta t_c, &
\! \! \textrm{for} \ Y = 0.15, \\
\tau_{ind}^a \approx 0.21\tau_{ed} \approx 1.91\Delta t_c, &
\! \! \textrm{for} \ Y = 0.5.
\end{array}
\eeq
Using the specific heat at constant volume $C_v = R/M(\gamma-1)$
instead of $C_p$ in eq.~(\ref{e:tinda}), i.e., assuming that burning
occurs at constant volume rather than at constant pressure, would
decrease the values of $\tau_{ind}^a$ by $\approx\!15\%$ to
$0.04\tau_{ed}$ and $0.18\tau_{ed}$, respectively. This shows that,
even though $\tau_{ind}^a$ is typically applicable in the context of
an autoignition process rather than propagation of a flame, for $Y =
0.5$ it provides a somewhat better approximation to the calculated
value of $\Delta t_c$ than $\tau_{ind}$ given in eq.~(\ref{e:tind2}).
Further details, including the effects of turbulent fuel heating on
$\tau^a_{ind}$, are given in \ref{AppB}.

Finally, in addition to the strong correlation between the values of
$S_T$ and $A_{0.15}$, the distribution shown in Fig.~\ref{f:CorrelAS}b
has another important property that is not present at higher values of
$Y$. The linear least squares fit for the case $Y = 0.15$, shown with
a solid line in Fig.~\ref{f:CorrelAS}b, has the form
\beq
\frac{S_T}{S_L} = 1.17 \frac{A_{0.15}}{L^2} - 0.08.
\label{e:fit}
\eeq
For $A_{0.15}/L^2 = 1$, this expression gives $S_T/S_L = 1.09$.  This
demonstrates that as the flame becomes less convolved, i.e., as
$A_{0.15}/L^2 \rightarrow 1$, it behaves progressively more like a
planar laminar flame which is manifested in $S_T \to S_L$. Linear
least squares fits for the two other values of $Y$, shown in
Fig.~\ref{f:CorrelAS}d,f, do not recover the value of $S_T/S_L = 1$ as
$A_Y/L^2 \rightarrow 1$.

\subsection{Relation between $A_Y$ and $\delta_T$}
%-------------------------------------------------------------------------------
\label{FA:Width}

Next consider the correlation of $A_Y$ with the second key global
characteristic of the turbulent flame, namely its width, $\delta_T$.
Fig.~\ref{f:CorrelADelta} shows the correlation between $A_Y/L^2$ and
$\delta_T/L$ for the same three values of $Y$ as in
Fig.~\ref{f:CorrelAS}.\footnote{Since both $A_Y$ and $\delta_T$
represent the same instantaneous configuration of the flame, no time
delay between them is expected. Consequently, no time shift was
applied to either quantity in Fig.~\ref{f:CorrelADelta}. As a check,
however, we verified that indeed the cross-correlation had its maximum
at the zero time lag.} The distribution of $\delta_T/L$ as a function
of $A_Y/L^2$ shows the same trend observed for $S_T/S_L$. In
particular, the correlation between the two quantities is also the
strongest for $Y = 0.15$ and decreases with increasing $Y$. The values
of $\delta_T/L$ and $A_{0.95}$ appear to be almost completely
uncorrelated. Note, however, that the correlation in the reaction
zone, and, in particular, in the region of peak reaction rate, is
weaker for $\delta_T$ than $S_T$, with the distributions of $\delta_T$
having a much larger scatter.

In order to illustrate the relation between the three key quantities
characterizing the instantaneous flame-brush structure, namely $A_Y$,
$\delta_T$, and $\Sigma_Y$, each data point in
Fig.~\ref{f:CorrelADelta} is colored according to the corresponding
value of $\Sigma_Y$ (cf. Fig.~\ref{f:CorrelASigma}). At lower $Y$, the
increase of the isosurface area is associated both with the increase
in the overall extent of the flame brush and with tighter packing of
the isosurfaces. With increasing $Y$, however, the correlation of
$A_Y/L^2$ with $\delta_T/L$ decreases while remaining quite pronounced
with $\Sigma_Y$. This shows that the area of the isosurfaces of higher
$Y$ changes primarily only due to their wrinkling on progressively
smaller scales, as discussed in \S~\ref{AreaDensity} (also see
\cite{Poludnenko}). Indeed, $\delta_T$ is determined predominantly by the
flame folding on the largest scales. Consequently, the weak
correlation between $A_Y$ and $\delta_T$ at high $Y$ demonstrates that
the increase of the areas of these isosurfaces cannot be associated
with such large-scale wrinkling. Instead, this increase must be
translated primarily into tighter packing of the isosurfaces causing
$A_Y$ to be correlated with $\Sigma_Y$.

\section{What determines $S_T$ in the thin reaction zone regime?}
%-------------------------------------------------------------------------------
\label{FlameSpeed}

\subsection{Can $S_T$ be explained by the increase in the flame surface area?}
\label{FS:Area}

The strong correlation of $A_{0.15}$ with $S_T$ and $\delta_T$, which
represent both the global energetics and the global structure of the
turbulent flame, show that, in the thin reaction zone regime,
isosurfaces of $Y$ close to the peak reaction rate accurately
characterize the overall evolution of the turbulent flame.
Consequently, for the reactive mixture considered here, the flame
surface area, $A_T$, and density, $\Sigma_T$, are represented by the
isosurface of $Y = 0.15$.

This conclusion is further supported by the limiting behavior of the
distribution of $S_T/S_L$, which approaches unity as $A_{0.15}/L^2
\rightarrow 1$. In contrast, at $Y = 0.5$, $S_T/S_L$ becomes
progressively less than $A_{0.5}/L^2$ at smaller isosurface areas.
Furthermore, at $Y = 0.95$, almost all values of $S_T/S_L$ lie below
the $S_T/S_L = A_{0.95}/L^2$ line, and they become almost half of
$A_{0.95}/L^2$ at larger isosurface areas. In $Le = 1$ reactive
mixtures, there are no mechanisms that can lower the local flame speed
below $S_L$, which would be necessary to produce values of $S_T/S_L$
lower than the increase of the flame surface area.  Consequently, such
properties of the distributions of $S_T/S_L$ confirm that isosurfaces
of higher $Y$ cannot represent the flame surface.

With this definition of the surface area of the turbulent flame, we
can now revisit the question of the relation between $A_T$ and the
speed of the turbulent flame in the thin reaction zone regime. This
relation can be represented by eq.~(\ref{e:Damk}) in which $Y = 0.15$,
similarly to eq.~(\ref{e:Damk0}) applicable in the wrinkled and
corrugated flamelet regimes.\footnote{In light of the fact that $S_T$
and $A_Y$ are best correlated when $A_Y$ is shifted in time with
respect to $S_T$, the question arises whether such a time shift should
also be applied in eq.~(\ref{e:Damk}). In principle, an argument could
be made that, due to the inherent delay in the response of $S_T$ to
the changes in the flame structure, the current value of $S_T$ is the
result of $A_Y$ that existed earlier in time and, thus, this earlier
value must be used to assess the relation between $S_T$ and $A_Y$. At
the same time, as was shown, the time lag $\Delta t_c$ is extremely
small for $Y = 0.15$. During this time, the change in the value of
$A_Y$ is negligible and applying the time shift in eq.~(\ref{e:Damk})
does not cause any appreciable change in the result.}
Fig.~\ref{f:CorrelIA} shows the ratio $I_{0.15} =
(S_T/S_L)/(A_{0.15}/L^2)$ for the distribution given in
Fig.~\ref{f:CorrelAS}b. Time-averaged values $\overline{I}_{0.15} =
\overline{(S_T/S_L)/(A_{0.15}/L^2)}$ for all three calculations are
listed in Table~\ref{t:Props}.\footnote{Using instead the expression
$\overline{I}_{0.15} = (\overline{S_T/S_L})/(\overline{A}_{0.15}/L^2)$
gives virtually the same result.} In particular, $\overline{I}_{0.15}
= 1.14$ in S3, and this value can be viewed as converged to within a
few percent with a faster than linear rate of convergence.

These results lead to the following two key conclusions:
\begin{enumerate}
\item{In the thin reaction zone regime, the instantaneous turbulent
flame speed is also primarily determined by the increase of the flame
surface area.}
\item{In addition, in the course of system evolution, $S_T$
periodically exhibits an exaggerated response to the increase in
$A_T$. For the system size and turbulence intensity considered here,
this excess burning velocity can become as high as $30\%$ of
$A_T/L^2$.}
\end{enumerate}

This exaggerated response in simulation S3 is represented by values of
$I_{0.15} > 1$ in Fig.~\ref{f:CorrelIA} and is illustrated in
Fig.~\ref{f:CorrelAS}a,b as the shaded gray area. Such a substantial
deviation of $I_{0.15}$ from unity brings up the following question:
what mechanism is responsible for raising $S_T$ beyond what can be
attributed to the increase in $A_T$?

\subsection{Potential causes of the excess values of $S_T$}
%-------------------------------------------------------------------------------
\label{FS:Causes}

As was discussed in \S~\ref{Intro}, Damk\"ohler's concept
\cite{Damkohler} implies that the burning velocity of a turbulent
flame in the presence of sufficiently intense turbulence is determined
by two processes: wrinkling of the flame surface by large-scale
turbulent motions increasing $A_T$, and the enhancement of the local
flame speed $S_n$ by the small-scale turbulence providing stronger
diffusive transport. While the simulations show that the effect of
large-scale turbulence is indeed present and has a dominant effect on
$S_T$, can the observed excess values of $S_T$ be ascribed to the
action of small-scale motions?

Consider the following important characteristic exhibited by the
distribution of $I_{0.15}$. Along with $I_{0.15}$ substantially larger
than one, the system also periodically develops values of $I_{0.15}$
within only a few percent of unity, which is comparable to the
uncertainty in estimating the instantaneous $A_{0.15}$ and, thus,
$I_{0.15}$ (see also Fig.~\ref{f:CorrelAS}a). This shows that at those
instances when $I_{0.15} \approx 1$, $S_n$ at all points of the flame
surface must be almost identical to $S_L$. Consequently, the
small-scale turbulence entering the flame, if it were directly
determining $S_n$, would have to periodically undergo transitions
between almost complete suppression, when $I_{0.15} \approx 1$ and
$S_n \approx S_L$, and significant enhancement, when $I_{0.15}$
becomes as high as $1.3$.

While we are unaware of any potential mechanisms that could cause such
periodic complete suppression of small-scale turbulence in the cold
fuel, our analysis also does not show any changes in the state of
turbulence that could serve as a manifestation of such suppression.
In order to determine whether there exists a connection between the
instantaneous turbulent field in the domain and the resulting burning
speed, we have analyzed the correlation of several turbulence
characteristics with both $S_T$ and $I_{0.15}$. In particular, we
considered the total r.m.s. velocity in the domain. Furthermore, to
characterize turbulence inside the flame brush and in the fuel
entering it, we also considered individual components of velocity,
$u_i$, as well as kinetic energy, $0.5\rho u_i^2$, averaged over the
volume of the flame brush and the cubic region of size $L$ located in
the fuel immediately ahead of it. We were not able to find virtually
any correlation between any of these quantities and either $S_T$ or
$I_{0.15}$. As an example, Fig.~\ref{f:CorrelIEkin} shows the
representative correlation scatter plot of $I_{0.15}$ as a function of
the normalized $x$-component of kinetic energy $E_{K,x} = (\rho
u_x^2)/(\rho_0 U_l^2)$ volume-averaged over the region discussed
above. Similarly to the case of the distribution of $I_{0.15}$
vs. $A_{0.15}/L^2$ in Fig.~\ref{f:CorrelIA}, the increasing trend seen
in the least squares fit is too weak to determine its statistical
significance based on the available data.

These considerations suggest that the enhancement of $S_n$ over $S_L$
by turbulent transport does not provide a plausible explanation of the
excess values of $S_T$ periodically developed by the flow.
Consequently, a different mechanism must augment the effect of the
increased $A_T$.

Instead of being the result of a uniform global increase of $S_n$
throughout the flame brush, accelerated burning can be caused by a
significant increase in $S_n$ in isolated regions of the flame
surface. In those regions, however, $S_n$ must be much larger than
$S_L$ in order to significantly raise the overall burning speed of the
turbulent flame.

Such local enhancement of $S_n$ cannot be caused by the local
inhomogeneities in the thermodynamic state of the fuel.\footnote{The
increase of $S_n$ due to the global fuel heating by turbulence was
taken into account in $S_L$, as discussed in \S~\ref{Simulations}.} In
particular, in a subsonic turbulent flow, such as the one considered
here, with the turbulent Mach number $Ma_F$ substantially less than
unity (see Table~\ref{t:Params}), local variations of density or
temperature are much too small to affect $S_n$ in any appreciable
way. For instance, it requires an increase of fuel temperature of over
$150$ K to double $S_n$. We do not observe temperature fluctuations of
such magnitude in the flow.

$S_n$ can also be increased by the substantially higher local
turbulence intensity associated with intermittency. While we do
observe intermittency in the flow field in our simulations, regions of
large velocity enhancement are statistically too rare, their spatial
extent is too small, and they are much too short lived to increase
$S_n$ and, thus, to make any significant contribution to the overall
increase in $S_T$. Potential effects of turbulence intermittency on
the flame have been studied by Pan \ea \ \cite{Pan} in the context of
the Rayleigh-Taylor-driven thermonuclear flames in the interior of a
white dwarf during a Type Ia supernova explosion. Their results
suggested the possibility that intermittency in the velocity field
could locally disrupt and broaden the flame, which potentially would
increase the local burning speed. It is not immediately clear,
however, to what degree their conclusions apply to the system
discussed here. The conditions in the white dwarf interior considered
in \cite{Pan} are substantially different from those in S1-S3.
Moreover, their analysis did not consider any feedback of the flame on
the turbulent field and, in particular, it did not include fluid
expansion due to heat release which invariably affects intermittency
of the flow. At the same time, the role of intermittency in the
dynamics of turbulent flames does require further investigation. In
particular, it would be important to extend the analysis of Pan \ea \
\cite{Pan} by including the feedback of the flame on turbulence under
conditions characteristic of chemical rather than thermonuclear
flames.

Instead, we suggest that significant local increase of $S_n$ does
indeed take place in the turbulent flame due to the flame collisions
resulting in the creation of regions of large flame curvature. We
discuss this process next.

\section{Accelerated burning caused by flame collisions}
%-------------------------------------------------------------------------------
\label{Mechanism}

\subsection{Local increase of $S_n$ in cusps}
\label{M:Collisions}

According to the theory of flame stretch (see \cite{Law} for a
review), when $Le = 1$, the flame is not affected significantly
\cite{Istratov} either by flow-induced strain or curvature. In
particular, a stationary spherical flame supported by a point source
of mass, which is an example of a curved unstrained flame, has an
internal structure, and thus local burning velocity, identical to that
of a planar laminar flame \cite{Law}. Such analysis, however, is
typically carried out for a flame with the curvature radius $r_c \gg
\delta_L$. Using the physical model and the Athena-RFX code described
in \S~\ref{ModelMethod}, we also performed simulations of an idealized
spherical and cylindrical flame propagating inward into the stationary
fuel. The results of these simulations are, in fact, consistent with
the theory.  Furthermore, they show that $S_n = S_L$ not only when
$r_c \gg \delta_L$, but up until the moment when the flame curvature
becomes $\approx \!  1/\delta_L$, i.e., until the flame effectively
collapses onto itself. At this point, $S_n$ increases substantially.
Beyond this moment, however, the flame itself effectively ceases to
exist and, therefore, the local flame speed looses its meaning. Such
an idealized situation is not representative of the conditions arising
in an actual turbulent flame. Nonetheless, regions of large flame
curvature $\sim \! 1/\delta_L$ are frequently created in the
high-speed turbulent flow considered here, and they naturally provide
a mechanism for a significant local enhancement of $S_n$.

The results presented in \S~\ref{AreaDensity} showed that intense
turbulence causes tight folding and packing of the flame inside the
flame brush. Consider the inverse of the surface density, which is a
measure of an average separation between surface elements. Given that
$\overline{\Sigma}_{0.15} = 0.67$ mm$^{-1}$ in simulation S3, the
average distance between individual flame sheets is
$1/\overline{\Sigma}_{0.15} \approx 1.49$ mm $\approx 4.7\delta_L$. At
the same time, $\Sigma_{0.15}$ can be as high as $0.97$ mm$^{-1}$
(Fig.~\ref{f:CorrelASigma}a), resulting in an even lower separation of
$\approx \!  3.2\delta_L$. This is comparable to the full flame width
$l_F \approx 2\delta_L$ (cf. Fig.~7 in \cite{Poludnenko}). Such
tightly folded flame configurations invariably result in frequent
collisions of individual flame sheets.

Fig.~\ref{f:isoV} gives an example of such collision in S3. The figure
shows the flame-brush structure at three times: $11.86\tau_{ed}$
(upper panel), as well as $0.1\tau_{ed} =$ 3 ${\mu}$s and
$0.2\tau_{ed} =$ 6 ${\mu}$s later (middle and lower panels,
respectively). Initially, a region with a highly convolved flame
develops in the flame brush (region A). At this time, $S_n \approx
S_L$ locally throughout region A and changes in the flame
configuration due to self-propagation can be neglected on timescales
considered in Fig.~\ref{f:isoV}. As turbulent motions continue to
bring individual flame sheets closer to each other, the curvature
radius of the flame becomes close to $\delta_L$, and the preheat zones
begin to overlap substantially over an extended region of the flame
surface. This marks the formation of two regions of high flame
curvature $\sim 1/\delta_L$ (regions B), which we will refer to as
``cusps.'' Regions C in the lower panel show that 3 $\mu$s later, the
flame sheets have merged and formed two extended reaction zones, which
suggests substantially accelerated burning in that area of the flame
brush. Note also a rapid decrease of the area of the $Y = 0.5$
isosurface represented with the thin black line. Comparison of regions
B and C shows that this isosurface propagated over the distance
$\approx \! 0.5$ mm over the time 3 $\mu$s, which implies the
propagation speed (but not the local burning speed) $\approx \!
55S_L$. During the time shown in Fig.~\ref{f:isoV}, two other highly
elongated regions of flame collision have formed near the upper face
of the domain. Thus, Fig.~\ref{f:isoV} illustrates that flame
collisions are ubiquitous in the turbulent flame, and they do produce
regions of large flame curvature. Moreover, their evolution in
Fig.~\ref{f:isoV} suggests substantial increase of both the local
burning and propagation speeds in the cusps.

Qualitatively, two different channels of cusp formation can be
identified. They are shown schematically in Fig.~\ref{f:CollTypes}. In
the first case, turbulent motions on a certain scale $\lambda$ with
velocity $U_{\lambda}$ stretch and fold the flame. As a result, a
narrow elongated structure is formed. When two flame sheets
effectively collide, the curvature of the flame surface becomes
$\sim\!1/\delta_L$ and the cusp is formed with a characteristic length
$l_c \sim \lambda$. We will refer to this channel as ``local flame
collisions,'' since the cusp is formed by flame regions located close
to each other both in 3D space and along the flame surface. In the
second case, turbulent motions fold the flame on a larger scale, thus
not increasing its curvature substantially. Two cusps are formed when
two smoothly curved flame sheets collide. We will refer to this
channel as ``nonlocal flame collisions,'' since the colliding flame
regions can be located far from each other along the flame surface and
they can, therefore, originate from different regions of the flame
brush. In this case, the characteristic length of the resulting cusp,
$l_c$, can be significantly smaller than the scale, $\lambda$, of the
motion that produced the cusp.

In an actual turbulent flame, cusps are often created under the
combined action of these two channels, an example of which is seen in
Fig.~\ref{f:isoV}. Region A is initially formed by the flame surface
being stretched and folded on a large scale, as in
Fig.~\ref{f:CollTypes}b. At the same time, smaller scale motions fold
the flame surface locally, as in Fig.~\ref{f:CollTypes}a, forming
smaller cusps seen in regions B.

While both local and nonlocal flame collisions are equivalent in terms
of the resulting cusp properties, they differ in one important
respect. In the first case, curvature of the flame surface at a
specific point increases gradually until it becomes large marking the
formation of a cusp. In the second case, the curvature remains small
at all points of the flame surface until the actual moment of
collision when two flame sheets merge and the region of large flame
curvature is abruptly formed. These distinctive characteristics of the
two processes will be important for the analysis of their relative
efficiency to form cusps at a given turbulent intensity and system
size.

\subsection{Structure and properties of cusps}
%-------------------------------------------------------------------------------
\label{M:Cusps}

Formation of cusps (or ``folds'') by the wrinkled flame in a turbulent
flow was first suggested by Karlovitz \ea \cite{Karlovitz} (also see
discussion in \cite{LewisElbe}). Their dynamical role was first
considered by Zel'dovich \cite{Zeldovich1} (also see
\cite{Zeldovich2}), who suggested them as a mechanism of flame
stabilization that prevents the unbounded exponential growth of the
flame surface under the action of the Landau-Darrieus instability
\cite{Landau}. Such regions of large curvature would have a high
propagation velocity, causing rapid annealing of the flame surface,
and thus would provide an efficient mechanism to counterbalance flame
instability \cite{Zeldovich1}. It must be noted that both Karlovitz
\ea \cite{Karlovitz} and Zel'dovich \cite{Zeldovich1} assumed that
cusps form due to the self-propagation of a wrinkled flame surface
following the Huygens principle, rather than as a result of flame
collisions by turbulence.

It was suggested early on both by Lewis and von Elbe \cite{LewisElbe}
and Zel'dovich \cite{Zeldovich1} that, due to the focusing of the heat
flux into fuel, $S_n$ ceases to be equal to $S_L$ near the tip of the
cusp, and the cusp region develops a structure similar to the tip of
the Bunsen flame.  The flame in
\cite{Karlovitz,LewisElbe,Zeldovich1,Zeldovich2}, however, was
considered to be a gasdynamic discontinuity, and so its internal
structure was ignored. As a result, the actual increase of the local
flame speed in the cusp could not be determined.

In order to analyze quantitatively the effect of flame collisions and
to determine the actual increase of $S_n$ in the resulting cusps,
consider an idealized model of the cusps observed in
Fig.~\ref{f:isoV}. In particular, consider two symmetrically located
planar flame sheets propagating with the speed $S_{L,0}$ and
approaching each other with the inclination angle to the mid-plane
$\alpha$. At the point of collision, the flame sheets merge and form a
cusp. Its properties, such as its structure, speed of propagation, and
effective burning velocity, are determined in this configuration only
by two parameters, $S_{L,0}$ and $\alpha$. Such cusps can be formed
through either of the channels shown in Fig.~\ref{f:CollTypes}.

Fig.~\ref{f:Cusp} illustrates the flame structure formed in this
situation. It shows the distribution of the fuel mass fraction in two
simulations performed for $\alpha = 1^{\circ}$ and $\alpha =
4^{\circ}$ with Athena-RFX using the same physical model as in
simulations S1-S3. The domain has the resolution $\Delta x =
\delta_L/32$ and zero-order extrapolation boundary conditions on all
sides. At $t = 0$, two intersecting flame sheets were initialized with
the exact structure of the planar laminar flame corresponding to the
fuel temperature $T_0$ and density $\rho_0$ (see
Table~\ref{t:Params}). Uniform pressure $P_0$ and zero velocities were
set in the domain at $t = 0$. After the initial transient stage, the
flame develops the structure shown in Fig.~\ref{f:Cusp}.  The
structure of the cusp itself does not change with time, provided that
the planar flame sheets extend sufficiently far from it.  Therefore,
Fig.~\ref{f:Cusp} can be viewed as a steady-state solution.

Fig.~\ref{f:Cusp} shows that flame collision at very low inclination
angles results in the formation of a highly elongated structure. It is
formed by rapid heating and subsequent ignition of a larger amount of
fuel than in the planar laminar flame due to the focusing of the
thermal flux from two approaching flame sheets. Consequently, the
extent of this structure is determined by the size of the region in
which the preheat zones of approaching flames overlap and, thus,
create the necessary enhanced thermal flux. Since the width of the
preheat zone is $\approx \! \delta_{L,0}$, very small values of
$\alpha$ are required for the cusp to be substantially broader than
$\delta_{L,0}$. This can further be seen in the distributions of $Y$,
$T$, and $\dot{Y}$ along the symmetry axis of the cusp shown in
Fig.~\ref{f:CuspStruct}.  The structure of the reaction zone in the
cusp very quickly approaches that of the laminar flame with increasing
$\alpha$, so that already at $\alpha \sim 4^{\circ}$ the two reaction
zones become very similar. At the same time, the preheat zone of the
cusp remains substantially broadened for much larger values of
$\alpha$.

There are three characteristic speeds in this problem. Sufficiently
far from the tip of the cusp, the flame locally propagates normal to
its surface with the laminar flame speed $S_{L,0}$, i.e., there $S_n =
S_{L,0}$. The cusp itself moves into the fuel with the phase velocity
$U_c$, which results simply from the two inclined planar surfaces
colliding with each other. This speed can be determined based on
purely geometrical considerations \cite{Zeldovich1},
\beq
U_c = \frac{S_{L,0}}{\sin\alpha}.
\label{e:CuspSpeed}
\eeq
This cusp propagation speed, normalized by $S_{L,0}$, is shown as a
solid line in Fig.~\ref{f:CuspSpeed}a along with $U_c/S_{L,0}$
determined as the velocity of the leftmost point of the $Y = 0.15$
isosurface in simulations for four values of $\alpha$.
Eq.~(\ref{e:CuspSpeed}) is within $\lesssim \! 1\%$ accuracy of the
computed values. $U_c$ approaches the laminar flame speed as $\alpha
\rightarrow 90^{\circ}$, i.e., when the two flame sheets cease to
advance toward each other. At the opposite limit of small $\alpha$,
$U_c$ can be quite large and, in principle, it could be infinite when
$\alpha \rightarrow 0^{\circ}$. In particular, at $\alpha \approx
0.5^{\circ}$, $U_c$ becomes larger than the sound speed in cold fuel
indicated with the horizontal dashed line in Fig.~\ref{f:CuspSpeed}a.
Note also that at $\alpha = 1^{\circ}$, the value of $U_c \approx
57S_{L,0}$ is comparable to the high propagation velocity $\approx \! 
55S_L$ of cusps in regions B - C in Fig.~\ref{f:isoV} estimated in
\S~\ref{M:Collisions}.

The third characteristic speed, which is the most important for us, is
the local burning speed in the cusp, $S_n$. The broadened reaction
zone at low flame-inclination angles (Figs.~\ref{f:Cusp} and
\ref{f:CuspStruct}b) causes more fuel to be consumed per unit flame
surface area than in the planar laminar flame, and, therefore, it
leads to a larger local flame speed $S_n > S_{L,0}$. Since the
reaction-zone width is the largest at the tip of the cusp, $S_n$ has
its maximum there and it gradually decreases to its laminar value
$S_{L,0}$ in the planar regions of the flame sufficiently far from the
cusp. The maximum value of $S_n$ at the cusp tip can be found as
\beq
S^*_n \equiv \max(S_n) = \frac{1}{\rho_0}\int\rho \dot{Y}dx =
-\frac{B}{\rho_0}\int\rho^2 Y \exp \Big(-\frac{Q}{RT}\Big) \ dx.
\label{e:CuspBurn}
\eeq
Here eq.~(\ref{e:ReacModel}) was used for $\dot{Y}$, and the integral
is taken along the symmetry axis of the cusp shown in
Fig.~\ref{f:Cusp}. Fig.~\ref{f:CuspSpeed}b shows the computed values
of $S^*_n$ normalized by $S_{L,0}$ for the same four inclination
angles as in Fig.~\ref{f:CuspSpeed}a.

Two key conclusions emerge from Fig.~\ref{f:CuspSpeed}. First, $S^*_n$
is substantially lower than $U_c$, being less by more than an order of
magnitude at the values of $\alpha$ considered here. Second and most
important, $S^*_n$ can indeed be much larger than $S_{L,0}$ at small
inclination angles.

\subsection{Connection between the increase of $S_n$ in cusps and $S_T$}
%-------------------------------------------------------------------------------
\label{M:Connection}

We are interested, however, not just in $S_n$ but, rather, in the
burning speed of an extended flame region, as this is a direct
equivalent of the turbulent flame speed. The total burning rate of the
flame configuration shown in Fig.~\ref{f:Cusp} is
\beq
S_c = \frac{\dot{m}_R}{\rho_0},
\label{e:SCusp}
\eeq
where $\dot{m}_R$ is the total mass of reactants converted into
product per unit time. If the flame were to propagate everywhere with
the speed $S_{L,0}$, then $S_c = A_YS_{L,0}$, where $A_Y$ is the
surface area of the flame in such a configuration. In this case all
isosurfaces of $Y$ would be parallel to each other and, thus, $A_Y$
based on any value of $Y$ could be used. Otherwise,
\beq
I_Y = \frac{S_c}{A_YS_{L,0}} > 1.
\label{e:DamkCusp}
\eeq
Here $I_Y$ is completely equivalent to the definition used in
eq.~(\ref{e:Damk}) in the context of the turbulent flame speed.
Indeed, substituting the definition of $S_T$ given by eq.~(\ref{e:ST})
into the definition of $I_Y$ given by eq.~(\ref{e:Damk}) shows that
$\dot{m}_R/\rho_0 = I_YA_YS_{L,0} = S_c$. Furthermore, as can be seen
in Fig.~\ref{f:Cusp}, at low $\alpha$ isosurfaces of $Y$ do not have
the same surface area. Therefore, similarly to the case of the
turbulent flame, here $I_Y$ is also the function of $Y$. For
consistency with previous results, we will consider $I_{0.15}$ based
on the isosurface of $Y = 0.15$ representing the peak reaction rate.

Since $S_n$ monotonically decreases away from the cusp, both $S_c$ and
$I_{0.15}$ depend on the size of the flame region being considered.
This size can be represented by the variable length $l_c$, illustrated
in Fig.~\ref{f:Cusp}, which is defined as the distance from the
leftmost point of the isosurface of $Y = 0.15$ in the direction of
cusp propagation. Considering different values of $l_c$ is equivalent
to varying the fraction of the total flame surface area in
Fig.~\ref{f:Cusp} represented by the planar section in which $S_n
\approx S_{L,0}$. Consequently, this allows one to vary the relative
contribution to $S_c$, and thus to $I_{0.15}$, of the cusp in which
$S_n > S_{L,0}$.

Fig.~\ref{f:DamkCusp} shows $I_{0.15}$, calculated using
eqs.~(\ref{e:SCusp})-(\ref{e:DamkCusp}), as a function of
$l_c/\delta_{L,0}$ for four values of $\alpha$. At smaller $l_c$ close
to the width of the reaction zone for a given $\alpha$ (see
Fig.~\ref{f:CuspStruct}b), $I_{0.15}$ essentially reflects only the
values of $S_n$ in the immediate vicinity of the tip of the cusp,
where $S_n$ is close to $S_n^*$. On the other hand, at larger $l_c$,
the surface area of the planar flame region, in which $S_n \approx
S_{L,0}$, is larger, and so this region represents a greater fraction
of the total surface area of the flame configuration shown in
Fig.~\ref{f:Cusp}. As a result, the contribution of the planar flame
starts to dominate that of the cusp, which causes $I_{0.15}
\rightarrow 1$ as seen in Fig.~\ref{f:DamkCusp}. Note that at lower
$\alpha$, the region in which $S_n > S_{L,0}$ extends progressively
further from the cusp tip. In particular, at $\alpha = 1^{\circ}$,
$I_{0.15}$ is substantially greater than unity even at $l_c >>
\delta_{L,0}$. At the same time, already at $\alpha = 4^{\circ}$,
$I_{0.15}$ drops to $\lesssim \!  1.04$ at $l_c \! \approx \!
5\delta_{L,0}$ which shows that $S_n$ recovers its laminar value
within a few laminar flame widths from the cusp tip.

These results show that the formation of a cusp due to the collision
of planar flame sheets produces values of $I_{0.15}$ substantially
larger than unity and comparable to those observed in the simulations
presented here. As a result of a higher local flame speed in a cusp,
its contribution to the global turbulent burning speed is
disproportionately large compared to the fraction of the instantaneous
flame surface area in it. A more complex flame configuration,
consisting of multiple planar flame sheets that approach each other
and collide forming multiple cusps, can then be considered. In this
system, for instance, if the flame surface density is increased, flame
sheets will get closer to each other causing flame collisions to
become more frequent and the resulting cusps to become more
numerous. Consequently, the fraction of the flame surface area
contained in cusps will increase and $I_{0.15}$ will grow, which is
demonstrated by the increase in $I_{0.15}$ with decreasing $l_c$.

This simplified model demonstrates the mechanism through which $S_T$
can produce an exaggerated response to the increase in $A_T$. Yet it
does not take into account the full complexity of an actual turbulent
flame. Such flame does not consist of perfectly planar flame sheets
that merge, but instead it is constantly wrinkled and folded on a
variety of scales (Fig.~\ref{f:CollTypes}). As a result, cusps will
form in a multitude of configurations. For instance, in addition to
the purely 2D situation considered here, there will also exist 3D
flame collisions which will result in even higher local flame speeds
in the cusps and, thus, will have an even larger effect on the
magnitude of $I_{0.15}$. The probability of the formation of such more
complex flame collisions, however, will rapidly decrease with the
increase in their complexity \cite{Driscoll,Khokhlov}. Nevertheless,
in order to predict a particular value of $I_{0.15}$ which can be
expected in the turbulent flow of a specific intensity, it is
necessary to understand the types of all cusps which can form, the
local flame speed of each type, its probability of formation, and,
thus, the contribution of each type to the exaggerated response of
$S_T$. Such detailed analysis is the subject for future studies.

\subsection{Effect of cusps on the diffusion velocities in the turbulent
flame}
%-------------------------------------------------------------------------------
\label{M:DiffVel}

Changes in the local flame structure (Figs.~\ref{f:Cusp} and
\ref{f:CuspStruct}) in cusps will result in modified local diffusion
velocities, $D|\nabla Y|$. This is the direct consequence of focusing
of the diffusive fluxes which alters the molecular and thermal
transport in regions of high flame curvature. It is, therefore,
instructive to compare the distribution of the diffusion velocities in
the idealized cusps considered above and in the actual turbulent
flame.

We calculated the surface-averaged normalized diffusion velocity,
\beq
\frac{V_D(Y)}{S_L} = \frac{\int_{A_Y}D|\nabla Y|dA}{A_Y S_L},
\label{e:DiffVel}
\eeq
for a set of discrete values of $Y$ in simulation S3 and in the cusp
formed at the low inclination angle $\alpha = 1^{\circ}$ shown in the
upper panel of Fig.~\ref{f:Cusp}. Here the diffusion coefficient $D$
is given by eq.~(\ref{e:CondDiff}) and the integration is performed
over an isosurface of a given $Y$. In S3, the values of $V_D(Y)$
change with time due to the fuel heating by turbulence. In order to
take this effect into account, we normalized the instantaneous
$V_D(Y)$ by the corresponding actual instantaneous laminar flame speed
$S_L$, as was discussed in \S~\ref{Simulations}. The resulting
distributions were subsequently time-averaged. In the idealized cusp
calculation, $S_L = S_{L,0}$ at all times, and the instantaneous flame
structure shown in Fig.~\ref{f:Cusp} represents a steady state. The
obtained $\overline{V_D(Y)/S_L}$ for both calculations is shown in
Fig.~\ref{f:DiffVel}. We also show two distributions of $V_D(Y)$
representing the exact planar laminar flame solutions constructed for
two fuel temperatures, namely $T_0 = 293$ K and $320$ K, which
correspond to the average fuel temperatures at the beginning and at
the end of S3. These were also normalized by the corresponding $S_L$.

Fig.~\ref{f:DiffVel} shows that the flame configuration in
Fig.~\ref{f:Cusp} (upper panel), which consists both of a cusp and the
extended planar flame regions, has diffusion velocities in the
reaction zone $\lesssim \! 10\%$ lower than in the planar laminar
flame. At the same time, a similar decrease of $V_D/S_L$ in a laminar
flame is produced as a result of fuel heating to $320$ K. On the other
hand, the time-averaged distribution of $\overline{V_D/S_L}$ inside
the turbulent flame is also similarly lower in comparison with the
base laminar case at fuel temperature $T_0 = 293$ K. This fact, along
with the $1\sigma$ standard deviation of the instantaneous values of
$V_D/S_L$ shown as the shaded gray region, indicates that the observed
decrease of the diffusion velocities in S3 is consistent with the
combined effect of fuel heating and the formation of cusps.

\subsection{Criterion for onset of the regime of flame evolution
influenced by cusps}
%-------------------------------------------------------------------------------
\label{M:Criterion}

Properties of cusps discussed above allow us to determine when the
evolution of a turbulent flame can be expected to become affected by
cusp formation, thereby, leading to values of $I_{0.15}$ substantially
larger than unity. The results presented in this paper show that such
high $I_{0.15}$ require a large number of cusps to exist or, in other
words, the flame must have large curvature $\sim\!1/\delta_L$ over a
significant fraction of its total surface. This means that the flame
must be tightly folded with a characteristic separation $1/\Sigma_T$
comparable to the full flame width $l_F$, which is indeed the
situation observed in the simulations discussed here.

Consider now the two channels of cusp formation discussed in
\S~\ref{M:Collisions} and illustrated in Fig.~\ref{f:CollTypes}.
Nonlocal flame collisions cannot create a tightly packed turbulent
flame since, by definition, they fold the flame only on large scales
causing flame sheets to move toward each other. Large curvature forms
only at the moments of collision after which the large speed of cusp
propagation, $U_c$, causes rapid annealing of the flame surface.
Consequently, a tightly packed flame can form only if local flame
collisions are an efficient process capable of folding the flame on
scales $\sim\!\delta_L$ not only in isolated regions but consistently
throughout the flame surface.

It was discussed in \S~\ref{M:Cusps} (Fig.~\ref{f:CuspSpeed}) that
$U_c$, unlike $S^*_n$, is much greater than $S_L$ even at large flame
inclination angles, and, thus, it increases gradually with
curvature. Therefore, $U_c$ is the primary factor responsible for
smoothing the flame surface while it is being wrinkled by
turbulence. In particular, as the turbulent motions $U_{\lambda}$
shown in Fig.~\ref{f:CollTypes}a fold the flame thereby progressively
increasing its curvature, they must overcome the straightening action
of the growing $U_c$. The formation of a cusp, as well as its
structure, are then determined by the balance between $U_\lambda$ and
$U_c$.

Next consider a section of an idealized curved flame front,
schematically shown in Fig.~\ref{f:CuspSchematic}, containing a cusp
formed through a local flame collision (Fig.~\ref{f:CollTypes}a). The
flame is perturbed with a wavelength $\lambda_c$ and an amplitude
$l_c$. This type of structure was considered by Zel'dovich
\cite{Zeldovich1} as a model of a flame formed under the action of
the Landau-Darrieus instability in order to analyze the stabilizing
effect of cusps. The same structure, however, can emerge under the
action of any destabilizing process, e.g., the Rayleigh-Taylor
instability or, in our case, turbulence.

In this case, the rate of decrease of the cusp amplitude is
\cite{Zeldovich1}
\beq
\Big(\frac{dl_c}{dt}\Big)_- = -S_L\Big(\frac{1}{\sin{\alpha}} + 1\Big).
\label{e:dCusp1}
\eeq
The angle $\alpha$ and, thus, $(dl_c/dt)_-$, will change with the cusp
amplitude. In order to relate $\alpha$ and $l_c$, a parabolic shape of
the flame was assumed in \cite{Zeldovich1}, and eq.~(\ref{e:dCusp1})
can then be rewritten as
\beq
\Big(\frac{dl_c}{dt}\Big)_- = -8S_L\frac{l^2_c}{\lambda^2_c}.
\label{e:dCusp2}
\eeq
The net rate of change of $l_c$ is determined by the balance of the
stabilizing process described by eq.~(\ref{e:dCusp2}) and a
destabilizing process, namely
\beq
\frac{dl_c}{dt} = \Psi - 8S_L\frac{l^2_c}{\lambda^2_c}.
\label{e:dCuspNet}
\eeq
In the turbulent flow, flame perturbations can be assumed to grow
linearly in time as they are being stretched by the turbulent speed
$U_{\lambda_c}$ characteristic of the scale $\lambda_c$. This gives
the growth rate $\Psi = U_{\lambda_c}$. By setting $dl_c/dt = 0$, the
limiting value of $l_c(\lambda_c)$ can be determined. It was shown by
Khokhlov \cite{Khokhlov} in the context of Rayleigh-Taylor-unstable
flames that nonlinear flame stabilization due to cusp propagation can
be expected to fail once $l_c \gtrsim \lambda_c$. Therefore, by
setting $l_c = \lambda_c$, $\Psi = U_{\lambda_c}$, and $dl_c/dt = 0$
in eq.~(\ref{e:dCuspNet}), the critical value of the turbulent
velocity on a given scale $\lambda_c$ can be found \cite{Khokhlov}
\beq
U_{\lambda_c} = 8S_L.
\label{e:Ucrit}
\eeq
This is the value of $U_{\lambda_c}$ that is needed to overcome the
stabilizing effect of cusps and, thereby, to deform the flame on scale
$\lambda_c$. This result was used to demonstrate that, in order for
the flame to be unstable at a wavelength $\lambda_c$ under the action
of turbulent motions, the turbulent speed on that scale must be
substantially larger than $S_L$ \cite{Khokhlov}.

This shows that given a specific turbulent cascade, the flame will be
stabilized on scales smaller than some critical wavelength
$\lambda_c^{crit}$ on which the condition in eq.~(\ref{e:Ucrit}) is
satisfied. In order for the cusps to have a pronounced effect on the
flame evolution, the flame must be folded on scales comparable to the
full flame width $l_F$. Therefore, by setting $\lambda_c^{crit} = l_F$
and using eq.~(\ref{e:Ucrit}), we find that the flame will be curved
on scales $\sim\! l_F$ when turbulent velocity on that scale becomes
$\sim \! 8S_L$. In a turbulent flow with the Kolmogorov energy
spectrum, the corresponding critical value of the integral turbulent
velocity then is
\beq
U_l^{crit} \approx 8S_L\Big(\frac{l}{l_F}\Big)^{\sfrac{1}{3}}.
\label{e:UlCrit}
\eeq
This gives the critical values of the Karlovitz number, $Ka^{crit}$,
and the Damk\"ohler number, $Da^{crit}$, written using their
traditional definitions \cite{Peters}
\beq
\begin{array}{l}
\displaystyle Ka^{crit} = \frac{\tau_F}{\tau_{\eta}} =
\Big(\frac{l_F}{L_G}\Big)^{\sfrac{1}{2}} =
8^{\sfrac{3}{2}} \approx 20 , \quad \\
\displaystyle Da^{crit} = \frac{\tau_T}{\tau_F} = \frac{lS_L}{l_FU_l} =
\frac{1}{8}\Big(\frac{l}{l_F}\Big)^{\sfrac{2}{3}}.
\label{e:KDCrit}
\end{array}
\eeq
Here $\tau_F = l_F/S_L$ is the characteristic flame time,
$\tau_{\eta}$ is the Kolmogorov time, $\tau_T = l/U_l$ is the
characteristic turbulent time on the integral scale, and $L_G = l
(S_L/U_l)^3$ is the Gibson scale.

Fig.~\ref{f:CuspDiagram} presents a traditional combustion regime
diagram \cite{Peters}. The orange region shows the range of the
regimes, in which $I_{0.15}$ is expected to be larger than unity,
according to eq.~(\ref{e:KDCrit}). The solid red square corresponds to
the simulations discussed here. We also determined the time-averaged
value of $I_{0.15}$ in a simulation similar to S2, but with
approximately half the turbulent intensity which, however, was still
above the $Ka^{crit} \approx 20$ line. In this calculation,
represented with the open red square in Fig.~\ref{f:CuspDiagram}, we
found $\overline{I}_{0.15} \approx 1.1$, compared with
$\overline{I}_{0.15} \approx 1.14$ found in simulation S3
(Table~\ref{t:Props}). Details of this calculation will be presented
in a separate paper.

Despite the simplicity of the cusp model used here,
eqs.~(\ref{e:UlCrit})-(\ref{e:KDCrit}) provide a rather accurate
criterion for the onset of the regime of the turbulent flame evolution
in which the contribution of cusps is expected to become important. In
particular, the decrease in the value of $\overline{I}_{0.15}$ between
the two turbulent intensities considered in Fig.~\ref{f:CuspDiagram}
suggests that below the $Ka^{crit} \approx 20$ line $I_{0.15}$, on
average, deviates from unity at most by a few percent.

\subsection{Effect of different turbulent intensities and system sizes}
%-------------------------------------------------------------------------------
\label{M:Regime}

For the given reaction-diffusion model and fuel properties, the only
two free parameters in the system considered here are the turbulent
intensity and system size represented by the turbulent integral
velocity, $U_l$, and scale, $l$. Consequently, the question arises,
how will the distributions of $S_T/S_L$ and, more importantly, of
$I_{0.15}$ as a function of $A_{0.15}/L^2$ shown in
Figs.~\ref{f:CorrelAS}b and Fig.~\ref{f:CorrelIA} vary with the change
in $l$ and $U_l$?

The results presented here and in \cite{Poludnenko} demonstrated the
following key property of the flame interaction with high-speed
turbulence. Turbulent motions are much more efficient at folding the
flame with increasingly greater curvature than at disrupting its
internal structure. This creates tightly packed turbulent flames
which, in regions of low curvature, retain locally the laminar
structure and velocity. Therefore, an increase in $U_l$ creates both
larger flame surface area, $A_T$, and density, $\Sigma_T$.

On the other hand, $l$ does not affect $\Sigma_T$. The analysis in
\S~\ref{M:Criterion} showed that flame folding on a given scale
is controlled only by the turbulent velocity on that scale rather than
by the total range of scales present in the system. Instead, a larger
system size results in a wider turbulent flame and, thus, in larger
values of $A_T$. This follows from the fact that, in the high-speed
regime, the turbulent flame width, $\delta_T$, is primarily determined
by $l$ (and, therefore, by the driving scale, $L$) rather than by
$U_l$, as was discussed in \cite{Poludnenko} based on the comparison
of the results of that work with the results of Aspden et
al. \cite{Aspden}. A similar conclusion was also reached by Peters
\cite{Peters} (cf. eq.~(2.175) therein).

At the same time, a given $U_l$ and $l$ is not characterized by a
single $S_T$, $A_T$, and $\Sigma_T$. Instead, a variety of different
flame configurations are realized in the course of flame evolution, as
was seen above (cf. Fig.~\ref{f:CorrelAS}b). These configurations have
different $A_T$, which is the primary factor determining the
distribution of $S_T$. More importantly, however, at a given $A_T$,
there are also different numbers and types of cusps present in the
flame. This creates a scatter in the distribution of $S_T/S_L$ vs.
$A_T/L^2$, which in the absence of cusps would lie on the $S_T/S_L =
A_T/L^2$ line. These cusps are the result of the combined action of
the two cusp-formation channels discussed in \S~\ref{M:Collisions}.
Consequently, this scatter reflects the dependence of the two types of
flame collisions on $A_T$ and $\Sigma_T$.

The efficiency of local flame collisions (Fig.~\ref{f:CollTypes}a) is
independent of either $A_T$ or $\Sigma_T$, since this process of cusp
formation is controlled only by the intensity of turbulent motions
which stretch and fold the flame at a certain point of its surface. As
a result, the probability of cusp formation per unit flame surface
area in this process does not increase with $A_T$ or $\Sigma_T$.
Consequently, if only local flame collisions were present in the
system, than the fraction of $A_T$ contained in cusps would grow
proportionally to the increase in $A_T$, and the resulting $I_{0.15}$
would not change.

On the other hand, the efficiency of nonlocal flame collisions
(Fig.~\ref{f:CollTypes}b) does depend not only on $U_l$, but also on
$A_T$ and $\Sigma_T$. To show this, the following qualitative model
can be used (see also \cite{Khokhlov}). In this process, cusps are
created due to collisions of flame sheets with very low curvature.
Therefore, consider an idealized structure consisting of planar flame
sheets discussed in \S~\ref{M:Connection}. If the flame surface
density of this configuration is $\Sigma_T$, then the average flame
separation is $1/\Sigma_T$. Flame sheets will move toward each other
with the speed $S_L + U_T$, where $U_T$ is some characteristic
turbulent velocity responsible for the advective transport of the
flame. Since $U_l$ is the largest velocity of coherent turbulent
motions, then $U_T$ can be approximated with $U_l$. Consequently, the
flame sheets will merge and annihilate in the time $t_c \sim
1/(\Sigma_T (S_L+U_l))$. The total number of collisions and, thus, the
total number of resulting cusps, $N_c$, will depend on the total
surface area of the flame sheets times the frequency of their
collisions, i.e., $N_c \propto A_T/t_c \propto A_T\Sigma_T(S_L +
U_l)$. Since $A_T$ and $\Sigma_T$ are well correlated
(\S~\ref{D:CorrelASigma}), $A_T$ can be viewed as simply proportional
to $\Sigma_T$, which gives\footnote{This expression is very similar to
the destruction term often used in the balance equation of the flame
surface density (see \cite{Duclos} for a review of a number of such
models). This is also analogous to the rate of collisions of gas
molecules, with $\Sigma_T$ playing the role of number density and the
turbulent intensity playing the role of temperature in determining the
speed with which constituents approach each other.} $N_c \propto
(S_L+U_l)A_T^2 \propto (S_L + U_l)\Sigma_T^2$. Therefore, the number
of cusps formed per unit flame surface area will be
\beq
\frac{N_c}{A_T} \propto \big(S_L+U_l\big)A_T \propto
\big(S_L + U_l\big)\Sigma_T.
\label{e:df}
\eeq

These considerations show that at a given $l$ and $U_l$, the
distribution of values of $I_{0.15}$ has two contributions: one due to
local flame collisions independent of $A_T$ and $\Sigma_T$, and the
other due to nonlocal flame collisions, which increases with $A_T$ and
$\Sigma_T$. In smaller systems and at lower turbulent intensities, or,
equivalently, at lower $A_T$ and $\Sigma_T$, the first process is
dominant. This can be seen in the distribution of $I_{0.15}$
vs. $A_{0.15}/L^2$ in Fig.~\ref{f:CorrelIA}, which shows almost no
dependence on $A_T$. Note, however, a weak increasing trend suggested
by the least squares fit. In particular, the lowest values of
$I_{0.15}$ become progressively larger with increasing $A_{0.15}$
approaching $\approx \! 1.1$ as $A_{0.15}/L^2 \to 5$. While the
statistical significance of this trend must be confirmed through
further studies, this trend would be indicative of the contribution of
nonlocal collisions. The analysis given above suggests that such
dependence of $I_{0.15}$ on $A_T$ will become pronounced in larger
systems or at larger turbulent intensities.

The fact that $I_{0.15}$ must depend on $A_T$ also follows from a
different argument. Consider first a limiting value of $\Sigma_T$. The
fact that the flame is not an infinitely thin surface means that such
a limit does exist. Once $\Sigma_T$ becomes large enough so that all
of the flame surface comes into contact with itself, $\Sigma_T$ cannot
increase further. This maximum value $\Sigma_T^{max}$ can be estimated
by assuming that the minimum flame separation is equal to the full
flame width $l_F$. Then\footnote{Note that $\Sigma_T^{max}$ is
different from $\Sigma_{max}$ typically used in combustion research
and discussed in \S~\ref{D:Isosurf}.}
\beq
\Sigma_T^{max} \equiv \Sigma_{0.15}^{max} = \frac{1}{l_F}.
\label{e:SigmaLim}
\eeq
Using the definition of $\Sigma_Y$ given by eq.~(\ref{e:Sigma}), the
corresponding limiting value of the flame surface area then is
\beq
\frac{A_T^{max}}{L^2} \equiv \frac{A_{0.15}^{max}}{L^2} =
\Sigma_T^{max}\delta_T = C_1\frac{L}{l_F} = (C_1C_2)\frac{l}{l_F}.
\label{e:ALim}
\eeq
Here, $C_1 = \delta_T/L$ and it typically is $\approx \! 2 - 4$ based
on the results of \cite{Poludnenko,Aspden}. The $C_2 = L/l$ and in the
case of the Kolmogorov turbulence $C_2 \approx 4$. In our case $l_F
\approx 2\delta_{L,0}$ (cf. Fig.~7 in \cite{Poludnenko}).  Therefore,
based on the values of $L$ and $l$ given in Table~\ref{t:Runs} for the
system considered here, $\Sigma_T^{max}
\approx 1.5 \ \textrm{mm}^{-1}$ and $A_T^{max}/L^2 \approx 7 - 8$.

The situation when all of the flame surface is in contact with itself
corresponds to the infinite local flame speed and, thus, infinite
$S_T$. This is equivalent to the complete domination of the nonlocal
process of cusp formation when the flame sheets collide simply because
they are very close to each other and not as a result of the folding
action of turbulence. Therefore, as $\overline{A}_T/L^2 \rightarrow
A_T^{max}/L^2$, both $\overline{S_T/S_L} \rightarrow \infty$ and
$\overline{I}_{0.15} \rightarrow \infty$. On the other hand, as
$\overline{A}_T/L^2 \rightarrow 1$, $\overline{S_T/S_L} \to 1$ and,
thus, $\overline{I}_{0.15} \to 1$. This shows that
$\overline{I}_{0.15}$ cannot be constant. Instead, it has to be a
monotonically increasing function of $A_T$, as suggested.

In summary, the following picture emerges. At a given $l$, lower
turbulent intensities cause the flame to be less convolved and, thus,
to develop less surface area. As a result, the overall distribution of
$S_T/S_L$ vs. $A_T/L^2$ would shift toward smaller values of $A_T/L^2$
and, thus, $S_T/S_L$. Furthermore, the efficiency of both local and
nonlocal flame collisions would decrease. As a result, cusps would
become infrequent representing only a small fraction of the total
flame surface. Consequently, the flame almost everywhere would
propagate locally with its laminar speed, thus causing the scatter in
the distribution of $S_T/S_L$ to decrease and values of $S_T/S_L$ to
collapse onto the $S_T/S_L = A_T/L^2$ line in the limit of $U_l \to
0$. Note, however, that while at $U_l < U_l^{crit}$ given in
eq.~(\ref{e:Ucrit}) local flame collisions practically cease to occur,
individual cusps can still form through nonlocal collisions. Since
these depend on $A_T$, $I_{0.15}$ would approach unity with decreasing
$U_l$ more slowly in larger systems that have greater $A_T$ than in
smaller systems.

At larger $U_l$, the behavior would be opposite. Not only would the
overall distribution of $S_T/S_L$ vs. $A_T/L^2$ shift to larger values
of $A_T/L^2$ and $S_T/S_L$, but also the efficiency of both local and
nonlocal flame collisions would increase, causing a much larger
scatter of the distribution. Furthermore, at high turbulent
intensities, flame configurations with very few cusps and, thus,
$I_{0.15} \approx 1$, would become increasingly less likely to
occur. Therefore, larger $U_l$ will cause not only a higher rate of
burning due to the increase in $A_T$, but it will also periodically
lead to a substantially more exaggerated response of $S_T$ to such
increase in $A_T$. Finally, dependence of $I_{0.15}$ on $A_T$ would be
more pronounced due to the increased contribution of nonlocal
collisions and, at a given $U_l$, larger systems would tend to have
higher values of $I_{0.15}$.

\section{Conclusions}
%-------------------------------------------------------------------------------
\label{Conclusions}

This work continued the analysis of the set of three numerical
simulations first presented in \cite{Poludnenko}. These calculations
model the interaction of a premixed flame with high-speed, subsonic,
homogeneous, isotropic turbulence in an unconfined system, \ie in the
absence of walls and boundaries. The turbulent r.m.s. velocity,
$U_{rms}$, is $\approx \! 35$ times larger than the laminar flame
speed, $S_L$. The resulting Damk\"ohler number based on the turbulent
integral scale and velocity is $Da = 0.05$.

It was demonstrated in \cite{Poludnenko} that this system represents
turbulent combustion in the thin reaction zone regime. Even in the
presence of such intense turbulence, the flame brush consists of the
highly convolved flame with its reaction-zone structure virtually
identical to that of the planar laminar flame and with the preheat
zone broadened by a factor $\approx \! 2$. The fact that turbulence is
unable to penetrate and disrupt the internal structure of the reaction
zone suggested that the flame must be propagating locally with the
speed $S_L$ \cite{Poludnenko}. This raised the following question: is
the turbulent flame speed, $S_T$, in the thin reaction zone regime
completely determined by the increase in the flame surface area,
$A_T$, as it is the case in the wrinkled and corrugated flamelet
regimes? Here we summarize the main findings of our study of this
issue.

Analysis of the area and density, $A_Y$ and $\Sigma_Y$, of the fuel
mass-fraction isosurfaces showed that the flame, folded inside the
flame brush in the presence of high-speed turbulence, cannot be viewed
as a thin uniform structure in which all isosurfaces are parallel to
each other. Different regions of the flame have quite different
response to the action of turbulence. In the higher-resolution
calculations S2 and S3, both $A_Y$ and $\Sigma_Y$, on average,
increase monotonically through the flame with increasing $Y$, which
leads to the distinctive inverted-S shape of the time-averaged
distributions of $A_Y$ and $\Sigma_Y$. Consequently, isosurfaces of
higher $Y$ are folded by turbulence on progressively smaller
scales. This causes substantially finer wrinkling of the flame surface
on the fuel side than on the product side, as observed in
\cite{Poludnenko}. Furthermore, in the reaction zone, $A_Y$ grows due
to both much tighter folding of isosurfaces and the increase in the
overall width of the flame brush, with the first process being more
dominant.  On the other hand, in the preheat zone, isosurfaces of $Y$
are packed primarily on small scales without contributing to the
increase of $\delta_T$.

These properties of the distributions of $\overline{A}_Y$ and
$\overline{\Sigma}_Y$ showed that, in the thin reaction zone regime,
the definition of the flame surface area must be revisited before the
relation between $A_T$ and $S_T$ can be considered. In particular, it
must be determined which value of $Y$ fully and accurately
characterizes the evolution and global properties of the turbulent
flame. To answer this question, we analyzed the correlation of $A_Y$
with quantities that characterize both the energetics and the global
structure of the turbulent flame, namely its speed, $S_T$, and width,
$\delta_T$. This analysis demonstrated that global properties of the
turbulent flame in the thin reaction zone regime are accurately
represented by the structure of the region of peak reaction rate. For
the reaction-diffusion model used in this work, this corresponds to $Y
\approx 0.15$. Therefore, the isosurface of this value of $Y$ must be
viewed as the flame surface and $A_T = A_{0.15}$. Correspondingly, the
flame surface density is $\Sigma_T =
\Sigma_{0.15}$.

Large values of $\Sigma_T$ developed in the course of the system
evolution, on one hand, and the absence of any broadening of the
reaction zone observed in \cite{Poludnenko}, on the other,
demonstrated the following important fact. High-intensity turbulence
is much more efficient at tightly packing the flame inside the flame
brush than at disrupting and broadening its internal structure (also
see \cite{Peters}).

The distribution of $S_T/S_L$ as a function of $A_T/L^2$ and, more
specifically, the ratio $I_{0.15} = (S_T/S_L)/(A_T/L^2)$ showed that,
in the thin reaction zone regime, the magnitude of $S_T$ is controlled
primarily by the increase of the flame surface area. In particular, on
average $I_{0.15} \approx 1.14$. As a result, a variety of different
flame configurations realized over time, with both $A_T$ and
$\Sigma_T$ varying by over a factor of $\approx\! 3$, cause $S_T/S_L$
also to vary over a broad range of values $\approx \! 1.6 - 5.5$. Such
significant changes in the burning rate demonstrate that, in order to
properly characterize the turbulent flame evolution in this regime, it
is not sufficient to consider only the time-averaged $\overline{S}_T$.
Instead, the full PDF of the instantaneous values of $S_T$ must be
determined.

At the same time, the distribution of $S_T/S_L$ exhibited another
important characteristic. In the course of the system evolution,
$S_T/S_L$ varies from within only a few percent of $A_T/L^2$ to as
high as $30\%$ larger than $A_T/L^2$. This led to the following key
conclusion of this work. In the thin reaction zone regime given a
sufficiently high turbulent intensity, $S_T$ periodically exhibits a
substantially exaggerated response to the increase in $A_T$. Such
accelerated burning in the turbulent flame means that an additional
mechanism must augment the effect of the increase of the flame surface
area.

Our analysis showed that tightly packed flame configurations, produced
by high-speed turbulence, have the average flame separation of only a
few full flame widths. This results in frequent flame collisions that
lead to the formation of regions of high flame curvature $\gtrsim
1/\delta_L$, or ``cusps.'' The resulting significant focusing of the
thermal flux over an extended region of the flame surface can
significantly increase the local burning speed, $S_n$, in the cusp
over its laminar value, $S_L$. These large values of $S_n$ cause the
contribution of cusps to the total $S_T$ to be disproportionately
large compared to the flame surface area they contain. This provides a
natural mechanism to accelerate burning in the turbulent flame above
what can be attributed to the increase in $A_T$, even in reactive
mixtures characterized by $Le = 1$. The increase of $S_n$ in cusps is
inherently local, and it does not require the flame to be broadened
and its speed increased by small-scale turbulent transport, in
agreement with the results of \cite{Poludnenko}.

Therefore, our results indicate that cusp formation is a crucial
process in the turbulent flame evolution responsible for controlling
$S_T$. It is capable of accelerating burning in addition to the two
processes originally suggested by Damk\"ohler \cite{Damkohler}, namely
the increase of $A_T$ and the enhancement of the local flame speed by
turbulence. Furthermore, both the criteria given by
eqs.~(\ref{e:Ucrit})-(\ref{e:KDCrit}) and the results of numerical
simulations show that effects of cusps become pronounced well within
the thin reaction zone regime and, in particular, at $Ka^{crit}
\gtrsim 20$ (Fig.~\ref{f:CuspDiagram}). Above this critical value of
$Ka$, increase of $S_T$ over $S_L$, on average, exceeds by $\gtrsim
10\%$ the corresponding increase of $A_T$. This shows that a
substantially accelerated turbulent burning can arise due to cusps at
much lower turbulent intensities than would be necessary to disrupt
the flame and increase its local speed by small-scale turbulence.

Above $Ka^{crit}$ (Fig.~\ref{f:CuspDiagram}), the flame evolution
likely remains substantially affected by cusp formation at all
turbulent intensities, as long as there is folding of the flame by
large-scale motions. This is the result of the fact that turbulence is
more efficient at packing the flame than at broadening it. In
particular, at high enough values of $U_l$, turbulence will eventually
break the internal flame structure, which will increase $S_n$. This
will, in turn, increase the right-hand side of eq.~(\ref{e:Ucrit}),
thereby, enhancing the stabilizing effect of cusps. On the other hand,
the flame width will also grow, increasing the critical flame
separation necessary for the onset of the regime influenced by cusp
formation. Since $S_n$ is determined by small-scale turbulence while
flame folding is governed by the faster large-scale motions, it is
unlikely that $S_n$ can grow fast enough to compensate for the
increase in both $l_F$ and the turbulent speeds which fold the
flame. Consequently, at higher $U_l$, i.e., even in the broken
reaction zone regime, cusps most likely continue to provide a
significant enhancement of $S_T$. This issue, however, requires
further investigation in future studies.

The effect of flame collisions on the turbulent flame speed suggests
that, even at high turbulent intensities, $S_T$ does not become
independent of the laminar flame speed. Analysis given in
\S~\ref{M:Criterion} and, in particular, eqs.~(\ref{e:Ucrit})-
(\ref{e:UlCrit}), showed that flame folding by turbulence, and thus
the efficiency of cusp formation, is controlled by $S_L$. For
instance, at the same $U_l$, the reactive mixture with higher $S_L$
will develop a less convolved flame with fewer cusps than the one with
lower $S_L$. Consequently, the exaggerated response of $S_T$ to the
increase of $A_T$ due to cusps will vary with $S_L$.

The results presented in this work show that for a large range of
turbulent intensities and system sizes, knowledge of $A_T$ is not
sufficient to predict the magnitude of $S_T$, even if the
reaction-zone structure remains unaffected by turbulence. For
instance, in the regime considered here, using $A_T$ as a guide would
result in errors as high as $30\%$. Such errors are quite substantial,
given that the flow evolution is typically very sensitive to the rate
of energy release. Therefore, it is particularly important to account
for the effects of cusps in subgrid-scale models that primarily focus
on determining the evolution of $A_T$ (e.g., \cite{Khokhlov}).
Furthermore, when constructing such modified subgrid-scale models,
flame propagation can no longer be viewed as a local process. In
particular, $S_n$ at each point of the flame surface is no longer
determined only by the local thermodynamic state of the flow or by
turbulent motions on scales $\lambda < \delta_L$. Instead, in the
presence of cusps, $S_n$ is also controlled by long-range velocity
correlations, which produce flame collisions and can span the full
size of the system.

The formation of cusps and the resulting rapid flame propagation in
certain regions are a crucial part of the turbulent flame-brush
evolution in the high-speed regime. Thus, properly capturing this
regime in numerical models requires the domain size to be larger than
the integral scale in order to accommodate the folding of the flame by
turbulence. Making the domain smaller than $l$ would significantly
hamper this process, while making the domain smaller than $\sim \!
\delta_L$ would completely eliminate it.

The increase of $S_n$ in cusps is discussed here for $Le = 1$. It is
well known, however, that when $Le \ne 1$, $S_n$ varies with the flame
curvature, even when it is $\ll \! 1/\delta_L$ \cite{Driscoll}.
Therefore, due to the large curvature in cusps, any imbalance between
thermal and diffusion fluxes can significantly exacerbate or suppress
the enhancement of $S_n$, depending on the value of $Le$. Therefore,
the role and properties of cusps in non-equidiffusive reactive
mixtures must be investigated in future studies. Furthermore, it is
important to extend the results presented in this work by considering
the effects of detailed chemistry, in particular, to address the role
of flame quenching and re-ignition at such high turbulent intensities
that are necessary for efficient cusp formation.

Finally, two possibilities exist for the flame evolution in the thin
reaction zone regime. If the turbulent intensity is not too high, an
equilibrium is established between flame-surface creation by
turbulence and its rapid destruction in cusps. This results in the
turbulent flame propagating in a steady state, which is the situation
observed here. On the other hand, if the turbulent intensity continues
to rise, eventually $\Sigma_T$ and $A_T$ can approach their limiting
values given by eqs.~(\ref{e:SigmaLim}) -(\ref{e:ALim}). In this case,
both $S_T/S_L$ and $S_T/A_T$ can become arbitrarily large, as was
discussed in \S~\ref{M:Regime}. Such singular behavior suggests that
in reality, unless turbulence substantially alters the flame
properties, the steady state must cease to exist and the system must
undergo a qualitative transformation in order to accommodate the rise
in $S_T$ or, equivalently, in the rate of energy release per unit
volume. Such qualitative change may indicate the transition from the
deflagration to a detonation. Detailed discussion of this nonsteady
regime will be presented in a separate paper.

\vspace{5mm}

\noindent{\bf Acknowledgments} We are deeply grateful to Forman
Williams for numerous valuable inputs that have greatly improved this
paper. We also thank Craig Wheeler, Vadim Gamezo, and James Driscoll
for stimulating discussions. This work was supported in part by the
Naval Research Laboratory, the Office of Naval Research, the Air Force
Office of Scientific Research under the grant F1ATA09114G005, and by
the National Science Foundation through TeraGrid resources provided by
NCSA and TACC under the grant TG-AST080006N. Additional computing
facilities were provided by the Department of Defense High Performance
Computing Modernization Program. We also gratefully acknowledge the
hospitality and support of the Nordic Institute for Theoretical
Physics, and in particular of Axel Brandenburg, during the ``Turbulent
Combustion'' program where parts of this work were done.

\clearpage
%--------------------------------------------------------------------------------
\begin{singlespace}

\begin{deluxetable}{ll}
\tablecaption{Nomenclature
\label{t:Nomen}}
\tabletypesize{\small}
\tablewidth{0pt}
\tableheadempty
\startdata
$A_Y \equiv A(Y)$                            &   surface area of the $Y$ isosurface                                       \\
$A_L$				             &   surface area of laminar flame  					  \\
$A_T \equiv A_{0.15}$		             &   surface area of turbulent flame					  \\
$B$                                          &   pre-exponential factor                                                   \\
$C_p$					     &   specific heat at constant pressure					  \\
$D$			 	             &   molecular diffusion coefficient (eq.~\ref{e:CondDiff})			  \\
$D_0$                                        &   molecular diffusion constant                                             \\
$D_T$				             &   turbulent diffusion coefficient (eq.~\ref{e:Damk1})			  \\
$Da$                                         &   Damk\"ohler number, $(l/U_l)/(l_F/S_{L,0})$ 				  \\
$I$				             &   stretch factor (eq.~\ref{e:Damk0})				          \\
$I_Y$				             &   ratio of $S_T/S_L$ to $A_Y/L^2$ (eq.~\ref{e:Damk})		          \\
$Ka$					     &   Karlovitz number, $(l_F/l)^{\sfrac{1}{2}}(U_l/S_L)^{\sfrac{3}{2}}$       \\
$Ka^{crit}$				     &   critical Karlovitz number for cusp formation (eq.~\ref{e:KDCrit})	  \\
$l$                                          &   integral scale                                                           \\
$l_c$					     &   characteristic cusp length						  \\
$l_F$                                        &   full width of laminar flame, $\approx\!2\delta_{L,0}$                    \\
$L$  		                             &   domain width, energy-injection scale                                     \\
$L_G$                                        &   Gibson scale, $l(S_{L,0}/U_l)^3$                                         \\
$Le$				             &   Lewis number, $\mathcal{K}/D$					          \\
$M$                                          &   molecular weight                                                         \\
$Ma_F$                                       &   Mach number in fuel, $U(\gamma P_0/\rho_0)^{-\sfrac{1}{2}}$              \\
$Ma_P$                                       &   Mach number in product, $U(\gamma P_0/\rho_{P,0})^{-\sfrac{1}{2}}$       \\
$n$                                          &   Temperature exponent                                                     \\
$P_0$                                        &   initial fuel pressure                                                    \\
$q$                                          &   chemical energy release                                                  \\
$Q$                                          &   activation energy                                                        \\
$S_L$                                        &   actual instantaneous laminar flame speed                                 \\
$S_{L,0}$                                    &   initial laminar flame speed                                              \\
$S_n$				             &   local flame burning speed					          \\
$S_n^*$					     &   maximum local burning speed in the cusp				  \\
$S_T$                                        &   turbulent flame speed                                                    \\
$t$					     &   time									  \\
$t_{ign}$                                    &   time of ignition                                                         \\
$t_{total}$                                  &   total simulation time from $t_{ign}$                                     \\
$T_0$                                        &   initial fuel temperature                                                 \\
$T_{P,0}$                                    &   initial post-flame temperature                                           \\
$\vect{u} = (u_x, u_y, u_z)$		     &	 flow velocity								  \\
$U$                                          &   turbulent velocity at scale $L$                                          \\
$U_c$					     &   speed of cusp propagation (eq.~\ref{e:CuspSpeed})			  \\
$U_l$                                        &   integral velocity                                                        \\
$U_l^{crit}$				     &	 critical value of $U_l$ for cusp formation (eq.~\ref{e:UlCrit})	  \\
$U_{rms}$                                    &   turbulent r.m.s. velocity                                                \\
$U_{\delta}$                                 &   turbulent velocity at scale $\delta_{L,0}$                               \\
$V_D$					     &	 diffusion velocity							  \\
$\vect{x} = (x,y,z)$			     &   spatial coordinate in the domain					  \\
$Y$				             &   fuel mass fraction						          \\
$\dot{Y}$			             &   reaction rate (eq.~\ref{e:ReacModel})				          \\
$z_{0,min}$				     &   left $z$-boundary of the flame brush (eq.~\ref{e:z0z1})		  \\
$z_{1,max}$				     &   right $z$-boundary of the flame brush (eq.~\ref{e:z0z1})		  \\
$z_{T,0}$                                    &   initial flame position along $z$-axis                                    \\
                                             &                                                                            \\
$\alpha$				     &   inclination angle between planar flame sheets				  \\
$\gamma$                                     &   adiabatic index                                                          \\
$\delta_L$                                   &   actual instantaneous thermal width of laminar flame                      \\
$\delta_{L,0}$                               &   initial thermal width of laminar flame                                   \\
$\delta_T$                                   &   turbulent flame width                                                    \\
$\Delta x$                                   &   cell size                                                                \\
$\Delta t_c$				     &   time lag of maximum cross-correlation					  \\
$\eta$				             &   Kolmogorov scale						          \\
$\varepsilon$                                &   energy-injection rate                                                    \\
$\kappa_0$                                   &   thermal conduction constant                                              \\
$\mathcal{K}$			             &   thermal conduction coefficient (eq.~\ref{e:CondDiff})		          \\
$\rho_0$                                     &   initial fuel density                                                     \\
$\rho_{P,0}$                                 &   initial post-flame density                                               \\
$\Sigma_Y \equiv \Sigma(Y)$                  &   surface density of the $Y$ isosurface (eq.~\ref{e:Sigma})                \\
$\Sigma_T \equiv \Sigma_{0.15}$              &   surface density of turbulent flame				          \\
$\tau_{ed}$                                  &   eddy turnover time, $L/U$                                                \\
                                             &                                                                            \\
$\overline{(\dots)}$                         &   time averaging
\enddata
\end{deluxetable}

%\clearpage
%--------------------------------------------------------------------------------

\begin{deluxetable}{cc}
\tablecaption{Input model parameters and resulting computed laminar flame
properties
\label{t:Params}}
\tabletypesize{\small}
\tablewidth{0pt}
\tableheadempty
\startdata
                 &   {\it Input}                                              \\
$T_0$            &   $293$ K                                                  \\
$P_0$            &   $1.01 \times 10^6$ erg/cm$^3$                            \\
$\rho_0$         &   $8.73 \times 10^{-4}$ g/cm$^3$                           \\
$\gamma$         &   $1.17$                                                   \\
$M$              &   $21$ g/mol                                               \\
$B$              &   $6.85 \times 10^{12}$ cm$^3$/(g s)                       \\
$Q$              &   $46.37$ RT$_0$                                           \\
$q$              &   $43.28$ RT$_0$ / M                                       \\
$\kappa_0$       &   $2.9 \times 10^{-5}$ g/(s cm K$^\textrm{n}$)             \\
$D_0$            &   $2.9 \times 10^{-5}$ g/(s cm K$^\textrm{n}$)             \\
$n$              &   $0.7$                                                    \\
                 &   {\it Output}                                             \\
$T_{P,0}$        &   $2135$ K                                                 \\
$\rho_{P,0}$     &   $1.2 \times 10^{-4}$ g/cm$^3$                            \\
$\delta_{L,0}$   &   $0.032$ cm                                               \\
$S_{L,0}$        &   $302$ cm/s
\enddata
\end{deluxetable}

%\clearpage
%--------------------------------------------------------------------------------

\begin{deluxetable}{lccc}
\tablecaption{Parameters of simulations\tablenotemark{a}
\label{t:Runs}}
\tabletypesize{\small}
\tablewidth{0pt}
\tablehead{
\colhead{}    &
\colhead{S1}  &
\colhead{S2}  &
\colhead{S3}}
\startdata
Mesh		          &       $64 \times 64 \times 1024$                                 &
		  	          $128 \times 128 \times 2048$                               &
		       	          $256 \times 256 \times 4096$                                   \\
$L$  		          &   &   $0.259$ cm $= 8 \delta_{L,0}$                              &   \\
$\Delta x$                &       $4.05\times10^{-3}$ cm                                     &
		                  $2.02\times10^{-3}$ cm                                     &
			          $1.01\times10^{-3}$ cm                                         \\
$\delta_{L,0}/\Delta x$   &       $8$                                                        &
		  	          $16$                                                       &
			          $32$                                                           \\
$z_{T,0}$                 &   &   $1.95$ cm $= 7.52 L$                                       &   \\
                          &   &                                                              &   \\
$\varepsilon$             &   &   $1.26\times10^9$ erg/(cm$^3$ s)                            &   \\
$U_{\delta}$              &   &   $4.53\times10^3$ cm/s $= 15 S_{L,0}$                       &   \\
$U$                       &   &   $9.07\times10^3$ cm/s $= 30 S_{L,0}$                       &   \\
$U_{rms}$                 &   &   $1.04\times10^4$ cm/s $= 34.48 S_{L,0}$                    &   \\
$U_l$                     &   &   $5.60\times10^3$ cm/s $= 18.54 S_{L,0}$                    &   \\
$l$                       &   &   $6.04\times10^{-2}$ cm $= 1.87 \delta_{L,0}$               &   \\
                          &   &                                                              &   \\
$\tau_{ed}$               &   &   $2.86\times10^{-5}$ s                                      &   \\
$t_{ign}$                 &       $3.0\tau_{ed}$                                             &
		                  $3.0\tau_{ed}$		                             &
		                  $2.0\tau_{ed}$		              	                 \\
$t_{total}$               &   &   $16.0\tau_{ed}$                                            &   \\
                          &   &                                                              &   \\
$Da$                      &   &   $0.05$                                                     &   \\
$L_G$                     &   &   $9.47\times10^{-6}$ cm $= 2.96\times10^{-4}\delta_{L,0}$   &   \\
$Ma_F$                    &   &   $0.25$                                                     &   \\
$Ma_P$                    &   &   $0.09$                                                     &
\tablenotetext{a}{ Parameters common to all simulations are shown only once in S2 column.}
\enddata
\end{deluxetable}

%\clearpage
%--------------------------------------------------------------------------------

\begin{deluxetable}{ccccccccccc}
\tablecaption{Time-averaged properties of the turbulent flame brush\tablenotemark{a}
\label{t:Props}}
\tabletypesize{\small}
\tablewidth{0pt}
\tablehead{
\colhead{}                                       &
\colhead{$\overline{\delta_T/\delta_L}$}         &
\colhead{$O(\overline{\delta_T/\delta_L})$}      &
\colhead{$\overline{S_T/S_L}$}                   &
\colhead{$O(\overline{S_T/S_L})$}                &
\colhead{$\overline{A}_{0.15}/L^2$}              &
\colhead{$O(\overline{A}_{0.15}/L^2)$}           &
\colhead{$\overline{\Sigma}_{0.15}$, mm$^{-1}$}  &
\colhead{$O(\overline{\Sigma}_{0.15})$}          &
\colhead{$\overline{I}_{0.15}$}                  &
\colhead{$O(\overline{I}_{0.15})$} }
\startdata
S1   &   20.03   &          &   5.19   &          &   3.91   &          &   0.74   &          &   1.31   &          \\
S2   &   17.39   &   2.21   &   3.91   &   2.19   &   3.23   &   2.84   &   0.67   &   4.07   &   1.20   &   1.50   \\
S3   &   16.66   &          &   3.55   &          &   3.12   &          &   0.67   &          &   1.14   &
\tablenotetext{a}{ Time-averaging for all variables is performed
over the time interval $[2\tau_{ed} - 16\tau_{ed}]$.}
\enddata
\end{deluxetable}

\end{singlespace}

\clearpage
%--------------------------------------------------------------------------------

%\clearpage
%--------------------------------------------------------------------------------

\begin{figure}[t]
\centering
\includegraphics[width=0.5\textwidth]{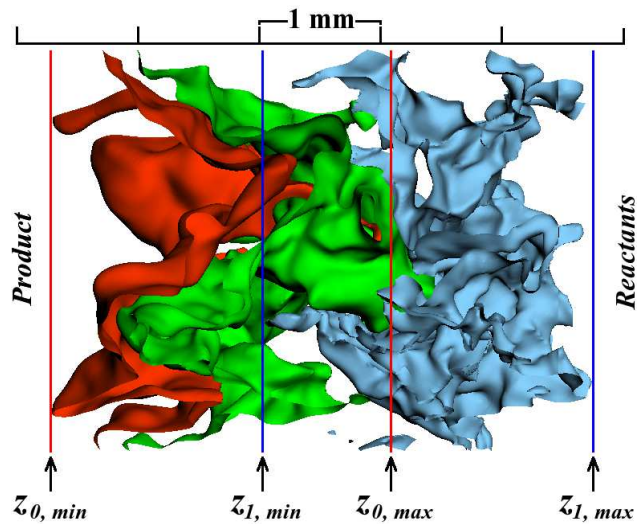}
\caption{Isosurfaces of $Y$ in simulation S2 at $t = 13\tau_{ed}$
(cf. Fig.3, middle row, left panel in \cite{Poludnenko}). Isosurface
values are 0.05 (red), 0.6 (green), 0.95 (blue). Red and green
isosurfaces bound the reaction zone of the flame. Green and blue
isosurfaces bound the preheat zone. The $z_{0,min}$ and $z_{1,max}$
mark the flame-brush bounds. The $z_{0,max}$ and $z_{1,min}$ indicate,
respectively, the maximum extents of product and fuel penetration into
the flame brush (see \ref{AppA} for further discussion of $z_{0,max}$
and $z_{1,min}$). (Reproduced from \cite{Poludnenko}.) }
\label{f:isoS}
\end{figure}

%\clearpage
%--------------------------------------------------------------------------------

\begin{figure}[t]
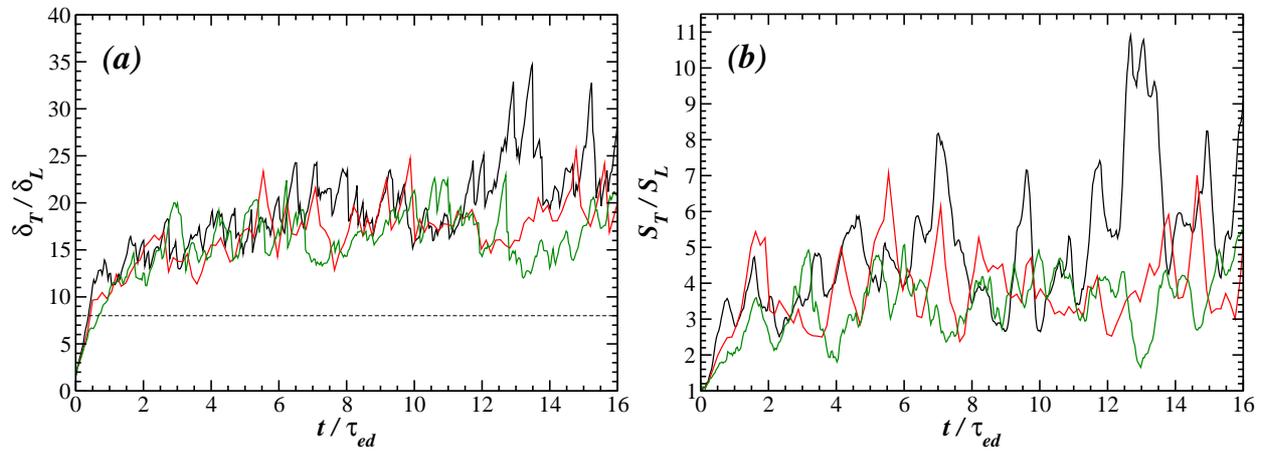

\centering 
\leavevmode
\includegraphics[clip, width=0.495\textwidth]{Lflame.v15.L8.fuel_heating.eps}
\includegraphics[clip, width=0.495\textwidth]{Vflame.v15.L8.fuel_heating.eps}
\caption{(a) Evolution of the normalized turbulent flame width
$\delta_T/\delta_L$. Note that domain width $L = 8 \delta_{L,0}
\approx 8\delta_L$ indicated with the horizontal dashed line. (b)
Evolution of the normalized turbulent flame speed $S_T/S_L$. In both
panels: black lines correspond to simulation S1, red to S2, and green
to S3. (cf. Fig.~4 in \cite{Poludnenko}.)}
\label{f:LV}
\end{figure}

%\clearpage
%--------------------------------------------------------------------------------

\begin{figure}[!t]
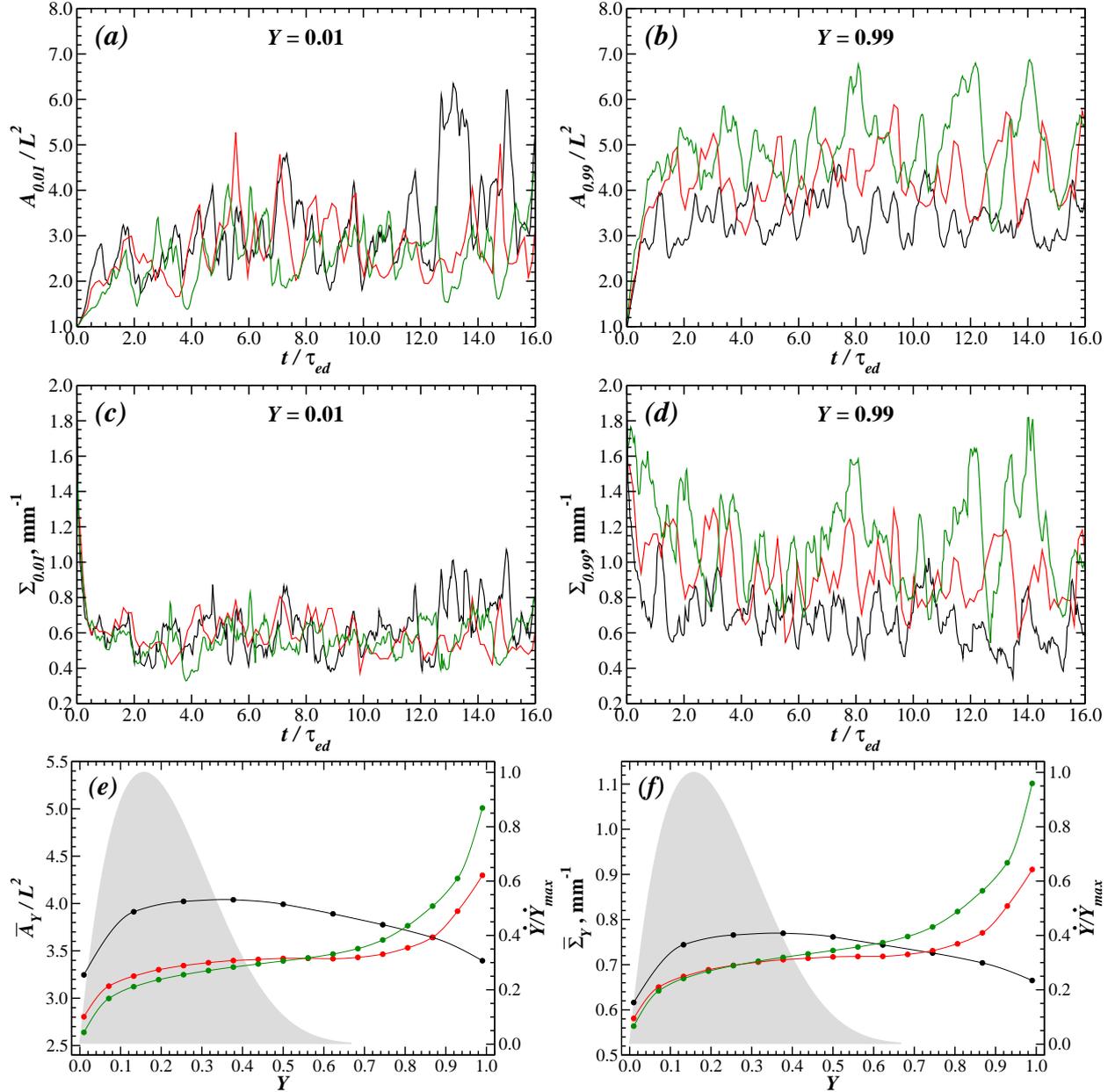

\centering 
\includegraphics[clip, width=.495\textwidth]{A_c0.01_vs_t.v15.L8.eps}
\includegraphics[clip, width=.495\textwidth]{A_c0.99_vs_t.v15.L8.eps}
\includegraphics[clip, width=.495\textwidth]{Sigma_c0.01_vs_t.v15.L8.eps}
\includegraphics[clip, width=.495\textwidth]{Sigma_c0.99_vs_t.v15.L8.eps}
\includegraphics[clip, width=.495\textwidth]{A_vs_c.v15.L8.eps}
\includegraphics[clip, width=.495\textwidth]{Sigma_vs_c.v15.L8.eps}
\caption{Evolution of the normalized flame surface area (a)
$A_{0.01}/L^2$ and (b) $A_{0.99}/L^2$. Evolution of the flame surface
density (c) $\Sigma_{0.01}$ and (d) $\Sigma_{0.99}$. (e) Time-averaged
normalized isosurface area $\overline{A}_Y/L^2$. (f) Time-averaged
isosurface density $\overline{\Sigma}_Y$. In panels (e) and (f) time
averaging is performed over the time interval
$[2\tau_{ed}-16\tau_{ed}]$, circles represent calculated values and
solid lines are the Akima spline fits. Shaded gray region shows the
distribution of the reaction rate, $\dot{Y}$, in the exact laminar
flame solution normalized by its peak value $\dot{Y}_{max} = 9.5
\times 10^4$ s$^{-1}$. In all panels: black corresponds to simulation
S1, red to S2, and green to S3.}
\label{f:ASigma}
\end{figure}

%\clearpage
%--------------------------------------------------------------------------------

\begin{figure}[!t]
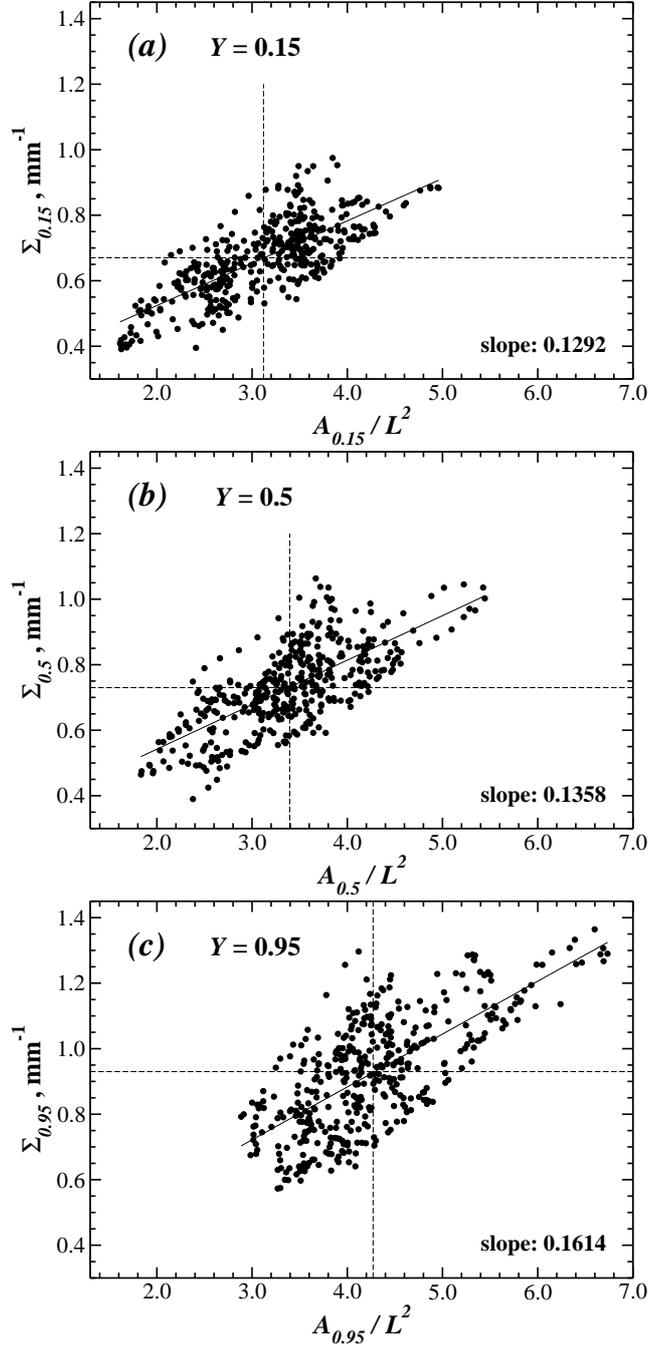

\centering 
\includegraphics[clip, width=.51\textwidth]{A_0.15_vs_Sigma_0.15.v15.L8.fuel_heating.eps}
\includegraphics[clip, width=.51\textwidth]{A_0.5_vs_Sigma_0.5.v15.L8.fuel_heating.eps}
\includegraphics[clip, width=.51\textwidth]{A_0.95_vs_Sigma_0.95.v15.L8.fuel_heating.eps}
\caption{Correlation between $\Sigma_Y$ and $A_Y/L^2$ in simulation S3
for three values of $Y$. Graphs exclude data from the time interval
$[0 - 2\tau_{ed}]$ during which the turbulent flame develops its
equilibrium state. Dashed lines show time-averaged values of
$\Sigma_Y$ and $A_Y/L^2$. Solid lines show the least squares fit with
its slope given in each panel.}
\label{f:CorrelASigma}
\end{figure}

%\clearpage
%--------------------------------------------------------------------------------

\begin{figure}[!t]
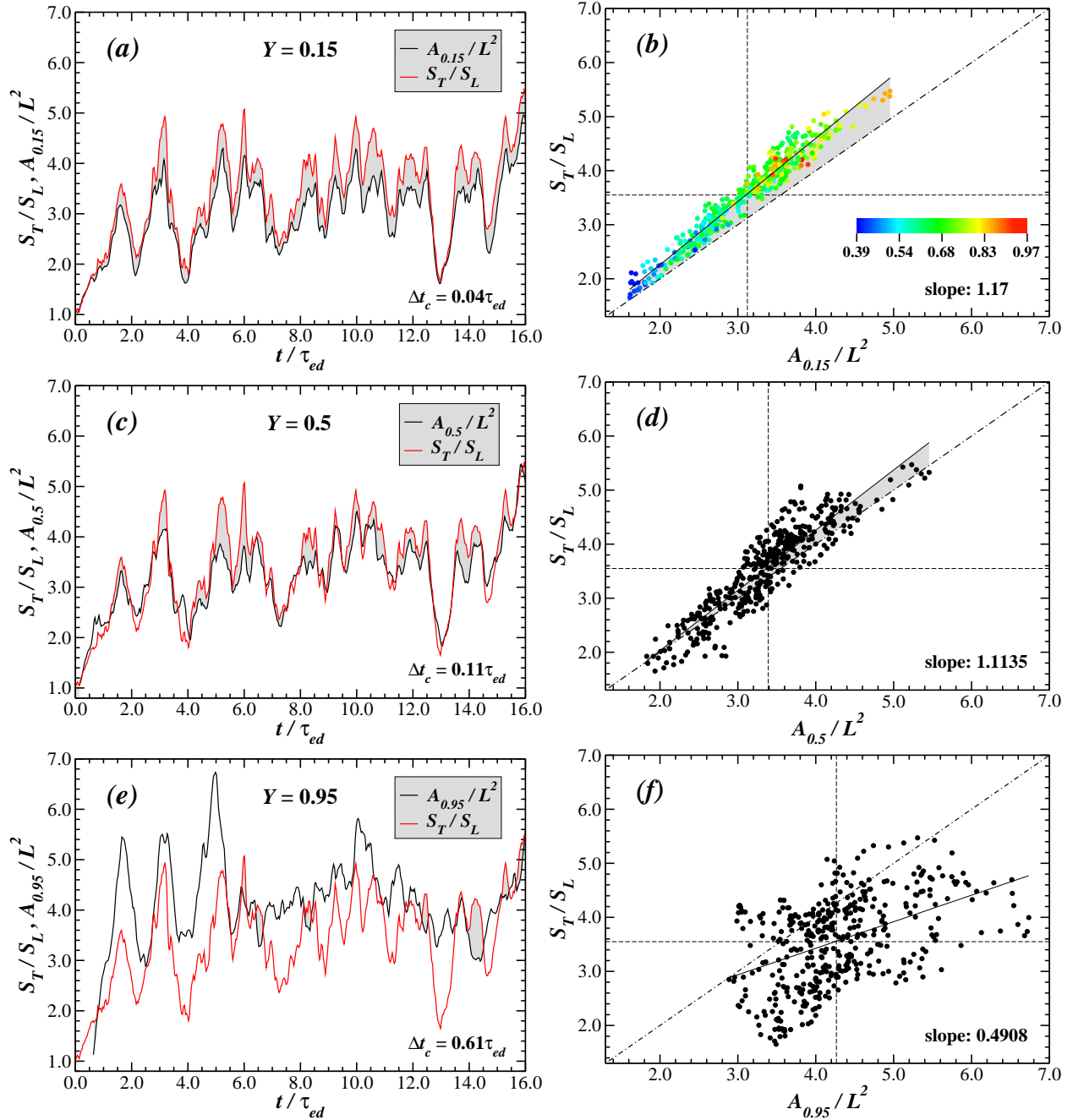

\centering 
\includegraphics[clip, width=.505\textwidth]{A_0.15_Vflame_vs_t.v15.L8.fuel_heating.eps}
\includegraphics[clip, width=.485\textwidth]{A_0.15_vs_Vflame.v15.L8.fuel_heating.eps}
\includegraphics[clip, width=.505\textwidth]{A_0.5_Vflame_vs_t.v15.L8.fuel_heating.eps}
\includegraphics[clip, width=.485\textwidth]{A_0.5_vs_Vflame.v15.L8.fuel_heating.eps}
\includegraphics[clip, width=.505\textwidth]{A_0.95_Vflame_vs_t.v15.L8.fuel_heating.eps}
\includegraphics[clip, width=.485\textwidth]{A_0.95_vs_Vflame.v15.L8.fuel_heating.eps}
\caption{Correlation between $A_Y/L^2$ and $S_T/S_L$ in simulation S3.
$A_Y$ is considered for three values of $Y$: (a, b) $Y=0.15$, (c, d)
$Y=0.5$, and (e, f) $Y=0.95$. In panels (a), (c), and (e) $A_Y/L^2$ is
shifted in time to the right by the time lag $\Delta t_c$ shown in
each panel. Graphs in panels (b), (d), and (f) are constructed based
on the data shown, respectively, in panels (a), (c), and (e) excluding
the time $[0 - 2\tau_{ed}]$ during which the turbulent flame develops
its equilibrium state. In panel (b) each data point is colored
according to the corresponding value of $\Sigma_{0.15}$ with the color
scale given in units $mm^{-1}$ (cf. Fig.~\ref{f:CorrelASigma}a). In
panels (b), (d), and (f) dashed lines show time-averaged values of
$A_Y/L^2$ and $S_T/S_L$, solid lines show the least squares fit with
its slope given in each panel, and dash-dot lines correspond to
$S_T/S_L = A_Y/L^2$. Shaded gray regions in panels (a)-(e) illustrate
the exaggerated response of $S_T$ to the increase of $A_Y$.}
\label{f:CorrelAS}
\end{figure}

%\clearpage
%--------------------------------------------------------------------------------

\begin{figure}[!t]
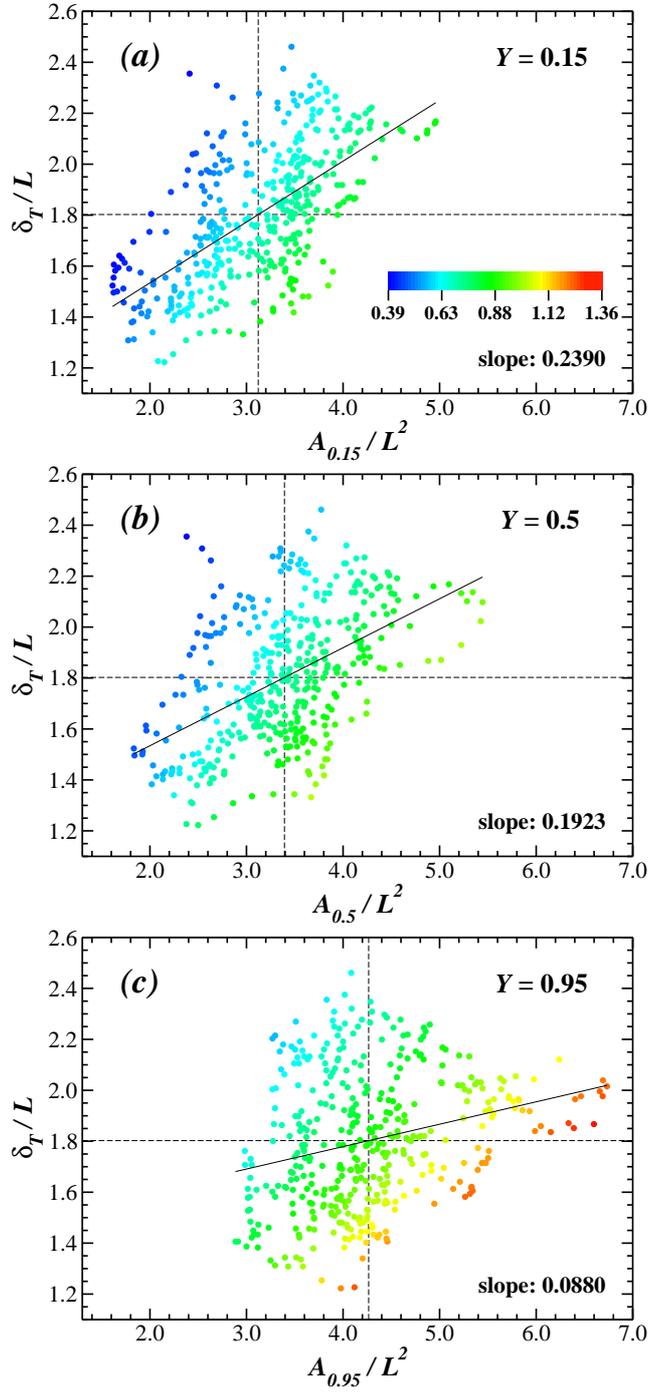

\centering 
\includegraphics[clip, width=.51\textwidth]{A_0.15_vs_Delta.v15.L8.eps}
\includegraphics[clip, width=.51\textwidth]{A_0.5_vs_Delta.v15.L8.eps}
\includegraphics[clip, width=.51\textwidth]{A_0.95_vs_Delta.v15.L8.eps}
\caption{Correlation between $A_Y/L^2$ and $\delta_T/L$ in simulation
S3 for three values of $Y$. Graphs exclude data from the time interval
$[0 - 2\tau_{ed}]$ during which the turbulent flame develops its
equilibrium state. Dashed lines show time-averaged values of $A_Y/L^2$
and $\delta_T/L$. Solid lines show the least squares fit with its
slope given in each panel. Each data point is colored according to the
corresponding value of $\Sigma_Y$ with the common color scale in units
$mm^{-1}$ shown in panel (a) (cf. Fig.~\ref{f:CorrelASigma}).}
\label{f:CorrelADelta}
\end{figure}

%\clearpage
%--------------------------------------------------------------------------------

\begin{figure}[t]
\centering
\includegraphics[clip, width=0.5\textwidth]{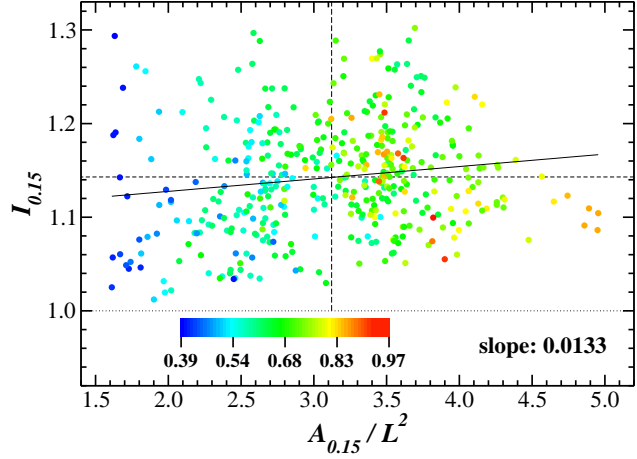}
\caption{Instantaneous values of $I_{0.15}$ as a function of
$A_{0.15}/L^2$ in simulation S3. Graph is constructed from data shown
in Fig.~\ref{f:CorrelAS}b. Each data point is colored according to the
corresponding value of $\Sigma_{0.15}$ with the color scale given in
units $mm^{-1}$. Dashed lines show time-averaged values. The solid
line gives the least squares fit with its slope given in the panel.}
\label{f:CorrelIA}
\end{figure}

%\clearpage
%--------------------------------------------------------------------------------

\begin{figure}[t]
\centering
\includegraphics[clip, width=0.5\textwidth]{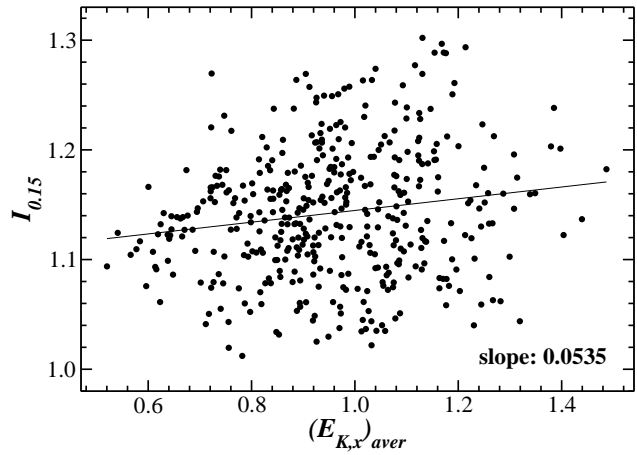}
\caption{Instantaneous values of $I_{0.15}$ in simulation S3 as a
function of the volume-averaged normalized $x$-component of the
kinetic energy inside the flame brush and in the fuel immediately
ahead of it. The solid line shows the least squares fit with its slope
given in the panel. See text for further details.}
\label{f:CorrelIEkin}
\end{figure}

%\clearpage
%--------------------------------------------------------------------------------

\begin{figure}[t]
\centering
\includegraphics[clip, width=0.39\textwidth]{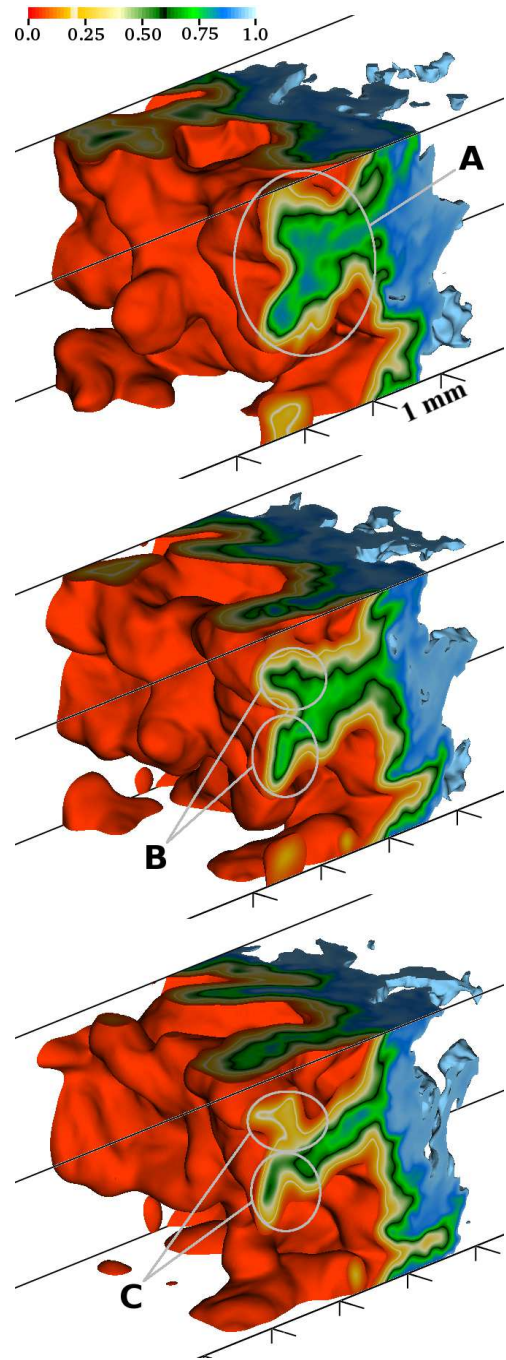}
\caption{Flame collision and the formation of cusps in the turbulent
flame. Shown is the flame-brush structure based on the isovolume of
$Y$ in simulation S3. Bounding isosurfaces represent $Y = 0.05$ and $Y
= 0.95$ and the flame brush is shown from the product side. Upper
panel corresponds to the time $t = 11.86\tau_{ed}$, while the middle
and lower panels show the flame structure, respectively, $0.1\tau_{ed}
=$ 3 ${\mu}$s and $0.2\tau_{ed} =$ 6 ${\mu}$s later. The thin black
line, corresponding to $Y = 0.6$, marks the boundary between the
preheat and reaction zones. The thin white line, corresponding to $Y =
0.2$, shows the location of the peak reaction rate. Regions A, B, and
C show the three main stages of the flame collision and the formation
of a cusp discussed in \S~\ref{M:Collisions}. Note also two elongated
regions of flame collision forming near the upper face of the domain.}
\label{f:isoV}
\end{figure}

\clearpage
%--------------------------------------------------------------------------------

\begin{figure}[t]
\centering
\includegraphics[clip, width=1.0\textwidth]{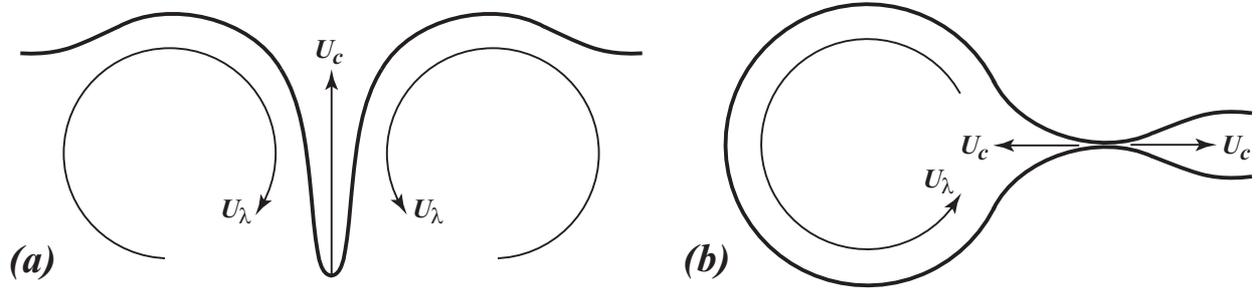}
\caption{Schematic of the two main channels of cusp formation. (a)
{\it Local flame collisions.} Turbulent motions with velocity
$U_{\lambda}$ on the scale $\lambda$ fold the flame, gradually
increasing its surface curvature at a specific point until it becomes
$\sim\!1/\delta_L$. (b) {\it Nonlocal flame collisions}. The flame is
folded by turbulent motions on a larger scale. The curvature of the
flame surface remains small at all points until the collision of the
two flame sheets causes an abrupt formation of two cusps.  Resulting
cusps propagate in both cases with the phase velocity $U_c$, which
depends on the local flame curvature.  See text for further details.}
\label{f:CollTypes}
\end{figure}

%\clearpage
%--------------------------------------------------------------------------------

\begin{figure}[t]
\centering
\includegraphics[clip, width=0.8\textwidth]{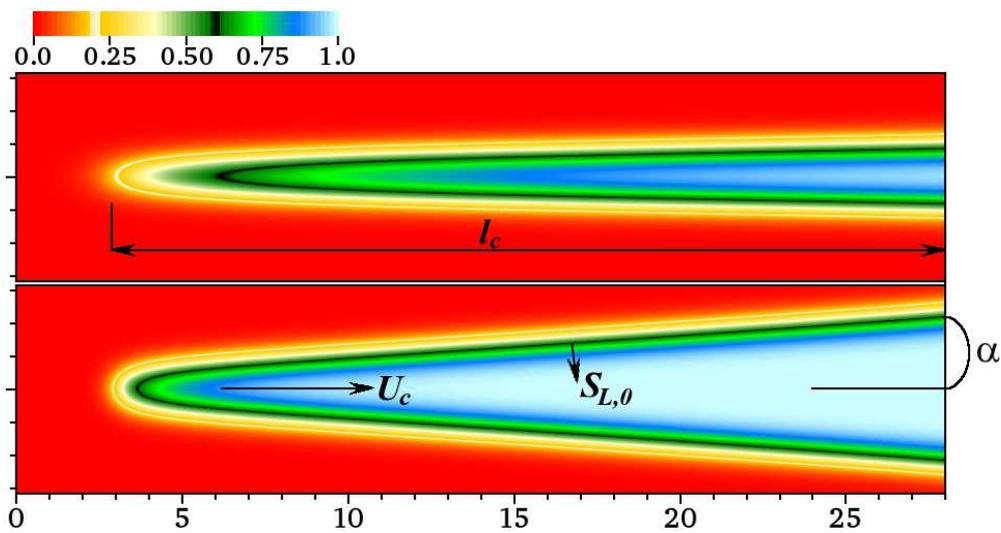}
\caption{Structure of a cusp formed by the collision of two planar
flame sheets. Shown is the distribution of $Y$ for the flame
inclination angle $\alpha = 1^{\circ}$ (upper panel) and $\alpha =
4^{\circ}$ (lower panel). Scale of the panel axes is given in units of
$\delta_{L,0}$. Away from the cusp tip, the flame propagates in the
direction normal to its surface with the laminar flame speed,
$S_{L,0}$, causing the tip to move to the right with the speed
$U_c$. Also shown is the variable length, $l_c$, of the collision
region (see text). Thin black line marks the boundary between the
reaction and preheat zones, while the thin white line indicates the
region of peak reaction rate. Note the substantial broadening of the
flame, and thus the reaction zone, near the tip of the cusp at $\alpha
= 1^{\circ}$ compared to $\alpha = 4^{\circ}$ (cf. Fig.~\ref{f:isoV}).}
\label{f:Cusp}
\end{figure}

%\clearpage
%--------------------------------------------------------------------------------

\begin{figure}[t]
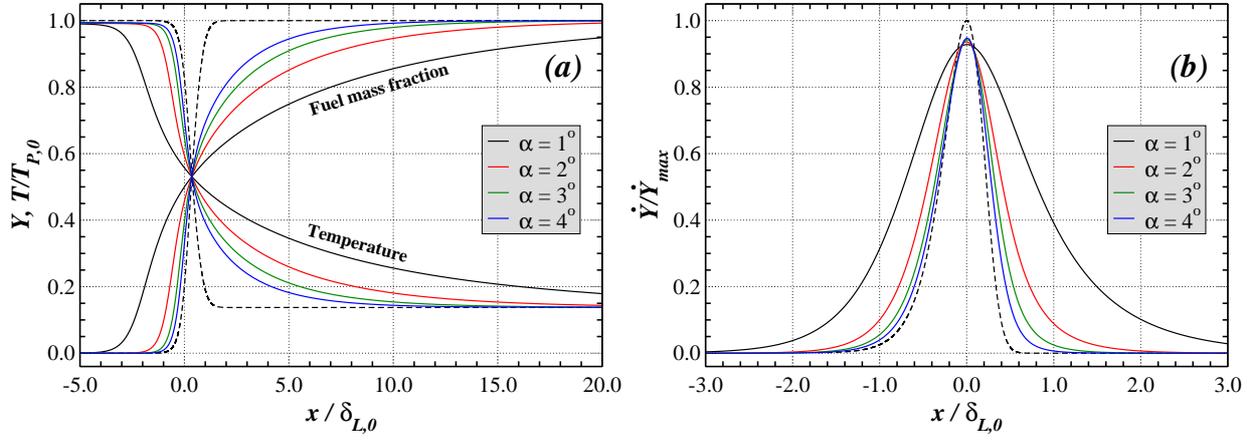

\centering
\includegraphics[clip, width=0.49\textwidth]{Cusp_Structure.eps}
\includegraphics[clip, width=0.49\textwidth]{Cusp_Reaction_Rate.eps}
\caption{Structure of a cusp formed by the collision of two planar
flame sheets for four flame inclination angles, $\alpha$. Shown are
distributions of (a) $Y$ and $T$, as well as (b) the reaction rate,
$\dot{Y}$, along the symmetry line of the cusp
(cf. Fig.~\ref{f:Cusp}). Dashed lines in both panels indicate the
exact planar laminar flame solution. Profiles of $T$ and $\dot{Y}$ are
normalized by their respective maximum values in the exact laminar
solution, i.e., $T_{P,0}$ (see Table~\ref{t:Params}) and
$\dot{Y}_{max} = 9.5 \times 10^4$ s$^{-1}$.}
\label{f:CuspStruct}
\end{figure}

%\clearpage
%--------------------------------------------------------------------------------

\begin{figure}[t]
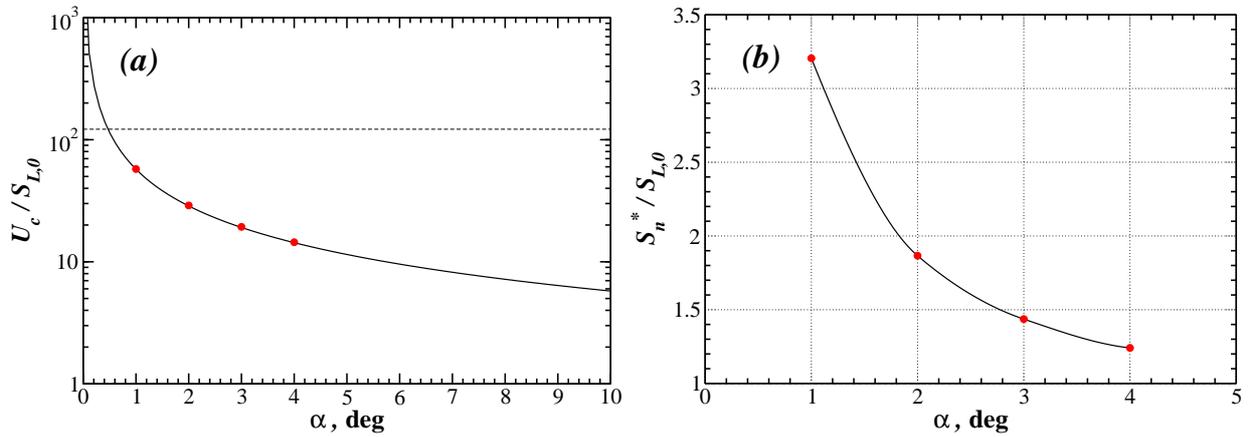

\centering
\includegraphics[clip, width=0.49\textwidth]{Cusp_Speed.eps}
\includegraphics[clip, width=0.49\textwidth]{Cusp_Burning_Speed.eps}
\caption{(a) Dependence of the normalized speed of cusp propagation,
$U_c / S_{L,0}$, on the flame inclination angle, $\alpha$. Solid line
corresponds to the analytic expression given in
eq.~(\ref{e:CuspSpeed}), red circles show the computed values.
Horizontal dashed line indicates sound speed in cold fuel. (b)
Dependence of the maximum normalized local flame speed in the cusp,
$S^*_n / S_{L,0}$, given by eq.~(\ref{e:CuspBurn}), on $\alpha$. Red
circles show the computed values, solid line is the Akima spline fit.}
\label{f:CuspSpeed}
\end{figure}

%\clearpage
%--------------------------------------------------------------------------------

\begin{figure}[t]
\centering
\includegraphics[clip, width=0.5\textwidth]{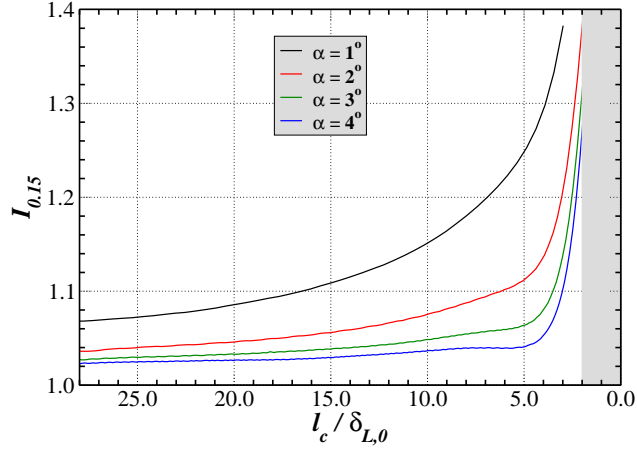}
\caption{Dependence of $I_{0.15}$, given by eq.~(\ref{e:DamkCusp}), on the
normalized length of the flame-collision region, $l_c/\delta_{L,0}$,
for four values of $\alpha$. The shaded gray area indicates the full
laminar flame width $l_F \approx 2\delta_{L,0}$. See text for further
details.}
\label{f:DamkCusp}
\end{figure}

%\clearpage
%--------------------------------------------------------------------------------

\begin{figure}[t]
\centering
\includegraphics[clip, width=0.6\textwidth]{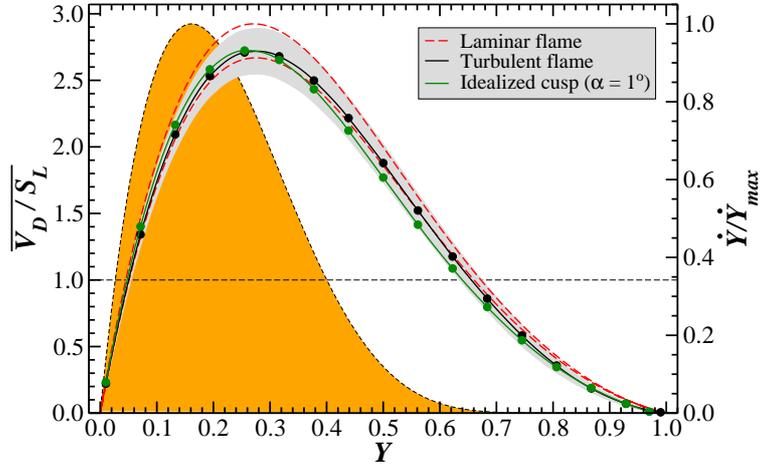}
\caption{Time-averaged distributions of the normalized diffusion
velocity in simulation S3 (black line) and in the idealized 2D cusp at
$\alpha = 1^{\circ}$ shown in the upper panel of Fig.~\ref{f:Cusp}
(green line). Circles are calculated values, solid lines are the Akima
spline fits. The shaded gray region is the $1\sigma$ standard
deviation of the instantaneous values of $V_D/S_L$ in S3. Dashed red
lines are the exact laminar flame solutions corresponding to fuel
temperatures $T_0 = 293$ K (upper line) and $320$ K (lower line). The
shaded orange region shows the distribution of the reaction rate,
$\dot{Y}$, in the exact laminar flame solution (fuel temperature
$T_0$) normalized by its peak value $\dot{Y}_{max} = 9.5 \times 10^4$
s$^{-1}$. See text for further details.}
\label{f:DiffVel}
\end{figure}

%\clearpage
%--------------------------------------------------------------------------------

\begin{figure}[t]
\centering
\includegraphics[clip, width=0.5\textwidth]{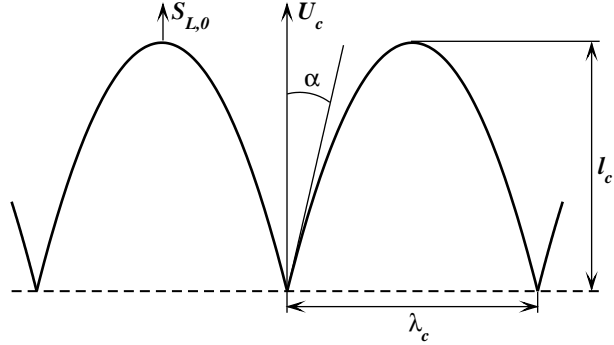}
\caption{Illustration of the idealized perturbed flame stabilized by the
propagation of a cusp with the speed $U_c$
(cf. Figs.~\ref{f:CollTypes}a and \ref{f:Cusp}).}
\label{f:CuspSchematic}
\end{figure}

%\clearpage
%--------------------------------------------------------------------------------

\begin{figure}[t]
\centering
\includegraphics[clip, width=0.5\textwidth]{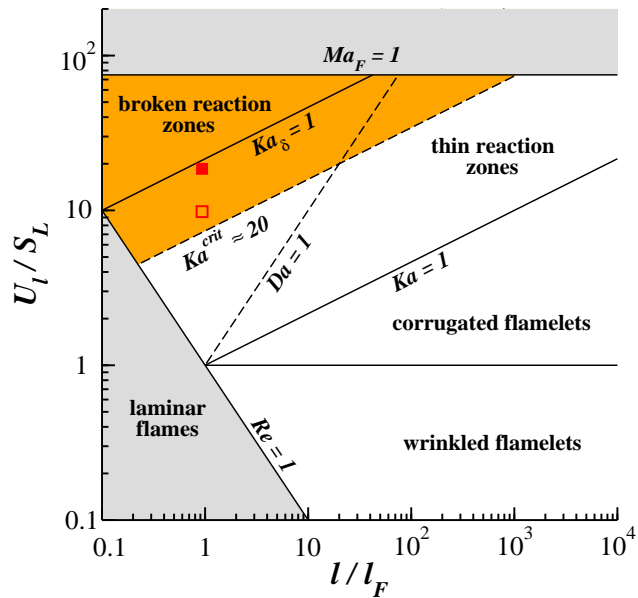}
\caption{Combustion regime diagram according to \cite{Peters}. The orange
region above the $Ka^{crit} \approx 20$ line shows, according to the
criterion discussed in \S~\ref{M:Criterion}, the range of the regimes
in which the formation of cusps is expected to substantially affect
the turbulent flame speed. The solid red square corresponds to the
simulation S3 presented in this work that gives $\overline{I}_{0.15}
\approx 1.14$. The open red square shows the regime with one half
the turbulent intensity in which the value of $\overline{I}_{0.15}
\approx 1.1$ was determined. Flamelets are typically believed to
exist in the regimes below the $Ka_{\delta} = Ka/100 = 1$ line
\cite{Peters}. The traditional form of the diagram was also modified
by adding the line $Ma_F = 1$ indicating the region of supersonic
turbulence in the cold H$_2$-air fuel under the atmospheric
conditions. See text for further details.}
\label{f:CuspDiagram}
\end{figure}

%\clearpage
%--------------------------------------------------------------------------------

\clearpage

\appendix
%-------------------------------------------------------------------------------
\section{Definition of the flame surface density}
%-------------------------------------------------------------------------------
\label{AppA}

In the definition of $\Sigma_Y$, given in eq.~(\ref{e:Sigma}), the
isosurface area, $A_Y$, was normalized by the total volume of the
flame brush, $\delta_T L^2$. Individual isosurfaces, however, do not
occupy all of this volume. For instance, this is the case for
isosurfaces of $Y = 0.05$ and $Y = 0.95$, which are used to specify
$\delta_T$ in eqs.~(\ref{e:deltaT})-(\ref{e:z0z1}), simply because, by
definition, they are separated at least by the full flame width,
$l_F$, or a larger distance if the flame becomes broadened. In our
case, $l_F \approx 2\delta_L$ (see Fig.~7 in \cite{Poludnenko}) is
a substantial fraction of $\delta_T$. Consider Fig.~\ref{f:isoS}. It
can be seen that neither the $Y = 0.05$, nor the $Y = 0.95$ isosurface
extend over the full width of the flame brush, and they indeed occupy
a smaller volume.

In order to consider this more quantitatively, we define $z_{0,max}$
and $z_{1,min}$ by analogy with $z_{0,min}$ and $z_{1,max}$
(eq.~\ref{e:z0z1}),
\beq
\begin{array}{l}
z_{0,max} = \min(z): Y(x,y,z) > 0.05 \ \forall \ (x,y,z > z_{0,max}), \\
z_{1,min} = \max(z): Y(x,y,z) < 0.95 \ \forall \ (x,y,z < z_{1,min}).
\end{array}
\label{e:z}
\eeq
In other words, $z_{0,max}$ is the $z$-coordinate of the rightmost
cell with pure product, while $z_{1,min}$, respectively, is the
$z$-coordinate of the leftmost cell with pure fuel. Thereby, they
effectively measure the furthest extent of product and fuel
penetration into the flame brush. Given the definitions of $z_{0,min}$
and $z_{1,max}$ (eq.~\ref{e:z0z1}), $z_{0,min}$ and $z_{0,max}$ bound
the volume confining the $Y = 0.05$ isosurface, while $z_{1,min}$ and
$z_{1,max}$ bound the $Y = 0.95$ isosurface. If $z_{1,min} <
z_{0,max}$, these two variables bound the region of macroscopic mixing
of product and fuel in the flame brush, \ie the region in which both
pure fuel and pure product can be found. On the other hand, values
$z_{0,max} \leq z_{1,min}$ exist only in a weakly wrinkled (nearly
planar) flame. Fig.~\ref{f:isoS} provides the illustration of all four
of these quantities.

Next, we define the position of the turbulent flame brush as
\beq
z_T = \frac{z_{0,min} + z_{1,max}}{2}.
\label{e:zT}
\eeq
Modified $\tilde{z}_{0,max}$ and $\tilde{z}_{1,min}$ are then defined
as offsets of $z_{0,max}$ and $z_{1,min}$ with respect to $z_T$
normalized by $\delta_T$, namely
\beq
\tilde{z}_{0,max} = \frac{z_{0,max} - z_T}{\delta_T}, \quad
\tilde{z}_{1,min} = \frac{z_{1,min} - z_T}{\delta_T}.
\label{e:zmod}
\eeq
Thus, $\tilde{z}_{0,max}$ and $\tilde{z}_{1,min}$ are the relative
measure of the extent to which pure fuel and pure product penetrate
into the flame brush. It follows from the definition (\ref{e:zmod}),
that both quantities take on values in the interval $[-0.5,0.5]$.  For
example, in a planar laminar flame, they are constant with
$\tilde{z}_{0,max} = -0.5$, $\tilde{z}_{1,min} = 0.5$. In the
turbulent flame brush, $\tilde{z}_{0,max} = \tilde{z}_{1,min} = 0$
would indicate that both fuel and product reach the midpoint of the
flame brush, they are confined to the left and right halves of the
brush, and there is no macroscopic mixing of fuel and product. This
would correspond to a weakly wrinkled flame. If $\tilde{z}_{0,max}
\approx 0.5$ and $\tilde{z}_{1,min} \approx -0.5$, then both pure
product and pure fuel can be found throughout the entire volume of the
flame brush.

\begin{figure}[!t]
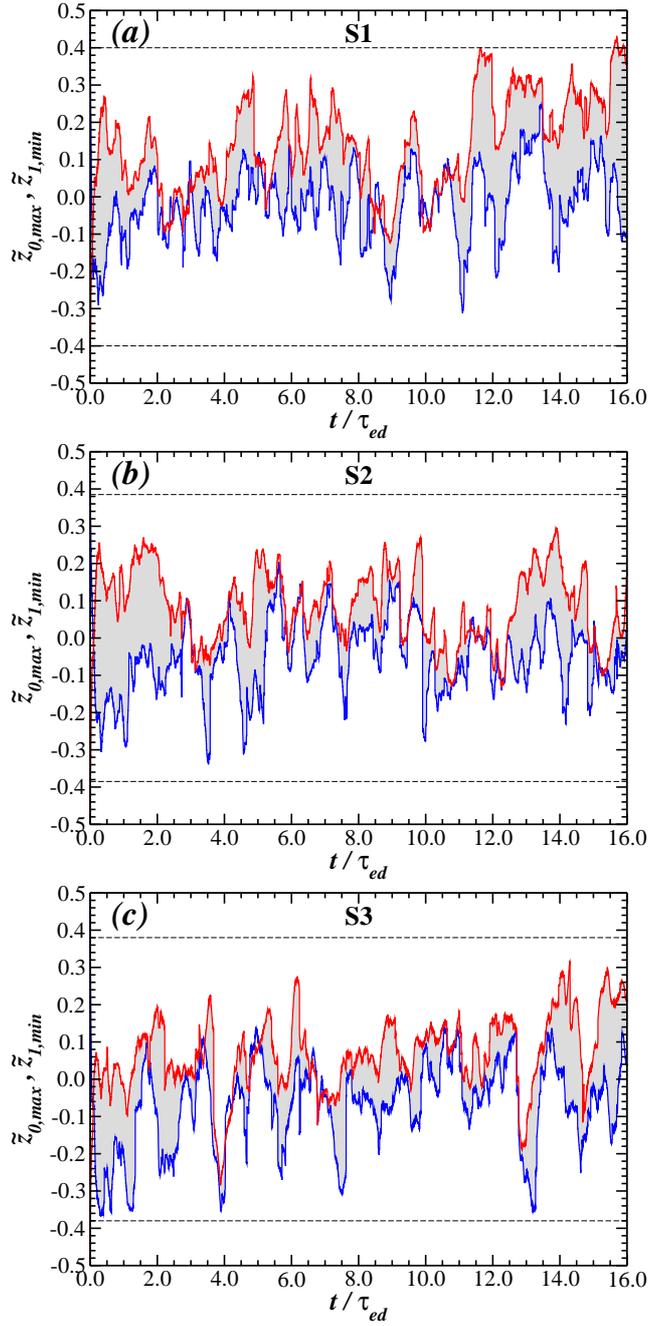

\centering
\includegraphics[clip, width=0.51\textwidth]{Extents.K64.v15.L8.eps}
\includegraphics[clip, width=0.51\textwidth]{Extents.K128.v15.L8.eps}
\includegraphics[clip, width=0.51\textwidth]{Extents.K256.v15.L8.eps}
\caption{Evolution of the normalized maximum extents of product and
fuel penetration into the flame brush $\tilde{z}_{0,max}$ and
$\tilde{z}_{1,min}$ (see \ref{AppA} for the definition and
Fig.~\ref{f:isoS} for the illustration) for simulations S1 (a), S2
(b), and S3 (c). The $\tilde{z}_{0,max}$ is shown with red line,
$\tilde{z}_{1,min}$ -- with blue line. Shaded regions mark the extent
of macroscopic mixing of pure fuel and product inside the flame
brush. Horizontal dashed lines show the average limiting values of
$\tilde{z}_{0,max}$ and $\tilde{z}_{1,min}$ given by the condition
(\ref{e:zmodCond}) and calculated using the values of
$\overline{\delta_T/\delta_L}$ listed in Table~\ref{t:Props}.}
\label{f:Extents}
\end{figure}

Fig.~\ref{f:Extents} shows the evolution of $\tilde{z}_{0,max}$ and
$\tilde{z}_{1,min}$ for simulations S1-S3. In all cases, both
parameters exhibit fairly similar behavior oscillating around zero.
As the system evolves, it undergoes recurring transitions between
periods of enhanced fuel-product mixing and episodes of their near
complete separation when the turbulent flame becomes more planar. The
correlation between these quantities and $S_T$ is less prominent
compared, for instance, with that between $A_Y$ and $S_T$, although it
is possible to associate some peaks and troughs with the corresponding
changes in $S_T$.

It follows from Fig.~\ref{f:Extents} that throughout the course of the
simulation, both the $Y = 0.05$ and $Y = 0.95$ isosurfaces indeed
occupy regions smaller than the full flame-brush volume. This is also
the case for other isosurfaces (cf. Fig.~\ref{f:isoS}).  Consequently,
an argument can be made that when calculating $\Sigma_Y$,
normalization should be performed not over the full volume of the
flame brush, but rather over the respective volume bounding a given
isosurface. For instance, for $Y = 0.05$, this volume is $(z_{0,max} -
z_{0,min})L^2 = (\tilde{z}_{0,max} + 0.5)\delta_T L^2$, which would
introduce an additional factor $1/(\tilde{z}_{0,max} + 0.5)$ in
eq.~(\ref{e:Sigma}). Such definition of $\Sigma_Y$ would be a more
accurate measure of how tightly a given isosurface is folded.
Fig.~\ref{f:Extents} shows that, on average, $\tilde{z}_{0,max}
\approx 0$ and, thus, the $Y = 0.05$ isosurface is confined to about
half of the total volume of the flame brush. Therefore, such modified
definition would result in approximately twice higher values of
$\Sigma_Y$. Ultimately, however, we are interested in determining how
the overall flame, rather than individual isosurfaces, is folded
inside the flame brush since only the flame as a whole has actual
physical significance. Consequently, the uniform normalization over
the total volume occupied by the flame appears to be a more physically
grounded choice.

Further motivation for our choice of normalization in the definition
(\ref{e:Sigma}) of $\Sigma_Y$ comes when we consider how
$\tilde{z}_{0,max}$ and $\tilde{z}_{1,min}$ would change for larger
system sizes. Since the $Y = 0.05$ and $Y = 0.95$ isosurfaces are
always separated at least by the full width of the flame, $l_F$, then
\beq
z_{1,max} - z_{0,max} \gtrsim l_F, \quad
z_{1,min} - z_{0,min} \gtrsim l_F.
\label{e:ineq}
\eeq
Using eqs.~(\ref{e:deltaT}), (\ref{e:zT}), and (\ref{e:zmod}), and the
fact that in our system $l_F \approx \! 2\delta_L$, inequalities
(\ref{e:ineq}) can be transformed into the following conditions which
must be satisfied at each moment in time,
\beq
\tilde{z}_{0,max} \lesssim 0.5 - \frac{2\delta_L}{\delta_T}, \quad
\tilde{z}_{1,min} \gtrsim -0.5 + \frac{2\delta_L}{\delta_T}.
\label{e:zmodCond}
\eeq
Horizontal dashed lines in Fig.~\ref{f:Extents} show the average
limiting $\tilde{z}_{0,max}$ and $\tilde{z}_{1,min}$ based on the
values of $\overline{\delta_T/\delta_L}$ listed in Table~\ref{t:Props}
for each simulation. In particular, $\tilde{z}_{0,max}$ must be
$\lesssim \! 0.4$ while $\tilde{z}_{1,min}$ must be $\gtrsim \!
-0.4$, which closely agrees with the data shown in
Fig.~\ref{f:Extents}.

It was discussed in \S~\ref{M:Regime} that $\delta_T$ increases with
the turbulent integral scale, or, equivalently, with the system size
and energy injection scale \cite{Poludnenko,Peters}. It then follows
from eq.~(\ref{e:zmodCond}) that as $\delta_T/\delta_L \to \infty$,
then limiting values of $\tilde{z}_{0,max} \to 0.5$ and
$\tilde{z}_{1,min} \to -0.5$.  Therefore, in larger systems, as
$\delta_T$ becomes large in comparison with $\delta_L$, the volume
bounding each isosurface becomes well approximated by the total volume
of the flame brush.  Consequently, in the limit of large values of
$\delta_T$, the uniform normalization for all values of $Y$ in
eq.~(\ref{e:Sigma}) becomes equivalent to the normalization by the
actual volume bounding a given isosurface.

%-------------------------------------------------------------------------------
\section{Induction time for the planar laminar flame}
%-------------------------------------------------------------------------------
\label{AppB}

\begin{figure}[!t]
\centering
\includegraphics[clip, width=0.6\textwidth]{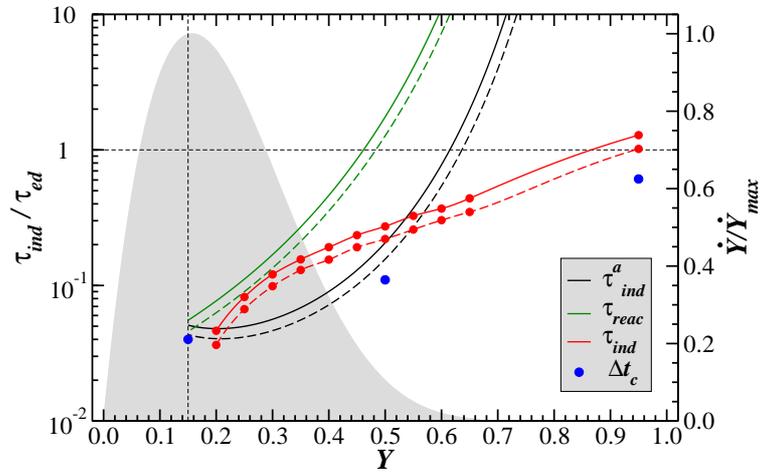}
\caption{Distributions of the actual induction times, $\tau_{ind}$, adiabatic
induction times, $\tau^a_{ind}$, and the characteristic reaction
times, $\tau_{reac}$, in the planar laminar flame. Also are shown the
correlation time lags $\Delta t_c$ (cf. Fig.~\ref{f:CorrelAS}). Red
circles show the calculated values of $\tau_{ind}$, red lines are the
Akima spline fits. Green and black lines represent
eqs.~(\ref{e:tinda}) and (\ref{e:treac}). Solid lines correspond to
the fuel temperature $T_0 = 293$ K, dashed lines to $320$ K. Shaded
gray region shows the distribution of the reaction rate, $\dot{Y}$, in
the exact laminar flame solution normalized by its peak value
$\dot{Y}_{max} = 9.5 \times 10^4$ s$^{-1}$. The vertical dashed line
indicates the region of peak reaction rate. The horizontal dashed line
indicates equality between the induction time and the large-scale eddy
turnover time, $\tau_{ed}$, during which turbulence completely
reorganizes the turbulent flame structure.}
\label{f:tind}
\end{figure}

Fig.~\ref{f:tind} shows the distribution of induction times
$\tau_{ind}$, calculated using the procedure described in
\S~\ref{FA:Speed}, throughout the planar laminar flame. The solid
red line corresponds to the fuel temperature $T_0 = 293$ K and density
$\rho_0$ (see Table~\ref{t:Params}), while the dashed red line was
obtained for the higher fuel temperature $320$ K and the same density.
The second temperature reflects the effect of fuel heating by
turbulence in the course of the simulation (\S~\ref{Simulations}).

For comparison, we also show the adiabatic induction time,
$\tau^a_{ind}$, calculated using eq.~(\ref{e:tinda}) based on the
exact planar laminar flame solutions. Solid and dashed line
correspond, respectively, to the same fuel temperatures as for
$\tau_{ind}$. Overall, throughout the reaction zone, i.e., for $Y =
0.2 - 0.6$, $\tau_{ind}$ and $\tau^a_{ind}$ differ by less than a
factor of $\approx\!2$. Moreover, near the peak reaction rate, i.e.,
at $Y \approx 0.2$, and at the boundary between the reaction and
preheat zones, i.e., at $Y \approx 0.55$, $\tau_{ind}$ and
$\tau^a_{ind}$ become virtually equal. In the preheat zone, however,
they begin to diverge substantially. This is to be expected, since
$\tau^a_{ind}$ considers only self-heating of a fluid element by
chemical reactions and neglects the diffusive transport of heat, which
is dominant in the preheat zone.

Despite the reasonable agreement in the reaction zone between
$\tau_{ind}$ and $\tau^a_{ind}$, $\tau^a_{ind}$ should not be viewed
as a model of $\tau_{ind}$, but rather as a useful approximation. In
particular, Fig.~\ref{f:tind} shows that $\tau^a_{ind}$ does not
capture the correct shape of the distribution of $\tau_{ind}$. The
$\tau_{ind}$ increases more slowly at larger $Y$ than $\tau^a_{ind}$
since slower heating of a fluid element by chemical reactions is
partially compensated by the larger influx of heat due to the thermal
conduction. Since $\tau^a_{ind}$ does not account for this, its rate
of growth with $Y$ is larger in the outer regions of the reaction
zone. At the same time, since energy balance of a fluid element in the
reaction zone is still dominated by self-heating rather than by
thermal transport, $\tau^a_{ind}$ remains close to the actual value of
the induction time given by $\tau_{ind}$.

We also show in Fig.~\ref{f:tind} the characteristic reaction time,
$\tau_{reac}$, defined based on the reaction rate
(eq.~\ref{e:ReacModel}) and neglecting the change in the fuel
concentration,
\beq
\tau_{reac}(Y') = \frac{1}{\rho' B}\exp \Big(\frac{Q}{RT'}\Big).
\label{e:treac}
\eeq
Similarly to $\tau^a_{ind}$, $T'$ and $\rho'$ for a given $Y'$ are
determined from the exact planar laminar flame solutions corresponding
to the fuel temperatures $T_0 = 293$ K (solid line) and $320$ K
(dashed line) and density $\rho_0$. Unlike $\tau^a_{ind}$, which is
typically applicable in the context of an autoignition process, this
quantity is often used to approximate the induction time in the
flame. Nonetheless, for all $Y > 0.15$, $\tau_{reac}$ provides a less
accurate approximation of $\tau_{ind}$ than $\tau^a_{ind}$.

Finally, blue circles show the correlation time lags $\Delta t_c$
calculated in \S~\ref{FA:Speed}. There is close agreement between
$\Delta t_c$ and both $\tau_{ind}$ and $\tau^a_{ind}$. Furthermore,
when fuel heating is taken into account (dashed lines), all three
quantities become even closer.

\clearpage

%-------------------------------------------------------------------------------

%--------------------------------------------------------------------------------
\end{document}